\documentclass[apj]{emulateapj}
\usepackage{epsfig}
\usepackage{appendix}
\usepackage{color, verbatim}
\usepackage[tbtags]{amsmath}
\usepackage{fancyvrb, graphicx, multirow}

\usepackage{hyperref}
\definecolor{darkblue}{rgb}{0,0,0.4}
\definecolor{darkgreen}{rgb}{0,0.5,0}
\hypersetup{colorlinks=true, urlcolor=blue, citecolor=darkblue, linkcolor=darkgreen}

\newcommand{\TRAINSET}{``Training Sets"}
\newcommand{\TRAINMOD}{``Trained Models"}
\newcommand{\TESTSET}{``Test Set"}

\newcommand{\HRK}{H}
\newcommand{\NZJG}{G10$'$}
\newcommand{\realhuge}{REAL}

\newcommand{\cosw}{$w$}
\newcommand{\Gtenpaper}{G10}
\newcommand{\Gtenmodel}{G10}
\newcommand{\Markpaper}{B14}
\newcommand{\Ktw}{K13}
\newcommand{\Nidealtest}{8000 }
\newcommand{\Nideal}{10 }
\newcommand{\Nreal}{20 }
\newcommand{\trainop}{Training Options}

\newcommand{\rawmu}{$\mu_{\rm fit}$}
\newcommand{\finalw}{$w$}
\newcommand{\finalOMM}{$\Omega_M$}
\newcommand{\finalmu}{$\mu$}

\newcommand{\CLresid}{$\Delta CL(\lambda)$}
\newcommand{\SCATTERresid}{$\Delta k(\lambda)$}
\newcommand{\FLUXresid}{$\Delta M_0$}
\newcommand{\TMscatter}{$\sigma_{\mu}$}
\newcommand{\resultw}{$-0.014 \pm 0.007$}

\begin{document}
\title{Cosmological Parameter Uncertainties from SALT-II Type Ia Supernova Light Curve Models}
\author{J.~Mosher\altaffilmark{1},
J.~Guy\altaffilmark{2,3},
R.~Kessler\altaffilmark{4},
P.~Astier\altaffilmark{2},
J.~Marriner\altaffilmark{5},
M.~Betoule\altaffilmark{2},
M.~Sako\altaffilmark{1},
P.~El-Hage\altaffilmark{2},
R.~Biswas\altaffilmark{6},
R.~Pain\altaffilmark{2},
S.~Kuhlmann\altaffilmark{6},
N.~Regnault\altaffilmark{2},
J.~A.~Frieman\altaffilmark{4,5},
D.~P.~Schneider\altaffilmark{7,8}
}

\submitted{ }
\journalinfo{Accepted by ApJ}


\altaffiltext{1}{Department of Physics and Astronomy, University of Pennsylvania, 209 South 33rd Street, Philadelphia, PA 19104, USA}
\altaffiltext{2}{LPNHE, CNRS/IN2P3, Universit\'e Pierre et Marie Curie Paris 6, Universi\'e Denis Diderot Paris 7, 
4 place Jussieu, 75252 Paris Cedex 05, France}
\altaffiltext{3}{LBNL, 1 Cyclotron Rd, Berkeley, CA 94720, USA}
\altaffiltext{4}{Kavli Institute for Cosmological Physics, University of Chicago, 5640 South Ellis Avenue, Chicago, IL 60637, USA }
\altaffiltext{5}{Center for Particle Astrophysics, Fermi National Accelerator Laboratory, P.O. Box 500, Batavia, IL 60510, USA}
\altaffiltext{6}{Argonne National Laboratory, 9700 South Cass Avenue, Lemont, IL 60439, USA}
\altaffiltext{7}{Department of Astronomy and Astrophysics, The Pennsylvania State University, University Park, PA 16802}
\altaffiltext{8}{Institute for Gravitation and the Cosmos, The Pennsylvania State University,University Park, PA 16802}

\begin{abstract}
We use simulated type Ia supernova (SN Ia) samples, including both photometry and spectra, 
to perform the first direct validation of cosmology analysis using the SALT-II light curve model. 
This validation includes residuals from the light curve training process, 
systematic biases in SN Ia distance measurements, and a bias on the dark energy equation
of state parameter $w$. 
Using the SN-analysis package {\tt SNANA}, we simulate and analyze realistic samples 
corresponding to the data samples used in the SNLS3 analysis:
$\sim 120$ low-redshift ($z < 0.1$) SNe~Ia,
$\sim 255$ SDSS SNe~Ia ($z < 0.4$), and
$\sim 290$ SNLS SNe~Ia ($z \le 1$). 
To probe systematic uncertainties in detail, we vary the input spectral model, the model
of intrinsic scatter, and the smoothing (i.e., regularization) parameters used during 
the SALT-II model training. 
Using realistic intrinsic scatter models results in a slight bias
in the ultraviolet portion of the trained SALT-II model, and $w$ biases
($w_{\rm input} - w_{\rm recovered}$) ranging from $-0.005 \pm 0.012$ to $-0.024 \pm 0.010$. 
These biases are indistinguishable from each other within the uncertainty;
the average bias on $w$ is \resultw. 
\end{abstract}

\section{Introduction}\label{sec:Intr}

In 1998, observations of type Ia supernovae (SNe Ia) revealed the accelerating expansion of the universe 
\citep{Riess:1998,Perlmutter:1999}, attributable to an unknown source of acceleration 
commonly called ``dark energy.''
SNe Ia based measurements allow the direct detection of dark energy, 
and as such remain valuable components of the quest to understand this mysterious phenomenon. 
In particular, over the past decade observational cosmology has focused on measuring the 
cosmological equation of state parameter $w$ \citep[e.g.,][]{2013PhR...530...87W}. 

The cosmological utility of SNe Ia is due to their nature as standardizable candles. 
Building on work by ~\citet{Pskovskii:1977}, ~\citet{PhillipsWL1} was the first to demonstrate 
that shapes of SN Ia light curves are correlated with their absolute luminosity. 
An additional correlation between SN Ia color and luminosity was shown by ~\citet{Tripp:WLColor},
resulting in a two-parameter luminosity correction.
Although other standardization methods exist, including infrared light curve 
shapes ~\citep{2012MNRAS.425.1007B, 2012PASP..124..114K} and 
spectral ratios ~\citep[e.g.,][]{2009A&A...500L..17B}, 
the ubiquity of optical SN Ia light curve data makes
shape and color standardization the most common technique. 

SN light curve analysis is the process of empirically
training an SN model with broadband photometry and spectra,
using the model to determine the SN light curve shape and color information, 
and deriving the best possible distance measurements from these SN parameters.
SN light curve analysis is one source of systematic uncertainty affecting our ability to
constrain the nature of dark energy.

A thorough study of systematic uncertainties arising from SN light curve analysis 
has been made by ~\citet{Guy:20103yr} (hereafter \Gtenpaper) 
specifically regarding the training of the SALT-II 
light curve model and its application to the 3-year Supernova Legacy Survey (SNLS) SN Ia 
cosmology sample ~\citep{Conley:SNLS3yrSYSSERR2011}. 
To quantify systematic uncertainties in SN Ia distance measurements, \Gtenpaper ~uses two main techniques: 
calculations and variation analysis. For instance, calculations are used to determine the model 
statistical uncertainties from the training covariance matrix, and variations are used to estimate 
uncertainties due to light curve model training procedures. 

The key results of these tests 
--- distance uncertainties $\sigma_{\mu}$ as a function of redshift ---  are shown in \Gtenpaper  (Figure 16). 
In general, \Gtenpaper ~finds light curve model-related fluctuations about the fiducial result to be small, 
with $\sigma_{\mu}<0.02$ at most redshifts. However, because these results are based on a limited training 
sample and test set, it is not clear that these results would hold in general for any training sample and test set. 
Nor is it possible to determine whether the distances themselves (and the associated cosmology parameter $w$) 
are biased. 

In this paper, we use detailed Monte Carlo (MC) simulations to reevaluate the systematic
uncertainties determined by \Gtenpaper ~for SALT-II and the 3-year SNLS cosmology sample.
In addition, the large statistics afforded by simulations allow us to 
rigorously determine biases in the Hubble Diagram (HD) and $w$ resulting from the full 
SN light curve analysis procedure.  

SN Ia modeling consists of a set of assumptions about the number of observable SN Ia parameters. As mentioned 
above, most SN Ia models assume that the family of SNe Ia may be described by two parameters -- 
light curve shape and color. 
Models define the SN Ia rest frame flux as a function of phase (rest-frame days since peak $B$-band magnitude), 
wavelength, and observable light curve parameters. 
Since SN Ia progenitors and explosion mechanisms remain ill-defined, models are empirically determined and must be ``trained'' 
from a subset of observed SN data for which initial light curve parameter values can be estimated. 
The training procedure consists of solving for the model parameters that best fit the training set of observed SN Ia data. 
For example, training of the magnitude-based MLCS2k2 model ~\citep{Jha2k2} 
includes solving for bandpass-dependent coefficients which 
relate magnitudes to light curve shapes. 
The training of the flux-based model SALT-II ~\citep{Guy:SALT2} includes solving for the 
coefficients of the spline basis functions used to represent SN Ia flux as a function of wavelength and phase.  

Once the best-fit model parameters have been found, the model is ready to be used for light curve fitting. 
For each SN in the data set, the trained model is used to compute synthetic observed magnitudes 
in conjunction with minimization routines to ``fit'' the most likely light curve parameters 
(e.g., stretch $x1$, color $c$, and an overall normalization factor $x0$). 
With the data set light curve parameters in hand, 
distances may be calculated and an HD constructed for cosmological parameter determination. 

Currently, the main source of the SN Ia HD systematic uncertainty is photometric calibration 
~\citep[e.g.,][]{2011ApJ...737..102S}. However, improved low-redshift SN Ia samples
~\citep{Hicken:CfALC, SDSSphot, Max:2011} and greater attention to calibration ~\citep[e.g.,][]{Betoule:2012}
are reducing the significance of this contribution, making it increasingly important 
to understand systematic uncertainties related to the training and implementation of the light curve model itself. 
Because the same trained model is used to fit all SN light curves, 
the model's statistical uncertainties introduce correlated uncertainties into the fitted light 
curve parameters. 

The SN Ia model, training procedure, and training set are all sources of uncertainties 
that affect light curve parameter measurement, and hence cosmology parameter measurement.
One approach to determine model-related systematic uncertainties is to compare cosmology results from the same 
set of SN Ia observations evaluated with different SN Ia models. 
The most notable studies of model-related systematic uncertainties have been undertaken by 
\citet{KesslerSDSScosmo:2009} (K09), comparing the  MLCS2k2 and SALT-II models, 
and \Gtenpaper, comparing SALT-II and SiFTO ~\citep{Conley:SIFTO}. 
K09 found a significant difference between MLCS2k2 and SALT-II derived cosmology parameters, 
which they attributed to the different handling of color in the two models and the fact that 
MLCS2k2 training is much more reliant on observer-frame observations in the ultraviolet region. 
\Gtenpaper ~found that SALT-II and SiFTO produced similar cosmology results. 

A second technique is to estimate light curve model-based distance uncertainties directly from the 
model training itself.   \Gtenpaper ~take this approach to estimate the sizes of systematic distance 
uncertainties caused by 1) finite training samples and 2) the wavelength dependence of the scatter 
between the model and the training data.  As shown in Figure 16 of \Gtenpaper, 
these uncertainties have been estimated as a function of redshift for the SALT-II model.

While these two approaches are used to estimate the model uncertainty, the true model bias cannot be 
determined from SN Ia observations. Therefore it is possible that systematic offsets in the trained model 
parameters may propagate as-yet unknown biases onto the final HD. 
Properly determining these biases is the focus of this work and will be instrumental to correctly 
interpreting SN Ia cosmology data, constructing future training sets, and designing the next generation 
of SN Ia models. 

\subsection{Training Test Overview}

In this work we use simulated SN Ia samples to directly evaluate systematic biases and 
uncertainties originating from the SN Ia light curve analysis, with a specific focus on model training 
and bias corrections.  
We use the SALT-II model exclusively: the most recent, most precise SN Ia cosmology results are 
based on SALT-II light curve fits ~\citep{2011ApJ...737..102S}, making this 
state-of-the-art model an ideal choice for these systematics studies. In addition, its automated 
training process makes it straightforward to generate the multiple training iterations needed
to evaluate biases.

By examining such quantities as training residuals, Hubble residuals, 
and best-fit cosmologies resulting from various training configurations 
and input training sets we will answer three key questions about SALT-II.
First, we will test the ability of the SALT-II framework to determine the model 
uncertainty from input data.
Second, we will quantify the biases in Hubble residuals and cosmology parameters
caused by ``regularization'', the training strategy used to manage phase or wavelength gaps in the training data. 
Third, we will measure biases resulting from mismatches between the underlying 
model and the assumed SALT-II model.\\

\begin{figure*}[t]
\begin{center}
\includegraphics[scale=0.45]{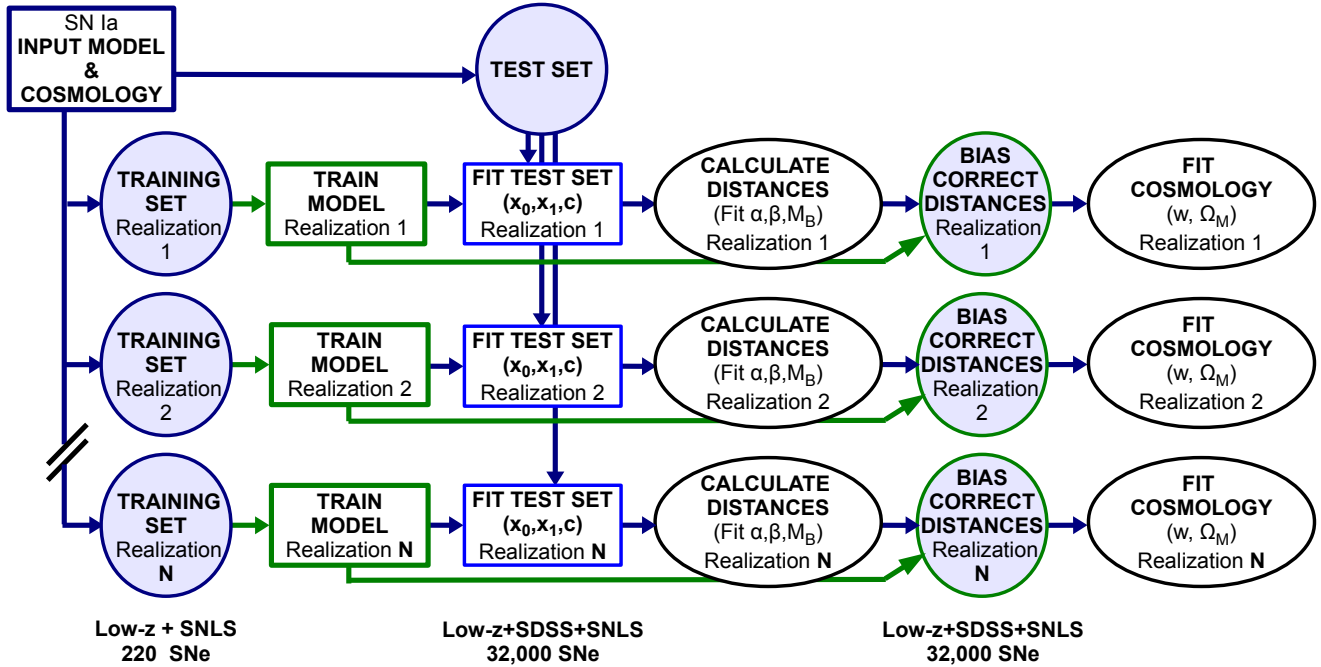}
\caption[../fig/training_schematic.eps]{
Schematic of SALT-II training test pipeline. From an input SN Ia model and cosmology,
$N$ sets of SN Ia training data and a single test set of SN Ia light curves are 
generated (Sections \S\ref{sec:Models}, \S\ref{subsec:SIMSEDSIMS}, \S\ref{subsec:TESTSETSIMS}).
Each set of training data is used to constrain a new SALT-II model (Section \S\ref{sec:TrainS2}). 
Then each new SALT-II model is used to determine test set SNe stretch and color. 
From this information, SALT-II model parameters $\alpha$, $\beta$, and $M_0$,
and test set SNe distances \rawmu ~are determined for each trained model (Section \S\ref{sec:fitting}).
The recovered test set stretch and color parameter distributions are used in 
conjunction with the trained models to generate and fit one bias correction set 
for each model (Section \S\ref{sec:MBCORR}).  
Using the bias corrections, corrected test set distances \finalmu ~are 
calculated for each model. The resulting HDs are fitted to obtain $N$ sets of best-fit 
cosmology parameters \finalw ~and \finalOMM.
}
\label{fig:trainschema}
\end{center}
\end{figure*}

Figure~\ref{fig:trainschema} provides an overview of the process this work 
uses to test SALT-II training.
 
From an SN Ia input model and cosmology, $N$ statistically independent realizations of SN Ia 
training sets (light curve photometry and spectra) are generated.
These observations (\TRAINSET) are used to train $N$ new SALT-II models 
(\TRAINMOD).
Finally, each trained model is used to fit a single set of SN Ia light 
curves (\TESTSET) generated from the SN Ia input model, 
resulting in $N$ sets of light curve parameters ($x_{0}, x_{1}, c_i$). 

We have adapted existing SN Ia MC simulation routines from the SN analysis 
software
package {\tt SNANA} ~\citep{SNANA} to enable accurate random realizations 
of both the light curve photometry and spectral data samples forming the basis of the most recent published 
SALT-II model training. There are two key parts to this change. First, we have added a spectral simulation component,
which uses an SN Ia  spectral energy distribution (SED)
model and a library of existing SN Ia spectral observations to produce spectra with realistic
signal-to-noise ratios, resolution, cadences, wavelength coverage, and galaxy contamination. Second, 
we have added several new intrinsic scatter models (described in \Ktw), as well as the machinery to
apply them to simulated light curves and spectra. 

In addition to measuring biases introduced by the training,
our multiple training set realizations are used to check estimates of the statistical uncertainty of 
SALT-II model parameters due to the finite sizes of existing training sets
 and directly measure uncertainties resulting from different choices of training parameters. 

Most importantly, because the underlying cosmology and parameter distributions of the 
simulated training sets are known, we are able to perform complete light curve analyses for each set 
of input models, from model training through HD construction and cosmology fits. 
We use this information to evaluate HD bias as a function of redshift, 
and biases in \cosw.
This is the first systematic study in which simulations are used to directly test uncertainties and biases 
resulting from SN Ia light curve analysis. The software used to run these light curve analysis tests
is part of the publicly available {\tt SNANA} supernova software package; these tools can be
adapted to test other SN Ia models and data samples. 
 
The outline of the paper is as follows.
A brief introduction to the SALT-II model and its training procedure  will be given in Section~\ref{sec:TrainS2}. 
Section~\ref{sec:Models} details the input SN Ia models which will be used for this work, 
followed by a discussion of our simulations and data analysis procedures, including redshift-dependent bias
corrections, in Section~\ref{sec:SimDet}. 
Section~\ref{sec:anal} describes the metrics with which training results will be evaluated. 
Section~\ref{sec:RESULTS} presents results of our two main training tests:
trainings with idealized training sets 
and trainings with realistic training sets.
Finally, the implications of the results of these tests for further SALT-II training, 
light curve model development, and training set observations are described in Section~\ref{sec:Disc}.

\section{Training SALT-II}\label{sec:TrainS2}

\begin{figure}
\includegraphics[scale=0.30]{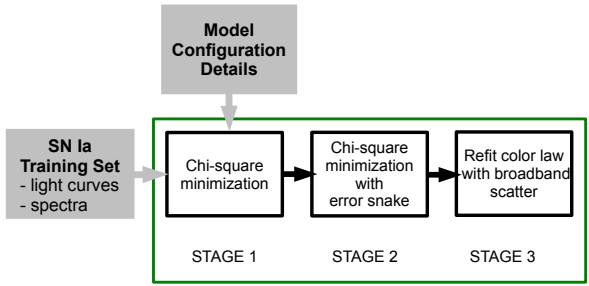}
\caption[../fig/pcafit_schematic.pdf]{
The three stages of the SALT-II model training process. 
Each stage calculates best-fit model 
parameters using successively improved estimates of model
uncertainties. 
}
\label{fig:pcafitschema}
\end{figure}

The SALT-II model describes SNe Ia with three components: two spectral time series $M_0$ and 
$M_1$, and a color law CL. 
The component $M_0$ is identified as the mean SN Ia  SED; 
component $M_1$ accounts for light curve width variations. The color law CL
incorporates any wavelength-dependent color variations that are independent of epoch. 
No assumptions about dust or extinction laws are made a priori.
These three components are assumed to be the same for all supernovae. 
The flux for an individual SN Ia is determined by the components 
described above
and three supernova-specific parameters: the overall flux scale $x_0$,
the light curve shape parameter $x_1$,  and the peak $B-V$ color $c$.
As a function of phase $p$ and wavelength $\lambda$ the flux is:
\begin{equation}
\protect \label{eq:S2function}
F(p,\lambda)=x_0~[M_0(p,\lambda)+x_1 M_1(p,\lambda)]~exp[c~{\rm CL}(\lambda)].
\end{equation}
More details on the SALT-II model may be found in ~\citet{Guy:SALT2} 
and \Gtenpaper. A cartoon schematic of the SALT-II training process is shown in 
Figure \ref{fig:pcafitschema}. The SALT-II model parameters and $\chi^2$ are
presented in Sections~\ref{sec:ConfigS2}-\ref{sec:pcafitreg}. 
Descriptions of the additional model uncertainties used to account for 
SN Ia intrinsic variability are given in Section~\ref{sec:ivar}.  

\subsection{SALT-II Model Parameters}\label{sec:ConfigS2}

Defining $\mathbf{M}$ to be the set of model parameters, 
solving for the best-fit model consists of varying $\mathbf{M}$
to minimize the chi-squared ($\chi^2$) corresponding to the difference between the 
training set SN fluxes, $f$, and corresponding 
model fluxes, $f_{\mathbf{M}}$. 
Training set fluxes obtained from spectra,$f_s$, 
are compared to model fluxes of the same wavelength, 
$\lambda_s$, and phase, $p_s$, 
adjusted for SN-appropriate recalibration parameters, $\mathbf{y_s}$.  
Training set fluxes obtained from broadband photometry, $f_p$, 
are compared with the integration of the model at phase $p_p$ 
over the appropriate broadband filter, $R_p$. 
In either case, the model is redshifted to 
match the redshift $z$ of the training set SN prior to flux comparison. 

The training $\chi^2$ is:
\begin{equation}
\protect \label{eq:modelchisq}
\begin{split}
\chi^2 = &\sum_{i=1}^{N_{SN}}\biggl[
\sum_{s=1}^{S_{\rm obs}} \left(\frac{(f_{s,i} - f_{\mathbf{M}}(\mathbf{x_i},z_i;p_s,\lambda_{s},\mathbf{y_s};\mathbf{M}))^2}{(\sigma_{i,s}^D)^2+(\sigma_{i,s}^M)^2}\right)
\\
& + \sum_{p=1}^{P_{\rm obs}} \left(\frac{(f_{p,i} - f_{\mathbf{M}}(\mathbf{x_i},z_i;p_p,R_{p};\mathbf{M}))^2}{(\sigma_{i,p}^D)^2+(\sigma_{i,p}^M)^2}\right) 
\biggr]\\
&+ \chi^2_{\rm REGUL}.
\end{split}
\end{equation}

Here $\mathbf{x_i}$ is the set of observable light curve parameters $x_0$, $x_1$, $c$ for the $i^{th}$ supernova, 
$\mathbf{y_s}$ is the set of recalibration parameters applied to the spectral flux $f_s$, 
$\sigma_{i,p}^D$ and $\sigma_{i,s}^D$ are the statistical uncertainties of the training SN fluxes, and 
$\sigma_{i}^M$ is the uncertainty in the model flux. 
An extra regularization term $\chi^2_{\rm REGUL}$, to be discussed
in Section~\ref{sec:pcafitreg}, is also included in the training $\chi^2$.
The summations are over all $S_{\rm obs} + P_{\rm obs}$ spectral and photometric observations of all 
$N_{\rm SN}$ training set SNe. 

Third-order b-spline basis functions are used to construct the $M_0$ and $M_1$
spectral time series. 
Closely resembling Gaussians, third-order b-spline basis functions are parameterized 
by knot values, which determine the local region of phase (or wavelength) space to which each 
basis function contributes, and control points, which determine the amplitude of each 
basis function. 
Only the control points are varied during the SALT-II $\chi^2$ minimization process.
The phase range $-$14 to $+$50 rest-frame days is spanned by 20 basis functions for
a phase resolution of 3.2 days. The wavelength range 2000 to 9200 \AA ~is spanned
by 100 splines, or 72 \AA ~per basis function. 
As in \Gtenpaper, the functional form of the color law is a polynomial with four free coefficients 
covering the wavelengths 2800 -- 7000 \AA. The best-fit color law is linearly extrapolated for 
wavelengths outside these regions. 

Table \ref{table:params} presents the total number of parameters used in a typical SALT-II 
training. Spectral recalibration parameters $\mathbf{y_s}$ are described in 
Section~\ref{sec:srecal}; the error snake and broadband scatter parameters $\mathbf{S}$ and 
$\mathbf{a}$ used to account for intrinsic SN Ia variability will be described in 
Section~\ref{sec:ivar}.

\begin{center}
\begin{deluxetable}{llc}[h]
\tablewidth{200pt}
\tabletypesize{\normalsize}
\tablecaption{Model Parameters\label{table:params}}
\tablehead{
  \colhead{Type}&
  \colhead{Origin}&
  \colhead{Number}
}
\startdata
& $M_0$ & 2002\\
$\mathbf{M}$ & $M_1$ & 2002\\
& CL & 4 \vspace{0.15cm}\\
& SN $x_0$ & $N_{\rm SN}$\\
$\mathbf{x}$ & SN $x_1$ & $N_{\rm SN}$\\
& SN $c$ & $N_{SN}$ \vspace{0.25cm}\\
$\mathbf{y}$ & spectral recalibration & $\propto \sum_{i=1}^{N_{\rm SN}} S_{i,\rm obs}$ \vspace{0.25cm}\\
$\mathbf{S}$ & $x_1$-related scatter & 60 \vspace{0.15cm}\\
$\mathbf{a}$ & $c$-related scatter & 4 \vspace{0.05cm}\\[-5pt]
\enddata
\parbox{6in}{\tablecomments{Number of model
parameters for a SALT-II training with $N_{\rm SN}$
input SNe. Parameters $\mathbf{M}$ and $\mathbf{x}$
are described in Section~\ref{sec:ConfigS2};
$\mathbf{y}$ in Section~\ref{sec:srecal}; 
$\mathbf{S}$ and $\mathbf{a}$ in section \S\ref{sec:ivar}.
}}
\end{deluxetable}
\end{center}

\subsection{Spectral Recalibration}\label{sec:srecal}
In general, SN Ia spectra are not as well calibrated as SN Ia photometry and spectral
flux uncertainties are not uniformly calculated. In order to maximize the benefits of
training with heterogeneous spectroscopic data, training set spectra are recalibrated to 
match the best-fit model of their input light curve data. 
The recalibration is performed by multiplying each input spectrum by the exponential of a polynomial whose 
order is determined by the wavelength range of the spectrum and the wavelength coverage of input light curve data.
As noted in Equation~\ref{eq:modelchisq} and Table~\ref{table:params}, the polynomial coefficients for each spectrum 
$\mathbf{y_s}$ are part of the best-fit model parameters and are determined iteratively during the minimization
process. 

\subsection{Regularization}\label{sec:pcafitreg}

If a region of wavelength and phase lacks spectral training data, the corresponding $M_0$ and $M_1$ components
are determined by the underlying spline interpolations and constrained solely by broadband photometry. 
The spline interpolation can produce high-frequency noise or ``ringing'' in the poorly constrained 
region. 
Extra ``regularization'' terms $\chi^2_{\rm REGUL}$ are added to the training $\chi^2$ (Equation~\ref{eq:modelchisq}) 
to disfavor models with unphysical rapid fluctuations in wavelength or phase. 
However, regularization can also oversmooth, thereby removing real features from the best-fit model.

As described in the Appendix of \Gtenpaper, two types of regularization are used for the SALT-II model:
gradient and dyadic. Gradient regularization $\chi^2_{\rm GRAD}$ penalizes changes in flux with 
respect to wavelength, but ignores changes with respect to phase:
\begin{multline}
\chi^2_{\rm GRAD} = \sum_{i=1}^{N_{\rm phase}}\sum_{j,l=1}^{N_{\lambda}}\frac{A_{\rm GRAD}}{n(p,\lambda)}\times\frac{1}{|\lambda_l - \lambda_j|}\times\\
[f_{\mathbf{M}}(p_i,\lambda_l) - f_{\mathbf{M}}(p_i,\lambda_j)]^2, 
\end{multline}
where $A_{\rm GRAD}$ is the gradient regularizaton weight, $n(p,\lambda)$ is the effective
number of spectral flux measurements constraining that region of phase space, 
and the sums run over all $N_{\rm phase}$ phase and $N_{\lambda}$ wavelength basis functions. 

SALT-II dyadic regularization $\chi^2_{\rm DYAD}$ is implemented as follows:
\begin{multline}
\chi^2_{\rm DYAD} = \sum_{i,k=1}^{N_{\rm phase}}\sum_{j,l=1}^{N_{\lambda}}\frac{A_{\rm DYAD}}{n(p,\lambda)} \times \frac{1}{|\lambda_l - \lambda_j||p_k - p_i|}\times\\
[f_{\mathbf{M}}(p_k,\lambda_l)f_{\mathbf{M}}(p_i,\lambda_j) - f_{\mathbf{M}}(p_k,\lambda_j)f_{\mathbf{M}}(p_i,\lambda_l)]^2, 
\end{multline}
where $A_{\rm DYAD}$ is the dyadic regularization weight, $n(p,\lambda)$ is the effective
number of spectral flux measurements constraining that region of phase space,
and the sums run over all $N_{phase}$ phase and $N_{\lambda}$ wavelength basis functions. 
Dyadic regularization favors model fluxes that can be decomposed separately as a function of phase 
times a function of wavelength 
(i.e., $f_{\mathbf{M}}(p_k,\lambda_l) = f_{\mathbf{M}}(p_k) ~f_{\mathbf{M}}(\lambda_l)$).

These terms are combined in the training $\chi^2$:
\begin{equation}
\chi^2_{\rm REGUL} = \chi^2_{\rm GRAD} + \chi^2_{\rm DYAD}. 
\end{equation}

\begin{figure}[h!]
\begin{center}
\includegraphics[scale=0.35]{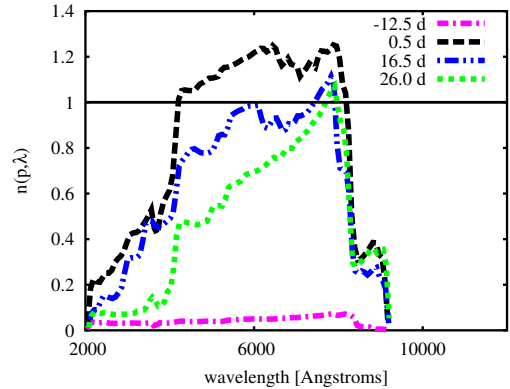}
\caption[]{
Effective number of spectral flux measurements constraining the 
\Gtenmodel ~SALT-II model shown as a function of wavelength. 
Four different phases are displayed: $-$12.5~days (dash-dotted line), 
0.5~days (dashed line), 16.5~days (dash-dot-dotted line), and 26.0~days(dotted line).
The solid line shows the value above which regularization weight 
$W(p,\lambda)$ is set to zero.}
\label{fig:regmasking}
\end{center}
\end{figure}

To limit the use of regularizations to those regions that need it the most, 
we set a cutoff value for $n(p,\lambda)$: in regions where $n(p,\lambda) \geq 1 $, 
$\chi^2_{\rm REGUL} = 0$. 
Figure~\ref{fig:regmasking} shows $n(p,\lambda)$ as a function of wavelength for four 
different phase values.  Near the epoch of peak luminosity, regularization is not used for 
wavelengths between 4000 and 8000~\AA. On the other hand, most wavelength bins 
are regularized at early and late times and wavelengths less than 4000~\AA ~are 
always regularized.

The overall strengths of the regularization terms are controlled
by the scaling factors $A_{\rm GRAD}$ and $A_{\rm DYAD}$. 
Nominal values for these parameters  are 
$A_{\rm GRAD} = 10$ and $A_{\rm DYAD} = 1000$. 

\begin{figure}
\includegraphics[scale=0.40]{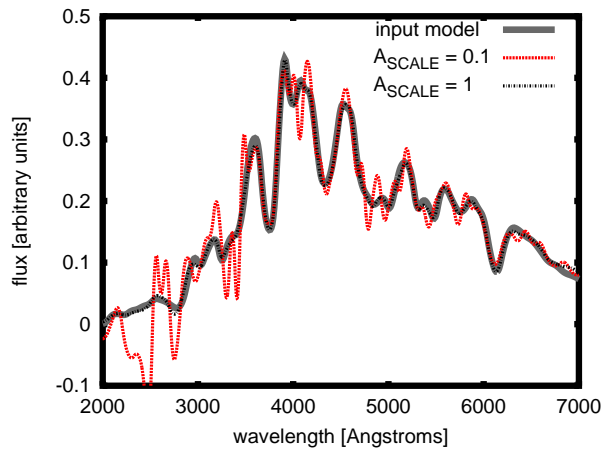}
\caption[../fig/extreme_regexamp.eps]{ 
Best-fit $M_0$ flux (phase=5~days) as a function of wavelength is shown for two different values of 
$A_{\rm SCALE}$: 1.0 (dashed line) and 0.1 (dotted line). For reference, the input model
is shown by the thick solid line. This example is for illustration only:
to achieve ringing of this magnitude, it was necessary to omit all training spectra
between phases 2~d and 8~d prior to running the SALT-II training. 
}
\label{fig:EXTRREGEXAMP}
\end{figure}

As an illustration of the importance of the scaling factor values, 
Figure~\ref{fig:EXTRREGEXAMP} demonstrates the results of model trainings 
where $A_{\rm GRAD}$ and $A_{\rm DYAD}$ are multiplied by an amount $A_{\rm SCALE}$. 

First, we generated a training set and removed all training spectra 
between phases 2 and 8. Then we ran the nominal SALT-II training twice:
once with $A_{\rm SCALE} = 1$ (i.e., nominal regularization), and
once with $A_{\rm SCALE} = 0.1$ (i.e., ten times weaker regularization).  
With $A_{\rm SCALE}=0.1$ the best-fit SALT-II model shows high-frequency oscillations 
around the input model which increase in amplitude toward the bluest wavelengths. 
When $A_{\rm SCALE}=1$,
the best-fit SALT-II model is almost identical to the input model. 
Only slight differences in feature strength are evident (e.g. 3000 \AA).

The choices of regularization type, scaling factor, and effective number threshold setting are 
discussed further in Section~\ref{sec:Regularization}.

\subsection{Intrinsic Variability}\label{sec:ivar}
To account for intrinsic variability in the SN Ia population 
the SALT-II model formalism incorporates two additional model uncertainties: 
a color-independent error scaling factor $S(p,\lambda)$ 
(hereafter ``error snake'') associated with the spectral time 
series and the $x_1$ parameter, and a broadband magnitude scatter 
$k(\lambda)$ associated with the color law and the $c$ parameter.
As illustrated by the training schematic shown in Figure \ref{fig:pcafitschema}, 
the calculation and subsequent incorporation of these added uncertainties 
require three successive rounds of chi-squared minimization before
the final model uncertainty $\sigma^{M}$ is 
obtained.\footnote{For sake of readability,
in this section we have omitted the SN-specific indices $i,s$ and $i,p$.}

After the model best-fit $M_0$ and $M_1$ basis parameters have been 
determined (``Stage 1'' in Figure \ref{fig:pcafitschema}) 
the error snake
is calculated by requiring light curve training data to have reduced chi-squared values of 1 
in any given phase-wavelength bin. 
The default error snake bin size is (6~days) $\times$ (1200~\AA), 
but it is increased if there are less than 10 data points in each bin.
Once the error snake has been calculated, the Stage 1 model uncertainty 
$\sigma_{1}^{M}$ is redefined as 
\begin{equation}
\sigma_{2}^{M} = S(p,\lambda) \times \sigma_{1}^{M}
\end{equation}
and the chi-squared minimization is repeated (``Stage 2'' in Figure \ref{fig:pcafitschema}).
It should be noted that the definition above, 
which has the error snake scaling the rms, is slightly different from that of 
\citet{Guy:SALT2} where the error snake scales the model variance.

A similar process takes place for the color law. 
The SALT-II model formalism assumes a phase-independent
broadband magnitude scatter which can be described as 
a function of the central filter wavelength 
$\lambda$ and a set of parameters $\mathbf{a}$ such that
\begin{equation}
\protect \label{eq:dispfunc}
k(\lambda) \equiv exp\left(\sum_{i=1}^{4} a_i \lambda^i\right).
\end{equation}
The best-fit scatter parameters $a_i$ are determined from the Stage 2 
light curve residuals and uncertainties. 
During this calculation, one free normalization parameter is
included for each SN and therefore the coherent contribution to 
intrinsic scatter is not included in $k(\lambda)$.

Subsequently, the model uncertainty is 
redefined to include $k(\lambda)$, such that the 
covariance of two measurements $i$ and $j$ of the same light curve is 
\begin{equation}
C_{ij}^{M} = \left(\sigma_{2}^{M}\right)_i^2 \delta_{ij} + 
k^2(\lambda; \mathbf{a}) ~f_{\mathbf{M},i} ~f_{\mathbf{M},j}
\end{equation}
and the final color law is fit (``Stage 3'' in Figure \ref{fig:pcafitschema}). 

One of the goals of this work is to test the ability of these model 
uncertainties to account for model mismatches. Results of these tests 
are described in sections Sections~\ref{sec:TCideal} and \S\ref{sec:TCreal}.

\subsection{Final $x_1$ distribution}\label{sec:traindistrib}
As discussed in Section 5 of ~\citet{Guy:SALT2}, the average value and scale of the
SALT-II stretch parameter $x_1$ are arbitrary. Therefore, the training rescales 
fitted $x_{1,i}$ values to satisfy the distributions $\langle x_1 \rangle =0$ and $\langle x_1^2 \rangle = 1$.
Although all simulations are performed with global stretch parameter $\alpha=0.11$,
this rescaling changes the expected value of $\alpha$. Analytic calculations of the
expected $\alpha$ values as a function of input model are described in Appendix~\ref{appxalpha}.

\section{SN Ia input models}\label{sec:Models}

As illustrated in Figure~\ref{fig:trainschema}, all training tests begin with the selection of a SN Ia input model. 
For this work three different input models have been used:
the SALT-II surfaces published in \Gtenpaper, 
a specially designed variant of the SALT-II surfaces published in \Gtenpaper, 
and a model based on the Hsiao templates \citep{Hsiao:2007}.
In Sections~\ref{sec:G10model}-\ref{sec:HRKmodel} we describe the motivation for the 
use of each input model and provide a general description of how the models were made. 
To match the observed intrinsic scatter, three intrinsic scatter models are combined with each
of our input models. Section~\ref{sec:scattermodels} provides details of the 
intrinsic scatter models we have selected. Finally, Section~\ref{sec:naminguses}
summarizes the model naming conventions used in this work. 

\subsection{\Gtenmodel ~Model}\label{sec:G10model}

This model is built from the \Gtenmodel ~spectral surfaces and the \Gtenmodel ~color law, 
combined as prescribed in Equation~\eqref{eq:S2function} for a range of stretch and color values. 
Because it contains some negative fluxes, especially at early times in the UV region and late times in the near-IR, 
this model is not suitable for our training purposes; however, it is used for internal cross checks. 

\subsection{\NZJG ~Model}\label{sec:NZJG}
A second model, hereafter \NZJG, uses the \Gtenmodel ~surfaces as a starting point. 
To ensure full SALT-II training compatibility, these surfaces were modified 
to remove negative fluxes, and then translated back onto the desired b-spline training basis.
Equation~\eqref{eq:S2function} was used to combine the output surfaces with the 
\Gtenmodel ~color law to add color variations. 

\subsection{\HRK ~Model}\label{sec:HRKmodel}
The third type of model we use is based on spectral templates presented in \citep{Hsiao:2007}, 
which we will refer to as the ``Hsiao templates.''
We use the Hsiao templates because they have real spectral features without being imprinted
by an underlying set of basis functions (as is the \NZJG ~model)
making this model a useful test of the ability of SALT-II training to reproduce arbitrary 
SN Ia models. 
We have added width variation by applying a stretch function to the flux, 
such that a supernova with stretch $s$ and peak time $t_0$ has flux at epoch $t$ and wavelength $\lambda$ 
\begin{equation}
F(t,t_0,s,\lambda) = F(t',t_0,s=1,\lambda),  
\end{equation}
where $t' = (t-t_0)/s$. 
As in the prior two models, the \Gtenmodel ~color law was used to add color variation, and
the same input range of $x_1$ and $c$ values was 
utilized\footnote{From ~\citet{Guy:SALT2}, $s= 1.00 + 0.091*x1 + 0.003*x1^2 -7.5e^{-4}*x1^3$}.

\subsection{Intrinsic Scatter Models}\label{sec:scattermodels}
Fits of SNe Ia data typically include an intrinsic scatter term on the order of $\sim 0.15$ mag, 
representing the amount of extra uncertainty required to be added in quadrature such that a fit
to a SN Ia derived HD obtains a reduced chi-squared value of $\sim  1$. 
Historically, this extra uncertainty has been associated with the peak $B$-band magnitude parameter $M_B$.
However, several recent papers have presented evidence suggesting that the SN Ia scatter varies 
as a function of wavelength, producing not just magnitude scatter but also color scatter 
\citep{Guy:20103yr, FoleyKasenHighV:2010, 2011A&A...529L...4C, 2011ApJ...740...72M, MAGSMEAR, 2013arXiv1306.4050S}
and that the assumptions we make about this scatter affect extinction laws and cosmology derived from our SN data sets. 
The underlying source of this scatter is a subject of ongoing study. 
In this work, we use intrinsic scatter models from ~\citet{MAGSMEAR} (hereafter \Ktw) to 
test the effects of different intrinsic scatter forms on SALT-II training. 
All of the \Ktw ~models are defined as wavelength-dependent perturbations to an underlying input model. 
By design, these perturbations average to zero so that the underlying SN Ia model is not changed.
Although all models used in this work are independent of redshift and epoch, 
the validity of this assumption is not well constrained with current data and 
remains an active area of research. 
The chosen models, COH, G10, and C11, are briefly described below.
We also employ a model with no intrinsic scatter, which we refer to as ``NONE.''
As this model is trivial, we omit a description.  
Combining these intrinsic scatter models with our base \HRK ~and 
\NZJG ~models provides eight total input models to choose from. 

\subsubsection{COH}
Coherent magnitude shifts are the simplest form of scatter to add to our base model. 
For each SN, a shift in flux $s$ is chosen from a Gaussian distribution of mean zero and standard deviation
equal to the desired scatter ($\sigma_{\rm COH}=0.13$ for this work). The underlying SN Ia flux is multiplied
by 1+$s$, then spectra and magnitudes are computed as usual. This scattering model produces a coherent shift
in magnitudes; as such it only changes the SN Ia parameter $M_B$ and produces no change in colors.

\subsubsection{G10}\label{subsub:Gtenscatter}
This scatter model is based on the wavelength-dependent magnitude dispersion measured from the \Gtenpaper ~SNLS3 training set during the SALT-II training process (Fig. 8 of \Gtenpaper). 
To create a perturbation that can be applied to our input models, 
independent scatter values are chosen from the measured dispersion at 800~\AA ~wavelength intervals 
and then  joined by a sine interpolation. 
This scattering model produces color scatter as well as magnitude scatter. 

\subsubsection{C11}
Our last scatter model is based on a covariance magnitude scatter model from the \citet{2011A&A...529L...4C} analysis of 
high-quality Nearby Supernova Factory spectra. The original scatter model was extended to the UV by \Ktw: we use the 
C11\_0 variant. 
As with the \Gtenmodel ~scatter model, the C11 scatter is determined for broadband 
wavelengths. To apply it directly to our input models, six correlated random magnitude shifts 
[$U'UBVRI$] are drawn from the covariance matrix
and then connected with the same sine interpolation used for the \Gtenmodel ~model. 
In contrast to the two previous models, the C11 model produces mostly color scatter with minimal magnitude scatter.

\subsection{Input Model Naming Conventions}\label{sec:naminguses}
To be clear about which model is being used to train SALT-II, 
a compound name will be used to describe the input. 
For example, an input data set created with the \NZJG ~base model altered by  G10 scatter
will be called \NZJG--G10, and an input data set created with the \HRK ~base model altered by COH scatter
will be called \HRK--COH. Two additional components are appended to the input model name to signify the types 
of simulations used (IDEAL or REAL, described in Section~\ref{sec:TCideal} and \ref{sec:TCreal}) 
for the training set and test set simulations. 
Table~\ref{table:trainopts} summarizes the training option naming conventions used in this work. 

\begin{center}
\begin{deluxetable}{cccc}
\tablewidth{0pt}
\tabletypesize{\normalsize}
\tablecaption{Training Options\label{table:trainopts}}
\tablehead{
  \colhead{input model}&
  \colhead{scatter model}&
  \colhead{training set}&
  \colhead{test set}
  }
\startdata
\HRK & NONE & IDEAL & IDEAL \\
\NZJG & COH & REAL & \realhuge \\
 & C11 &   &   \\
 & G10 &   &  
\enddata
\parbox{4in}{\tablecomments{Main training options used in this work. 
Particular training tests will be identified by a concatenation of options.
For instance, a training test based on the \HRK ~input model with C11 intrinsic scatter 
trained with an ideal training set and tested on a realistic test set
would be designated \HRK-C11-IDEAL-\realhuge.}}
\end{deluxetable}
\end{center}

\section{Simulation and Analysis Methods}\label{sec:SimDet}
As illustrated in Figure~\ref{fig:trainschema}, using a set of SN Ia light curves 
and spectra to measure cosmology parameters ~is a multi-step
process. In our case, we begin with an SN Ia model (``INPUT MODEL'') from which we
simulate $\mathbf{N}$ SN Ia training set realizations and a single SN Ia test set. 
These simulations are the only aspect in which our light curve analysis technique differs
from analyses based on real data. Following the order of
Figure~\ref{fig:trainschema} from left to right, five steps are implemented:
(1) the training set is used to train a SALT-II light curve model; (2) the trained
light curve model is used to fit the cosmology test set, determining the object-specific 
model parameters $x_0$, $x_1$, $c$ for each SN in the set; (3) the ensemble of fitted
SN parameters is used to determine global SN Ia parameters $\alpha$, $\beta$, and $M_B$.
These global parameters are used to calculate the initial SN distances \rawmu.
(4) Redshift-dependent bias corrections are determined and applied to inital distances
in order to obtain final distances \finalmu.
(5) Cosmology parameters are determined by fits from the HD. 
In the following sections we describe the methods used for each of these steps.

\subsection{Simulation Overview}\label{sec:simoverview}
As shown by the shaded circles in Figure ~\ref{fig:trainschema}, three different sets 
of SN Ia simulations are generated as part of our training test procedure: 
(1) the model training sets, which match real data statistics; 
(2) the high-statistics SN Ia test set, which represents the data set from
which cosmological parameters are measured; 
(3) the high-statistics bias correction sets. 
This section provides a brief overview of our simulations. 
We begin by describing those methods shared by the three sets of simulations,
including general simulation techniques and the SN Ia data samples after which our 
simulations are patterned. Then, we move through our training and test set simulations, 
describing methods specific to each. Section~\ref{subsec:LCSIMS} presents more information
on our bias correction simulations. 
We close by outlining the tests we have performed to evaluate our simulations. 

\subsubsection{General Techniques}\label{sec:gensimstuff}
All of our simulations are performed with the {\tt SNANA} MC code ~\citep{SNANA}, 
and all are based on a flat $\Lambda$CDM cosmology with $\Omega_k=0$, 
$w = -1$, $\Omega_M = 0.3$, and $\Omega_{\Lambda} = 0.7$. 
SN Ia models may be constructed analytically from a set of model parameters 
and associated stretch and color parameters (e.g. the SALT-II model, Equation~\ref{eq:S2function}), 
or from a spectral time series lookup table.\footnote{The main advantage of a lookup table 
is to allow arbitrary flexibility in the choice of SN model.} 
Both techniques are used in this work. More details of the latter
technique will be given in Section~\ref{subsec:SIMSEDSIMS}. 

For a given SN Ia simulation, we begin by selecting stretch and color parameter values.
Parent distributions of the stretch ($x_1$) and color ($c$) parameters are chosen to be
asymmetric Gaussians characterized by a mean value and two standard devations $\sigma_-$ and
$\sigma_+$. We choose to use the same stretch and color distributions for all simulated surveys
within a given training test, rather than use the separate distributions for each survey that
best match the observed data (e.g. \Ktw, Section 3). 
This choice is made to simplify the interpretation of our training test results. 

The goals of this work --- to verify the systematic uncertainty estimates of the SALT-II training 
procedure, to test the impact of regularization on HD biases, and to measure training-related
HD bias as a function of redshift --- will be satisfied so long as the simulations provide an 
approximate representation of the data. Comparisons between data and simulations will be 
presented in Section~\ref{subsubsec:LCSIMS}.

Once the color and stretch are selected, the SN Ia spectra are generated. 
All of our SN Ia models are based on time sequences of rest-frame spectra. 
To generate observer-frame magnitudes for a specific epoch, the appropriate rest-frame spectrum 
is redshifted and its flux is integrated over the appropriate filter-response curves. 
To account for non-photometric conditions and varying time intervals between observations
due to bad weather, actual observing conditions are used when available. 
For each simulated observation, the noise is determined from the measured 
point-spread function (PSF), Poisson noise 
from the source, and sky background. Contamination from the host-galaxy background is included as
needed. The simulated flux in CCD counts is based on the observed mag-to-flux zero points 
and a random fluctuation drawn from the noise estimate. 
When available, realistic models of survey search efficiencies are used to determine 
whether a particular simulated SN Ia epoch would be observed. 
More details of {\tt SNANA} light curve simulations can be found in ~\citet{SNANA}
and in Section 6 of ~\citet{KesslerSDSScosmo:2009}. 

\subsubsection{The SN Ia Data Samples}\label{subsec:SNIadatasets}
To generate simulated samples similar to the real data used in \Gtenpaper ~
we pattern our simulations after three key data sets: 
the SDSS-II sample, 
the SNLS3 sample, 
and the nearby sample.

\paragraph{SDSS-II}
The SDSS-II SN survey discovered and spectroscopically confirmed $\sim$500
type Ia SNe during its three fall operating seasons from 2005--2007 ~\citep{FriemanSDSSTECH:2008}. All 
data were acquired on the SDSS 2.5~m telescope ~\citep{SDSStelescope} 
with the SDSS camera ~\citep{SDSScamera} 
and $ugriz$ filters ~\citep{1996AJ....111.1748F, Doi:2010}.
This set of homogenous observations bridges the gap in redshift  between
nearby and high $z$ SN Ia samples, 
making it interesting for both cosmology and light curve training. 
Contamination from host-galaxy background is observed to have a small effect on low-redshift
SDSS SNe; thus we include it in our simulations. Although light curves have 
only been released for 146 of these SNe ~\citep{SDSSphot}, 
our simulated SDSS-II data use cadence and imaging noise information drawn from 
all three seasons of SDSS-II observing conditions. 
The entire SDSS-II SN light curve data set is published in ~\citet{SDSS3yeardata}.

\paragraph{SNLS 3-year}
The SNLS 3-year data set ~\citep{Conley:SNLS3yrSYSSERR2011}
consists of 279 SN Ia discovered and spectroscopially confirmed 
during the first three years of the SNLS survey. 
All data were acquired on the 3.6~m Canada--France--Hawaii 
Telescope (CFHT) using the MegaCam imager with $g_M r_M i_M z_M$ filters. 
These homogeneously observed and reduced light curves cover the 
redshift range from $\sim 0.2$--$1.0$ . 
The photometric noise contribution from the host galaxy is negligible for this
sample and has not been included in our simulations. 

\paragraph{Nearby}
To approximate the \Gtenmodel ~SALT-II training, our simulation's
nearby sample is patterned after the Cal\'an/Tololo SN Ia survey \citep{1996AJ....112.2408H}
and the CfA monitoring campaign ~\citep{1999AJ....117..707R, Jha:CfA2006, Hicken:CfALC}. 
These data cover the redshift range $0.02 - 0.13$ and have been reported either in the
Landolt system ~\citep{Landolt:1992} or in the native system. 
Because the bulk of the ~\Gtenmodel ~training spectra come from the 
Cal\'an/Tololo survey and the ~\citet{Jha:CfA2006} compilation, we simplify our simulations
by considering only these components, using \citet{1990PASP..102.1181B} 
$UBVRI$ filters. Nearby-SN statistics approximate those of the \Gtenpaper
~data: simulated nearby SNe make up half of the training set and one-fifth of the test set.

Since we do not have the observing conditions (mainly PSF and sky noise) needed for the {\tt SNANA} simulations, 
each observed SN in the nearby sample is used to define an observational sequence, 
including observational redshift, time of peak brightness, and cadence.

\subsection{Training Set Simulations}\label{subsec:SIMSEDSIMS}

Training set simulations are generated from one of the eight sets of input model-intrinsic 
scatter pairs described in Section~\ref{sec:Models}. To be 
consistent with the \Gtenpaper ~analysis, ~the training set SNe are 
restricted to nearby and SNLS ($z<0.7$) components. 
Training set simulations require simulated SN Ia light curves and spectra.
In this case, both light curves and spectra are generated from a four-dimensional 
flux lookup table. 
The four dimensions are phase (71 bins, spanning $-20$ to $+50$ days), 
wavelength (721 bins, spanning 2000~\AA ~to 9200~\AA),
stretch $x_1$ (25 bins, spanning $x1 = -3$ to $+3$), 
and color $c$ (41 bins, spanning $c = -0.3$ to $+0.5$ mags). 

The lookup table is created from one of our input models (Section~\ref{sec:Models}).
Intrinsic scatter is included when the lookup tables are constructed.
When a table entry is generated for a given $x_1$, $c$ pair a random set of intrinsic scatter
parameters is chosen. The scatter corresponding to the chosen parameters is applied to the flux
as a function of wavelength, $F(\lambda)$, along with the color and stretch adjustments. 
In other words, each stretch--color pair in the lookup table is associated with a random 
realization drawn from the intrinsic scatter model. 

For a given selection of phase, stretch, and color, light curve photometry is obtained from the 
lookup table by integrating the fluxes over wavelength with the appropriate broadband filters. 
Whereas phases are interpolated, exact grid points are used for stretch and color.
The parameter distributions from which the training set simulations 
are generated are listed in Table ~\ref{table:gausspars}.

Finally, three quality cuts are applied:
we require training set light curves to have at least one observation before $-2$ days, 
one observation after $+10$ days, and ten observations with S/N$>1$.

\subsubsection{Spectrum Simulation Details}\label{subsubsec:SPECSIMS}
Using a set of simulation libraries\footnote{Referred to as ``SIMLIBS'' in the {\tt SNANA} manual.}
the {\tt SNANA} MC code is able to produce light curves with S/Ns
and cadences. 
We have developed a similar set of SN spectral libraries to enable the generation of SN spectra with 
realistic S/N and cadences. 
Spectroscopic information for a survey is stored in a file called a ``{\tt SPECLIB}.''
For SNe with associated observed spectra, each SPECLIB entry 
captures the relevant details of one observed spectrum,
including rest-frame epoch, beginning and ending observer-frame wavelengths, 
and S/N as a function of observed wavelength. 
SNe without associated spectra are also included in the SPECLIB,
to ensure that SNe with no spectroscopic data are accurately accounted for in the simulation.

At the start of a simulation, the subset of library entries matching the simulation redshift range is selected. 
From this subset, each simulated SN is matched with the library SN closest in redshift. 
A simulated SN spectrum is generated for each observation date in the library SN entry. 
The simulated spectrum's observed phase, wavelength range, and S/N are calculated to match 
the library SN spectrum upon which it is based. 
All valid library entries are used once before a library entry is reassigned to an additional MC SN. 
Two spectral libraries are used for this paper:
a ``low-redshift'' library and an SNLS library. 

Although the SNLS3 training set only included those spectra judged to have negligible galaxy contamination, 
we have included galaxy contamination capabilities in our spectral simulator to check the impact of 
small spectral biases.
For each simulated SN a galaxy contamination fraction at peak brightness can be chosen from a parent distribution.
A galaxy template is then normalized to the appropriate flux and added to the SN spectra. 
We focus on contamination from elliptical and normal spiral galaxies, 
using as templates the sb and elliptical spectra\footnote{
The template files {\tt elliptical\_template} and {\tt sb\_template} were downloaded from the website
http://www.stsci.edu/hst/observatory/cdbs/cdbs\_kc96.html}
from the Kinney-Calzetti Spectral Atlas of Galaxies ~\citep{1994ApJ...429..582C, 1996ApJ...467...38K}. 

\subsection{Test Set Simulations}\label{subsec:TESTSETSIMS}
Test set SN light curves are simulated identically to training set light curves:
the same models, lookup tables, and parameter distributions are used. This setup mirrors
what happens in real SN Ia cosmology experiments: both the model training set and 
the cosmology data are drawn from the same set of objects. 

There are three differences between the test set
simulations and the training set simulations:
(1) the test set simulations do not include spectra,
(2) no quality cuts are applied to the test set simulations,
and (3) the number of simulated test set SNe is two orders of magnitude larger 
($\sim$36,000 SNe compared with 220 SNe in the training set).
We want to calculate HD and best-fit
cosmology parameter biases due to the training and fitting process only. Therefore, we
dramatically increase the size of the training set to limit the impact of its 
statistical uncertainties on our results. A summary of the average numbers of SNe in each of our 
simulated data sets is given in Table~\ref{table:simSNnumbers}.

\begin{center}
\begin{deluxetable}{lcccc}[h]
\tablewidth{200pt}
\tabletypesize{\normalsize}
\tablecaption{Numbers of Simulated SNe\label{table:simSNnumbers}}
\tablehead{
  \colhead{Simulation}&
  \colhead{Nearby}&
  \colhead{SDSS}&
  \colhead{SNLS}&
  \colhead{$N$}
}
\startdata
Training & 120 & ... & 120 & 240 \\
Test     & 5900 & 12240 & 13870 & 32010\\
Bias     & 5900 & 12240 & 13870 & 32010
\enddata
\end{deluxetable}
\end{center}
 
\subsection{Simulation Tests}\label{subsubsec:LCSIMS}

\begin{figure}[h]
\epsscale{0.25}
\centering
\begin{tabular}{@{}cc@{}}
 \includegraphics[width=0.22\textwidth]{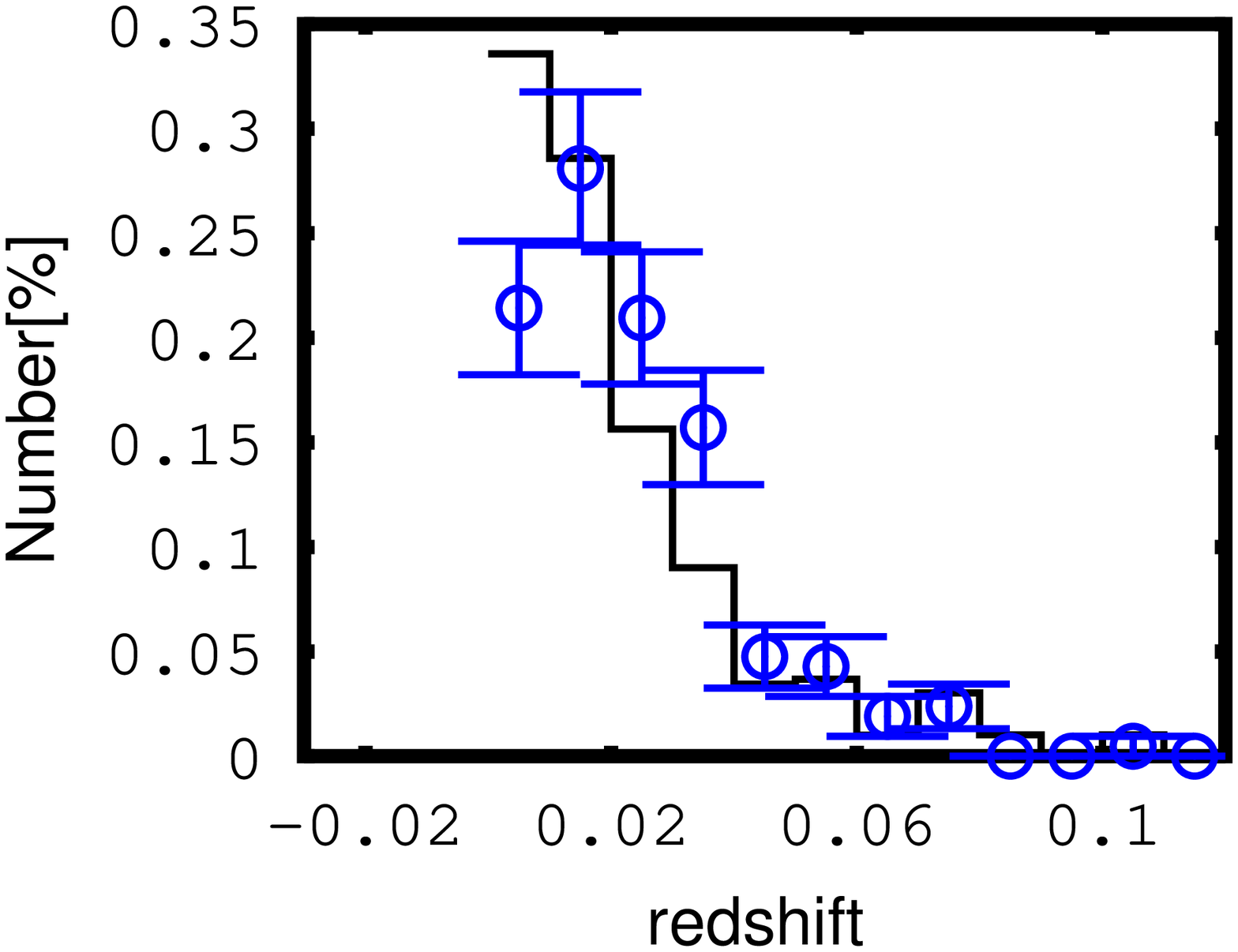}
 \includegraphics[width=0.22\textwidth]{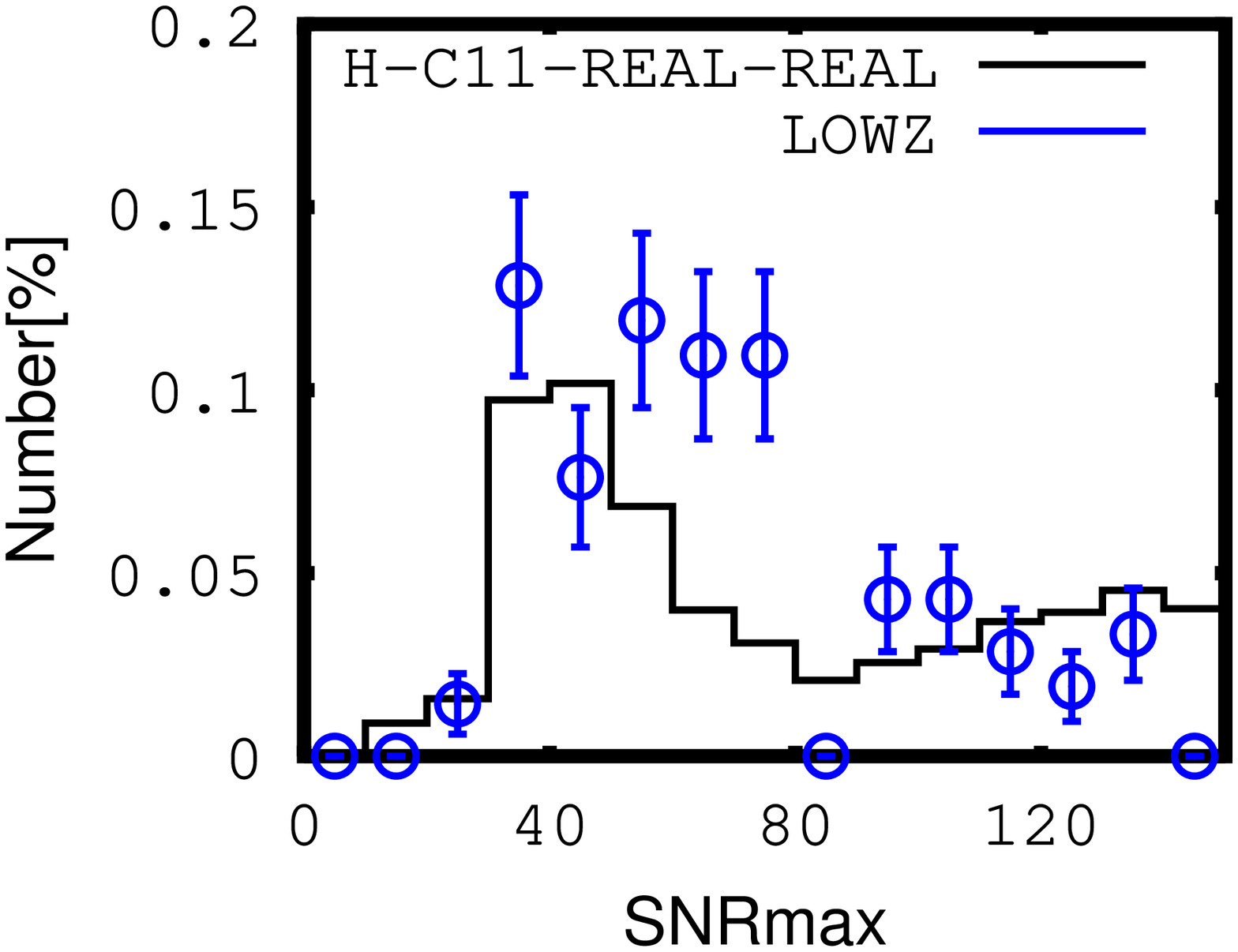}\\
 \includegraphics[width=0.22\textwidth]{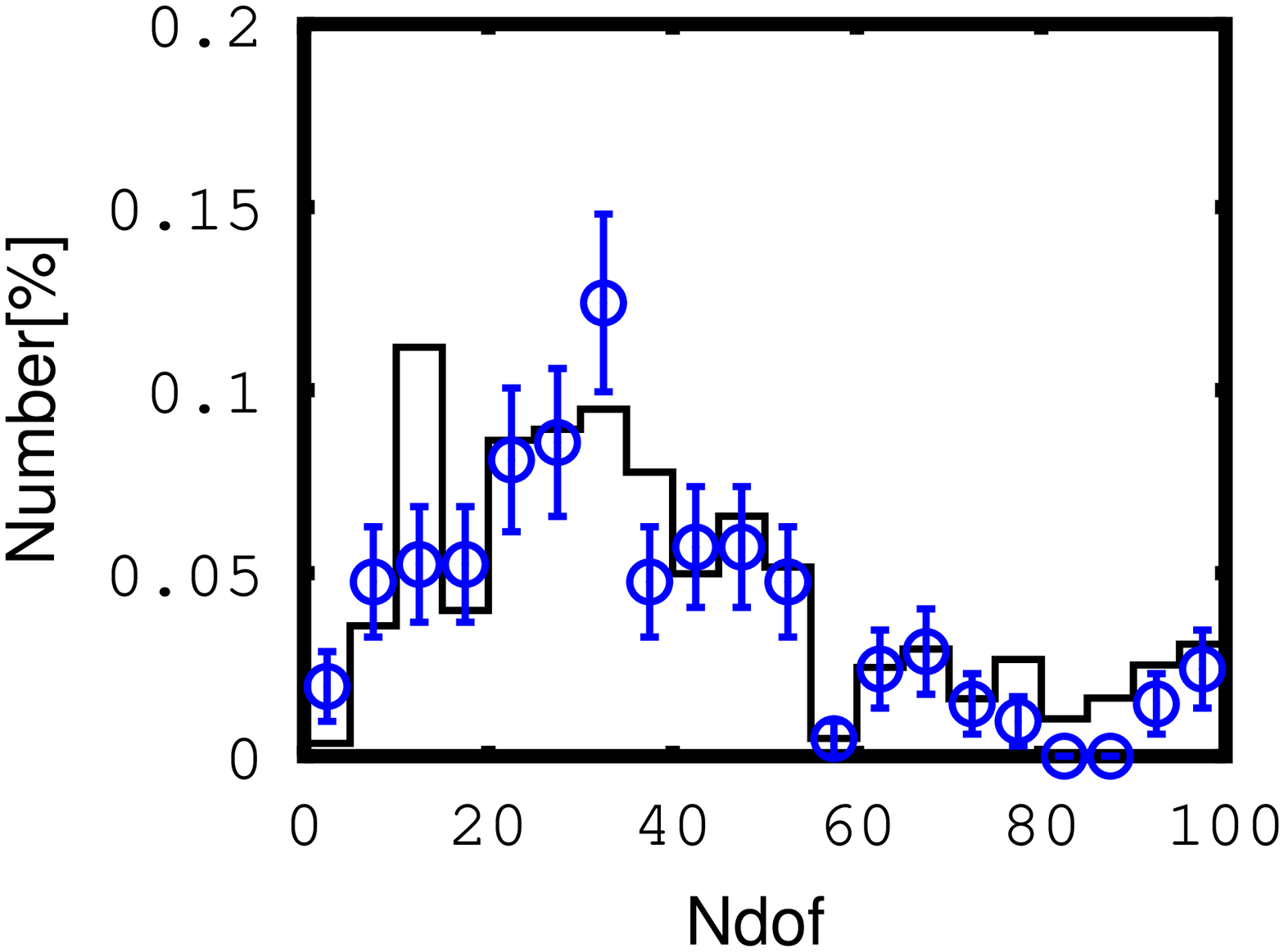}
 \includegraphics[width=0.22\textwidth]{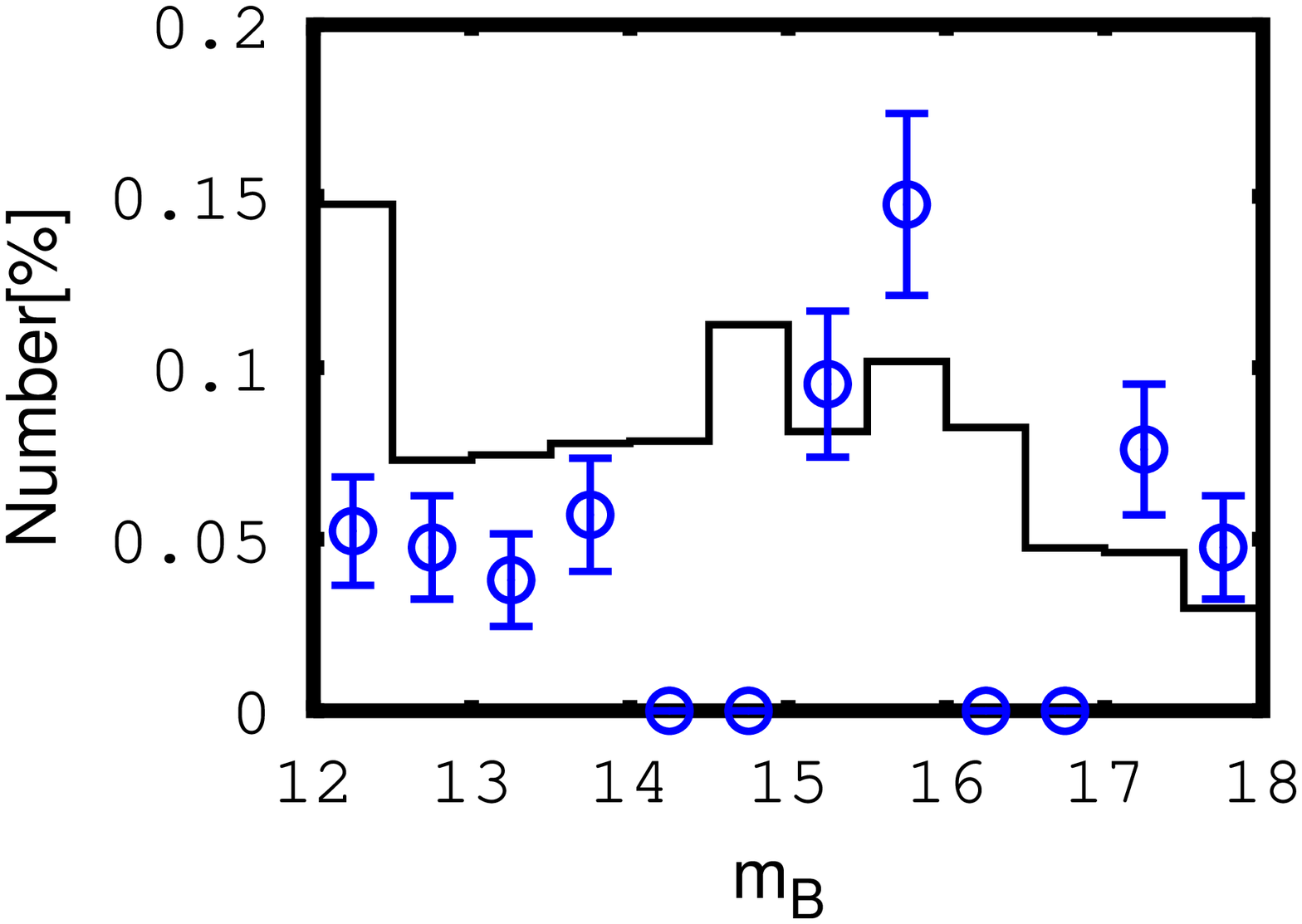}\\
 \includegraphics[width=0.22\textwidth]{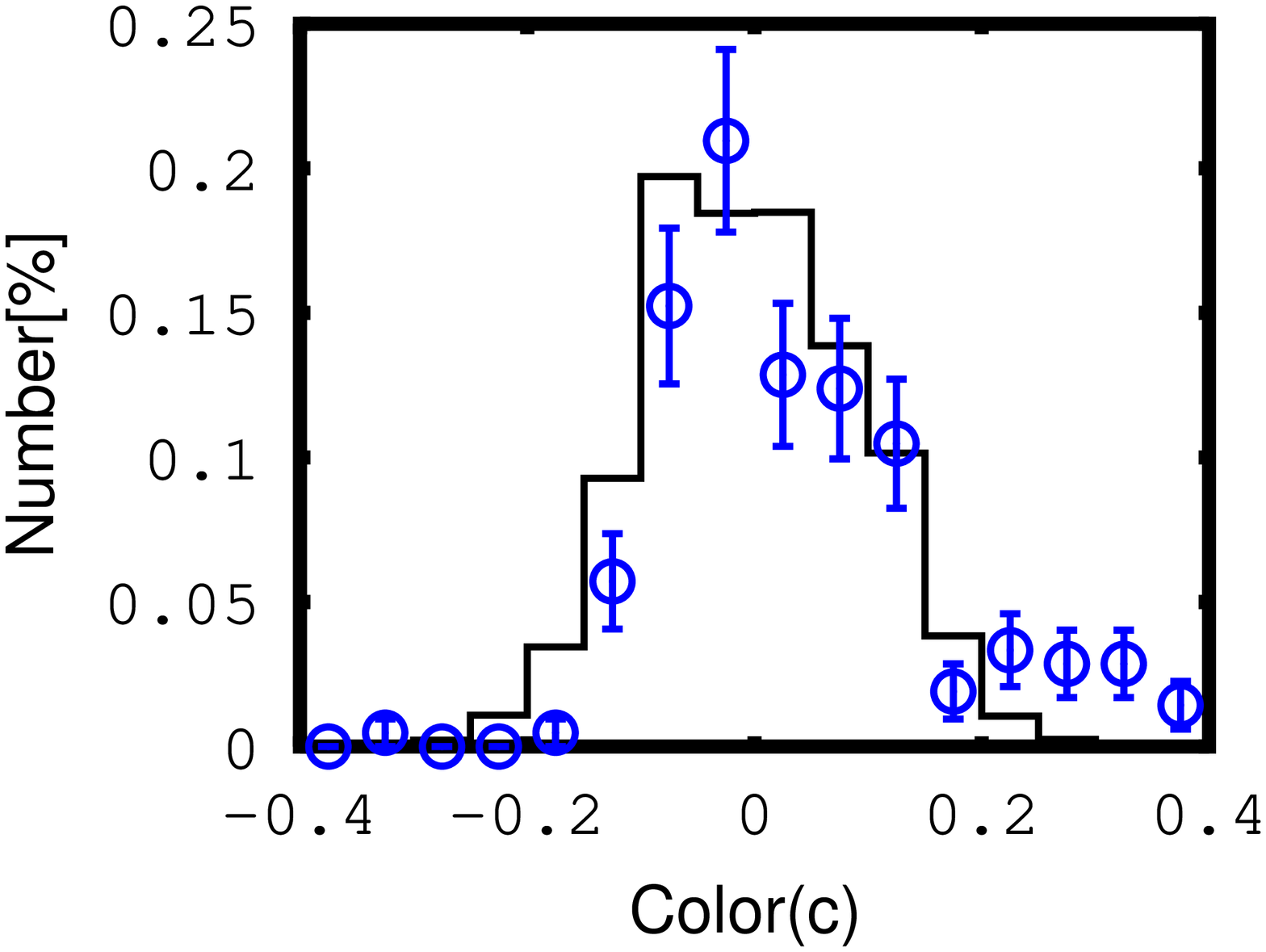}
 \includegraphics[width=0.22\textwidth]{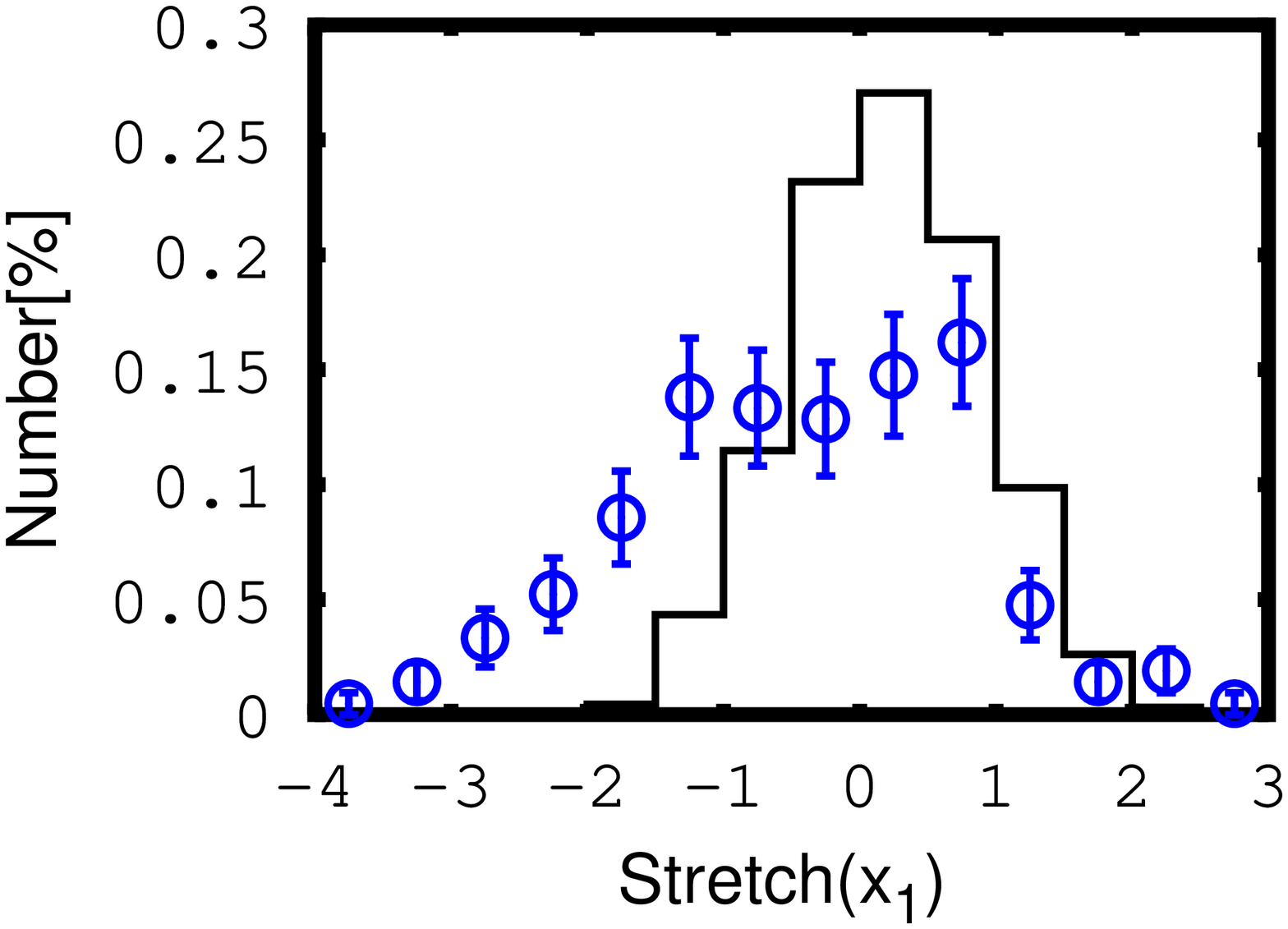}\\
 \includegraphics[width=0.22\textwidth]{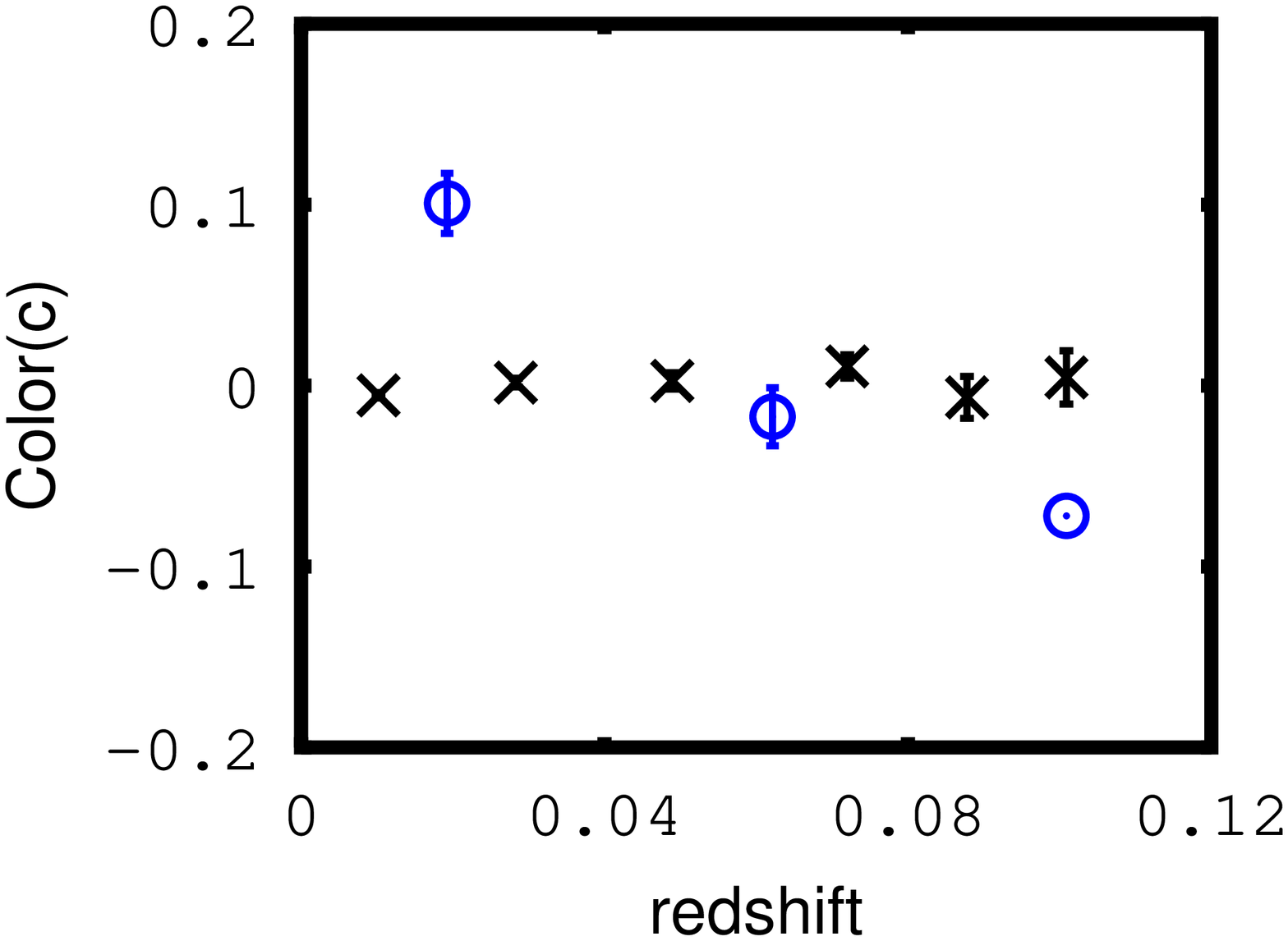}
 \includegraphics[width=0.22\textwidth]{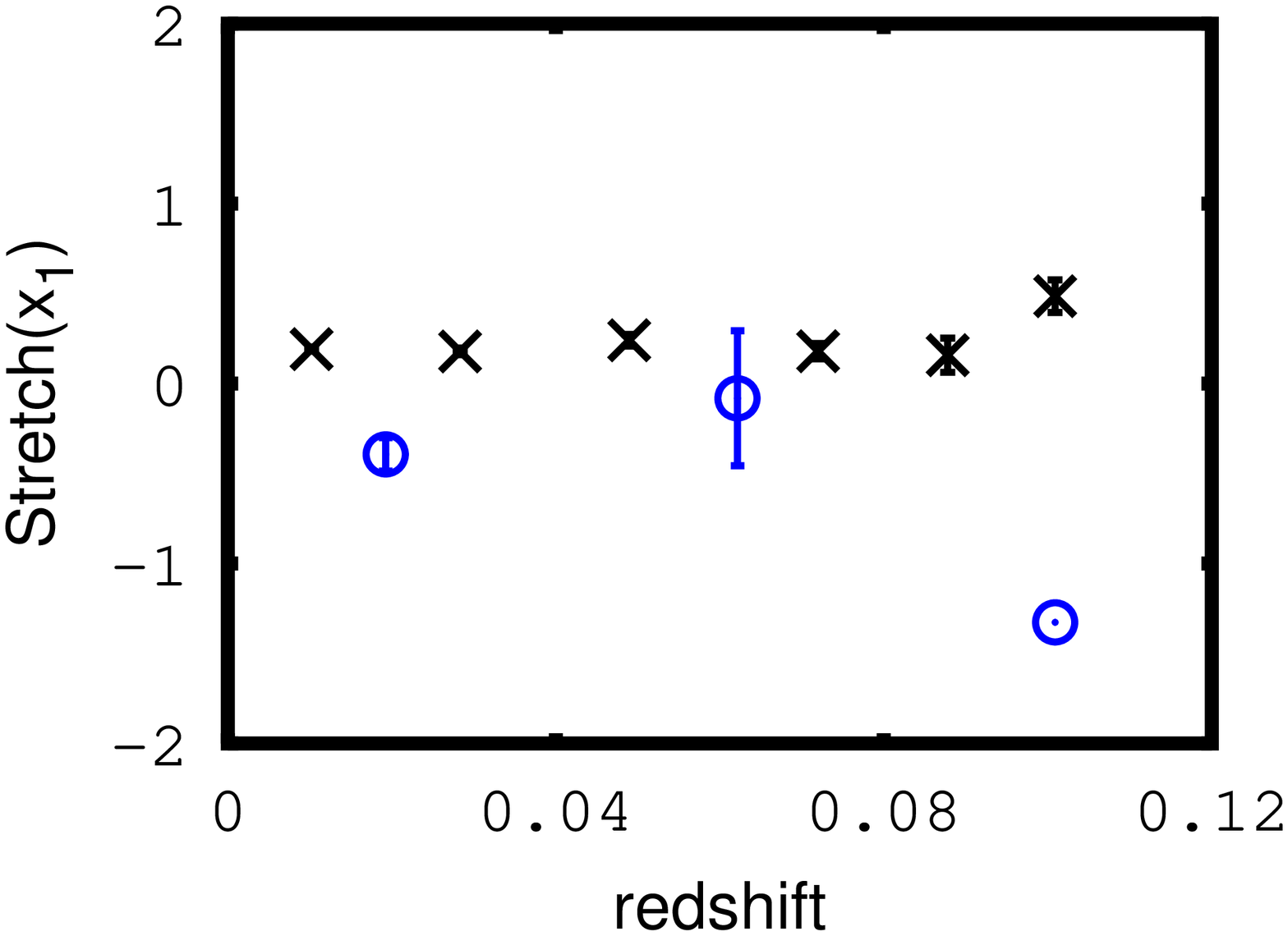}\\
\end{tabular}
\caption{Comparison of fitted parameter distributions for JRK07+CFA3 data (blue circles) 
and H-C11-REAL-REAL Test Set MCs (black histograms/stars). 
All fits have been performed with the \Gtenmodel ~light curve model.
From top to bottom and left to right, the distributions shown are (1)~redshift, 
(2)~maximum fitted $S/N$, (3)~fit degrees of freedom, (4)~fitted observer-frame $B$-band
magnitude ($m_B$), (5)~SALT-II color ($c$), and (6)~SALT-II stretch ($x_1$).
The bottom  row shows mean SALT-II color ($c$) and shape parameters ($x_1$) 
as a function of redshift.}
\label{fig:REALDATA_LOWZ}
\end{figure}

\begin{figure}[h]
\epsscale{0.3}
\centering
\begin{tabular}{@{}cc@{}}
  \includegraphics[width=0.22\textwidth]{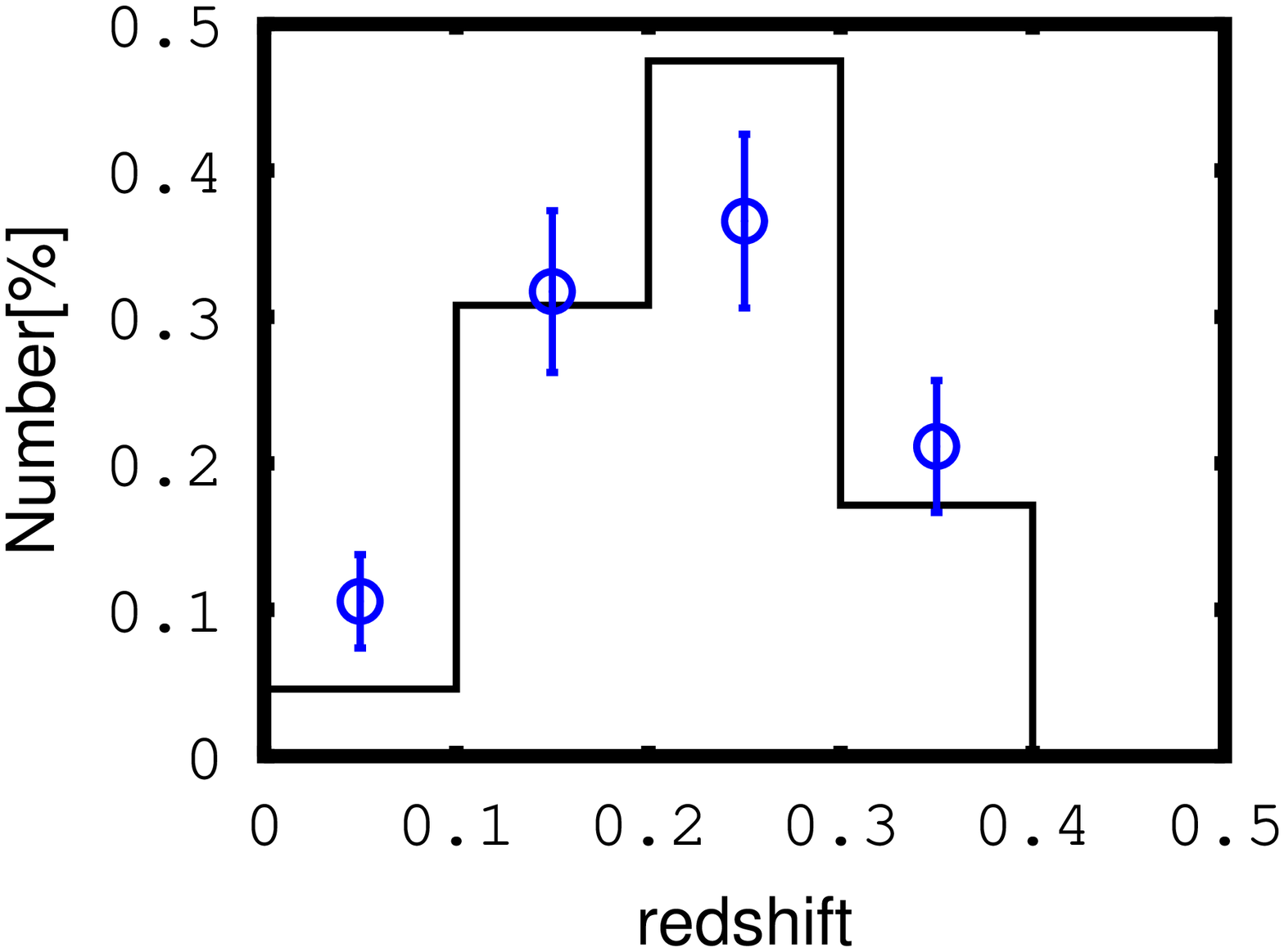} &
  \includegraphics[width=0.22\textwidth]{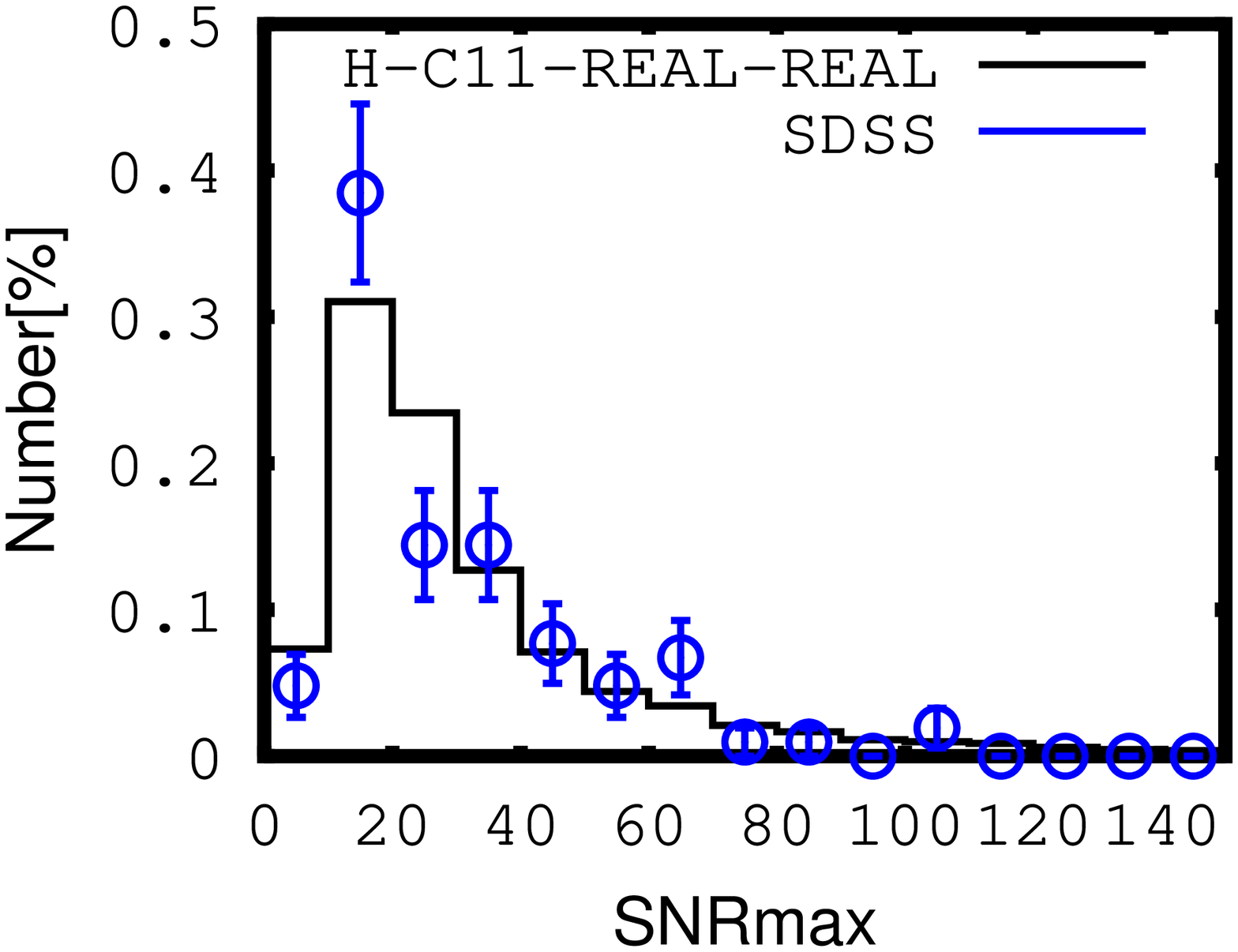}\\
  \includegraphics[width=0.22\textwidth]{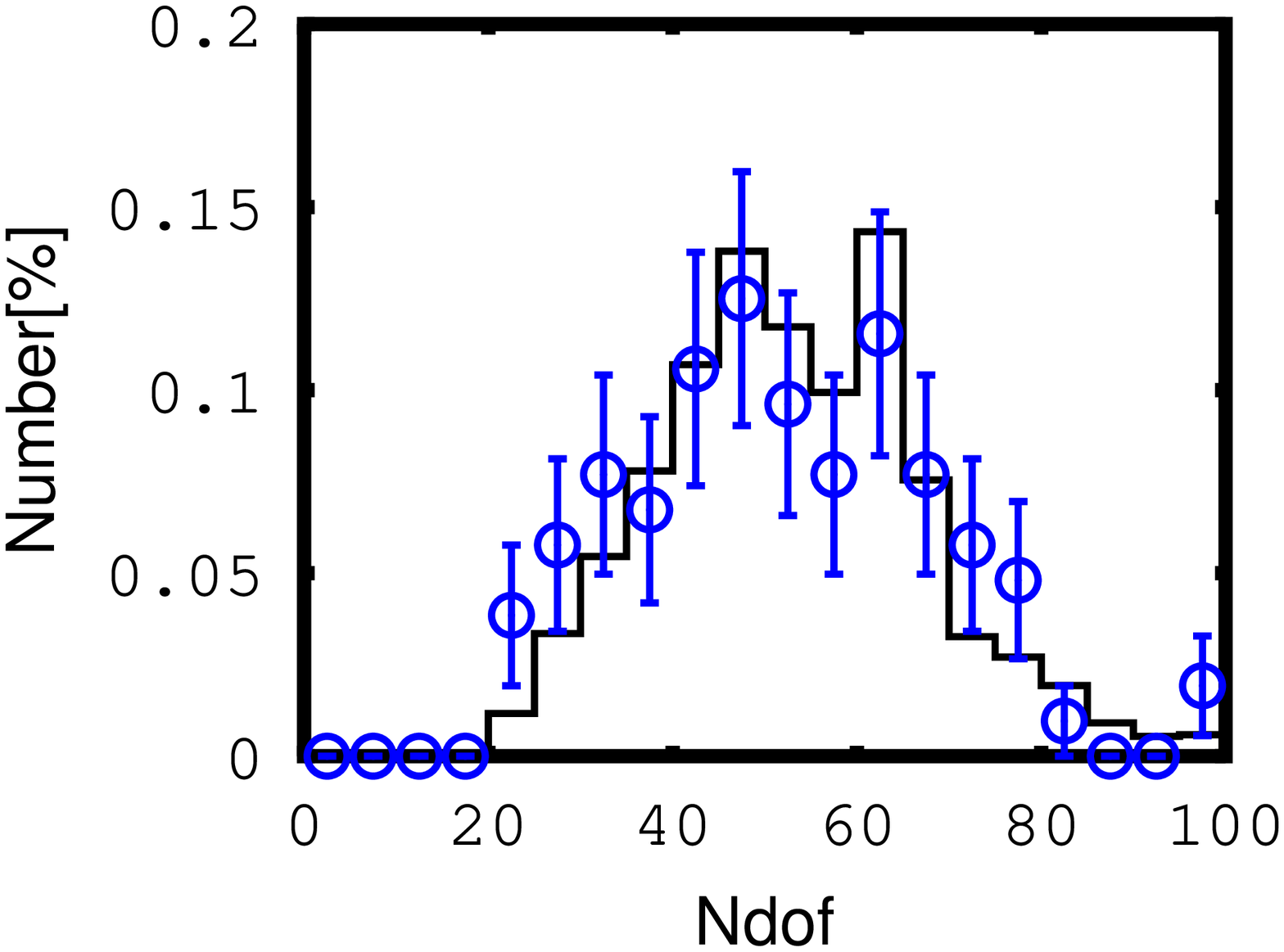} &
  \includegraphics[width=0.22\textwidth]{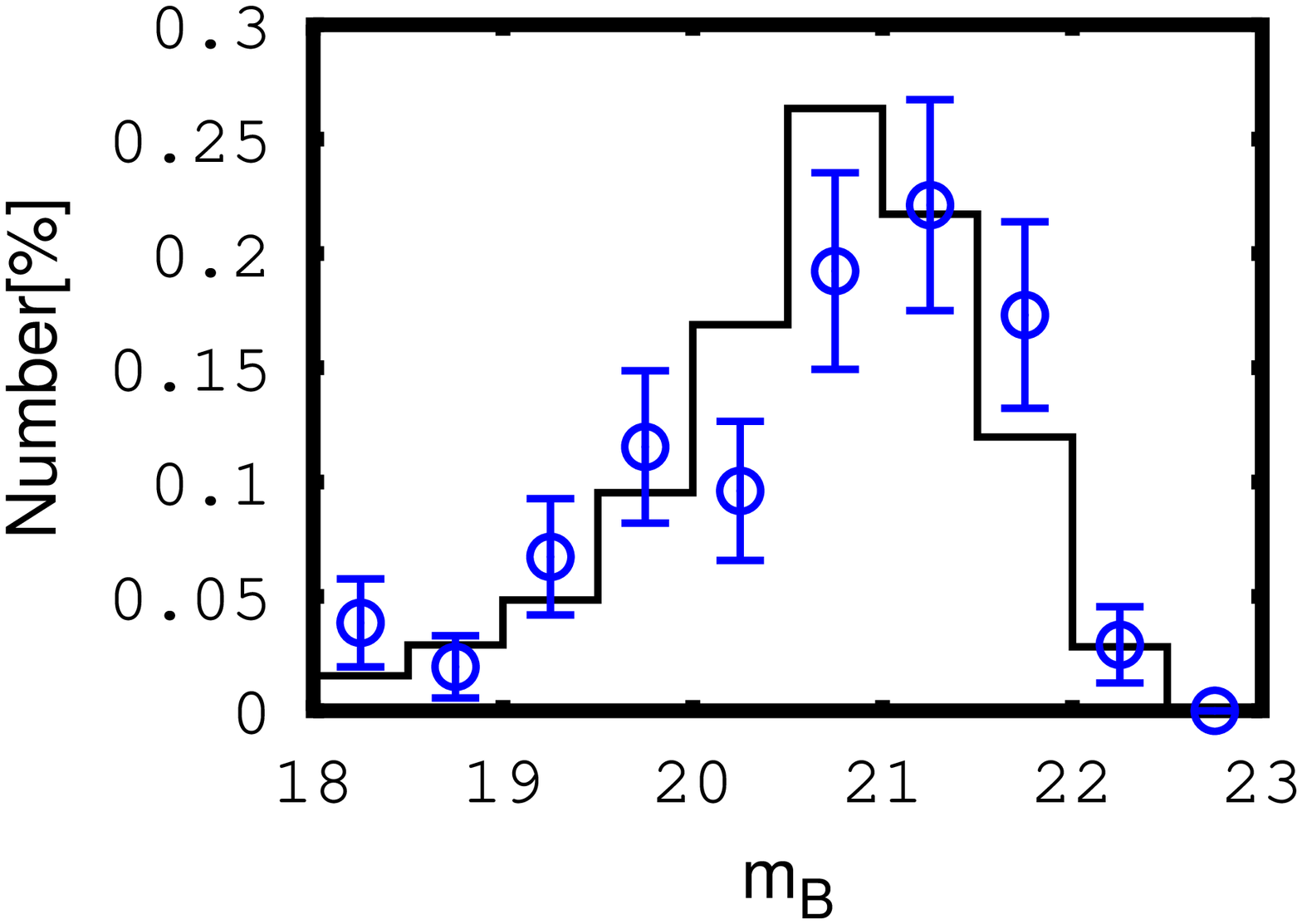}\\
  \includegraphics[width=0.22\textwidth]{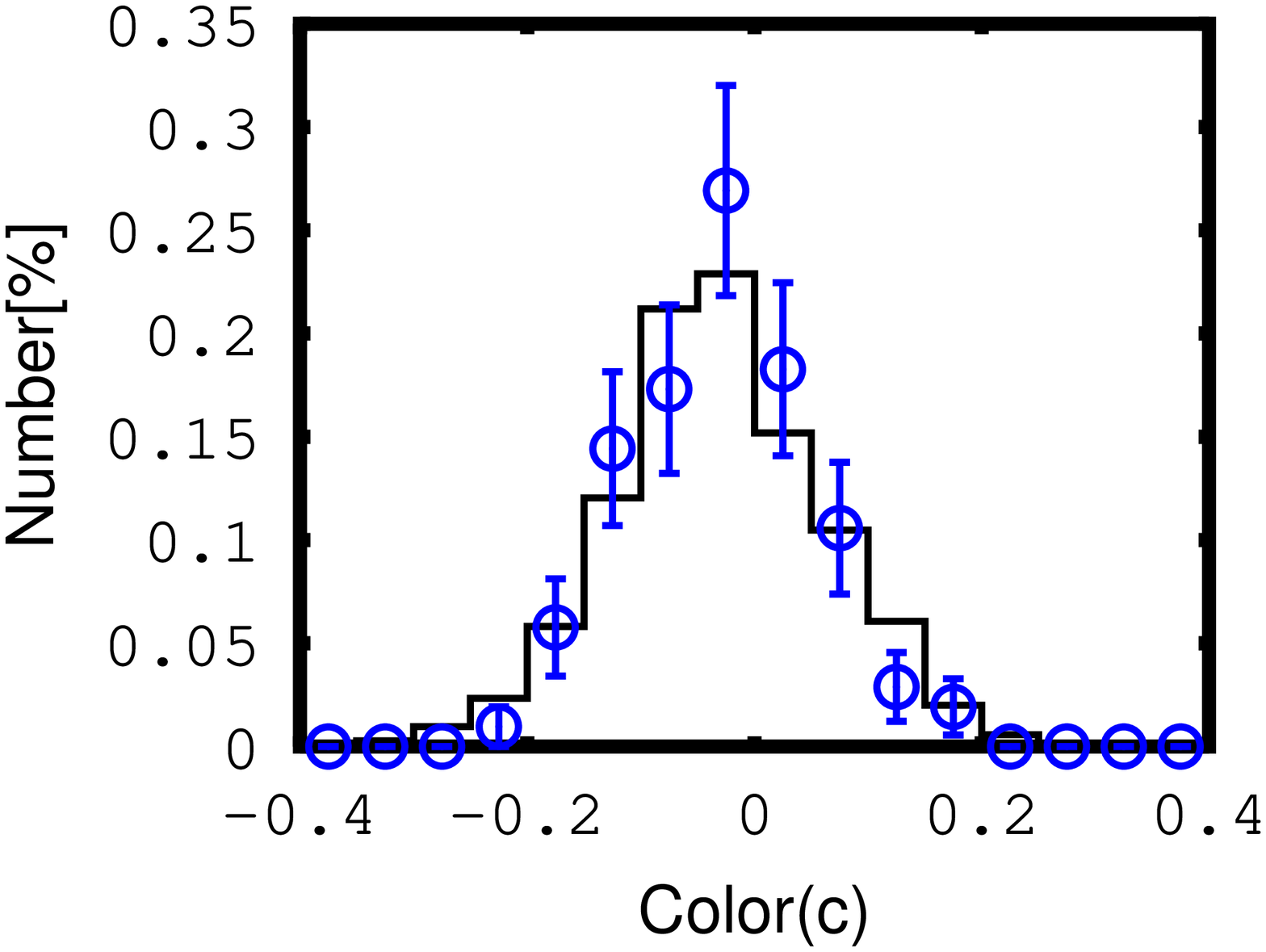} & 
  \includegraphics[width=0.22\textwidth]{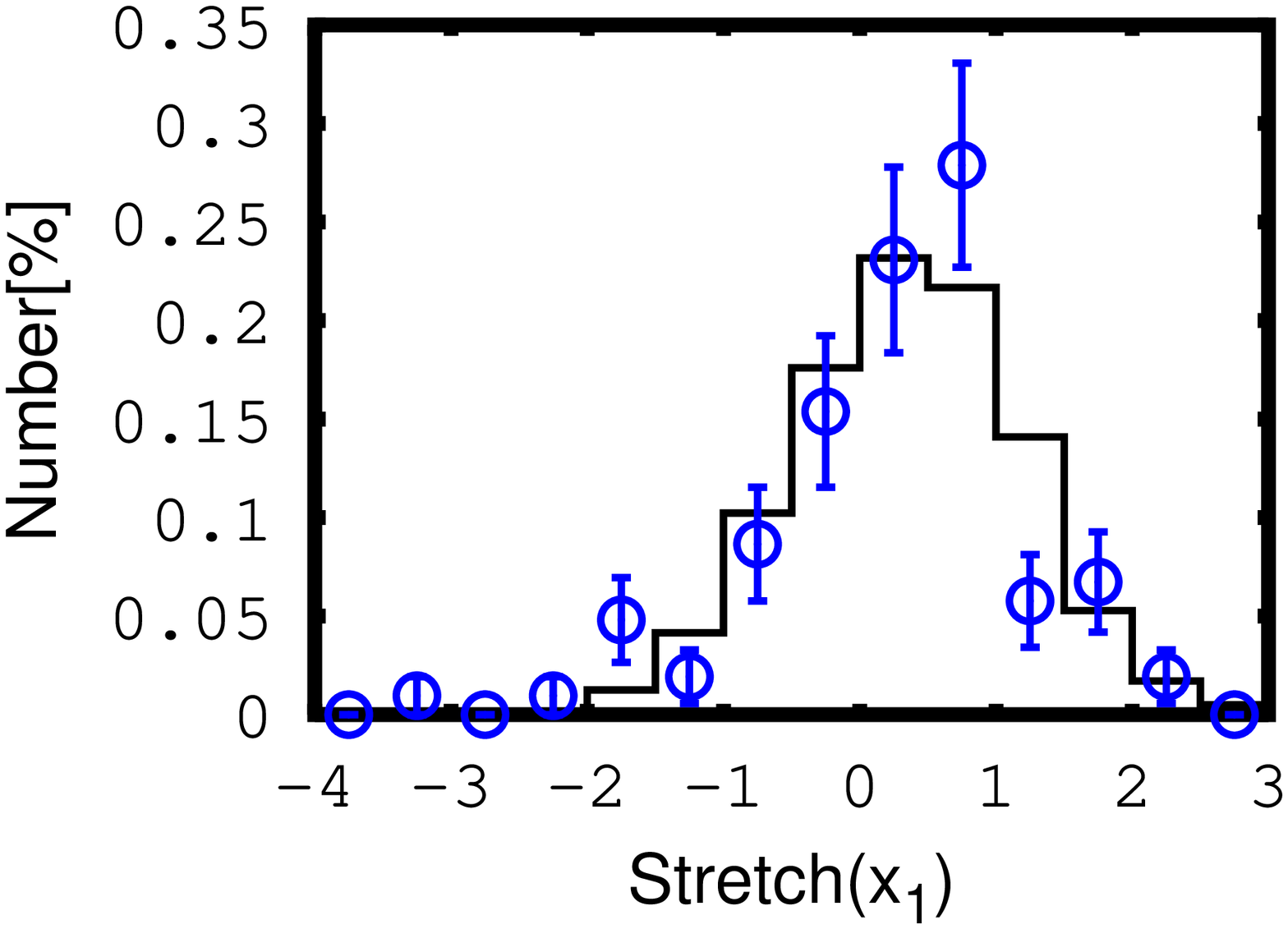}\\
  \includegraphics[width=0.22\textwidth]{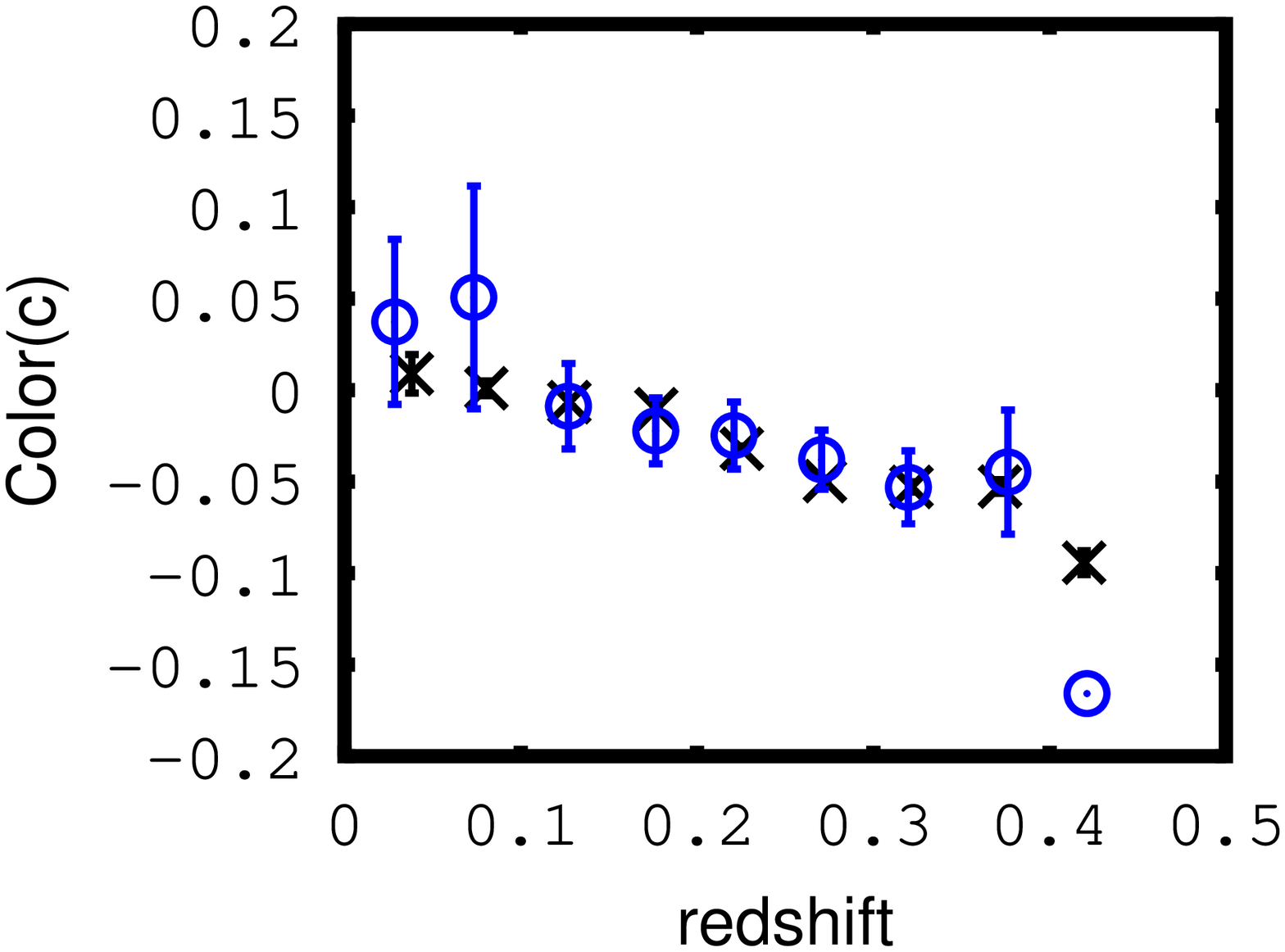} &
  \includegraphics[width=0.22\textwidth]{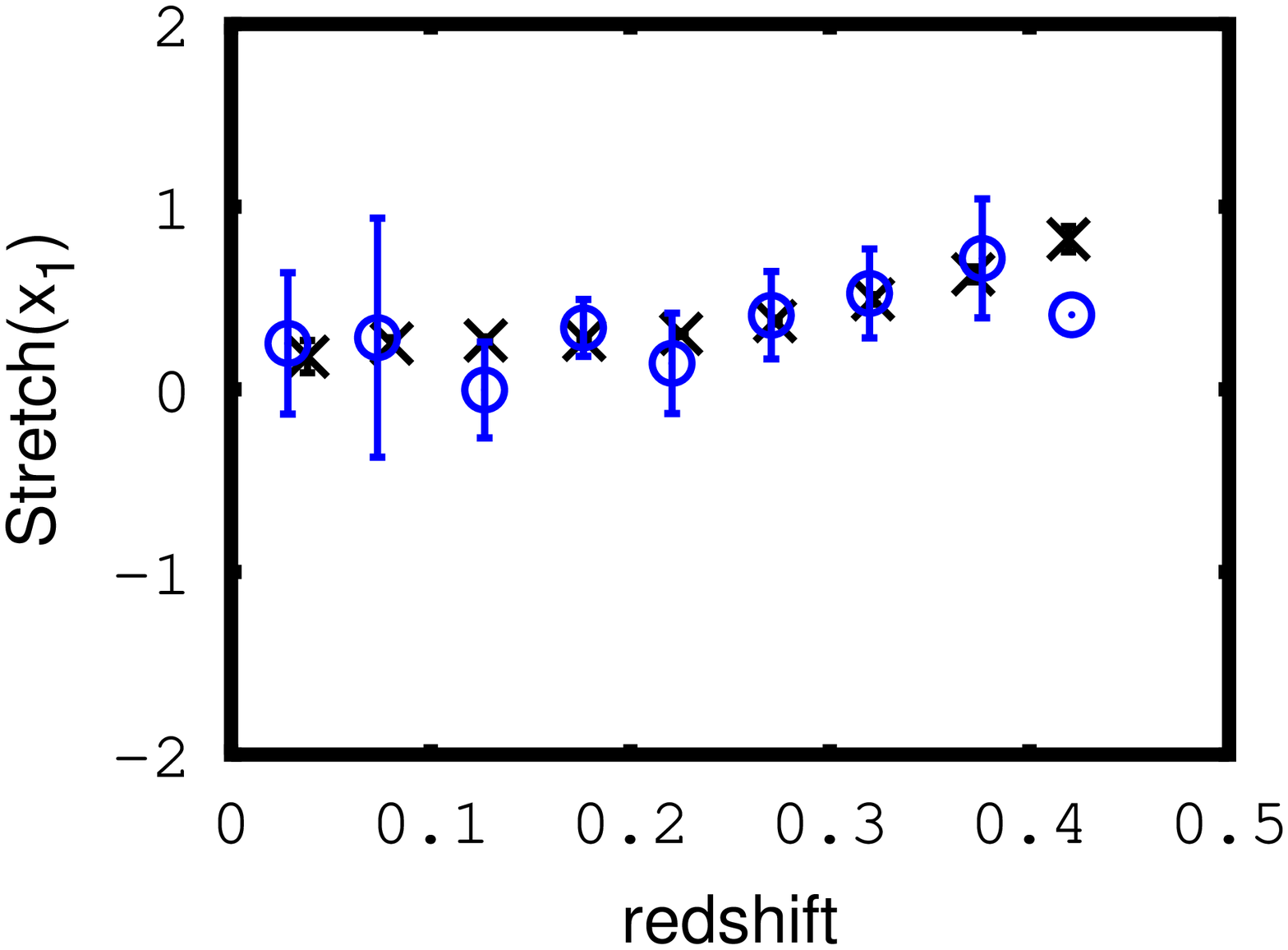}\\
\end{tabular}
\caption{Same as Fig~\ref{fig:REALDATA_LOWZ}, with SDSS-II data replacing the nearby data.}
\label{fig:REALDATA_SDSS}
\end{figure}

\begin{figure}[h]
\epsscale{0.25}
\centering
\begin{tabular}{@{}cc@{}}
 \includegraphics[width=0.22\textwidth]{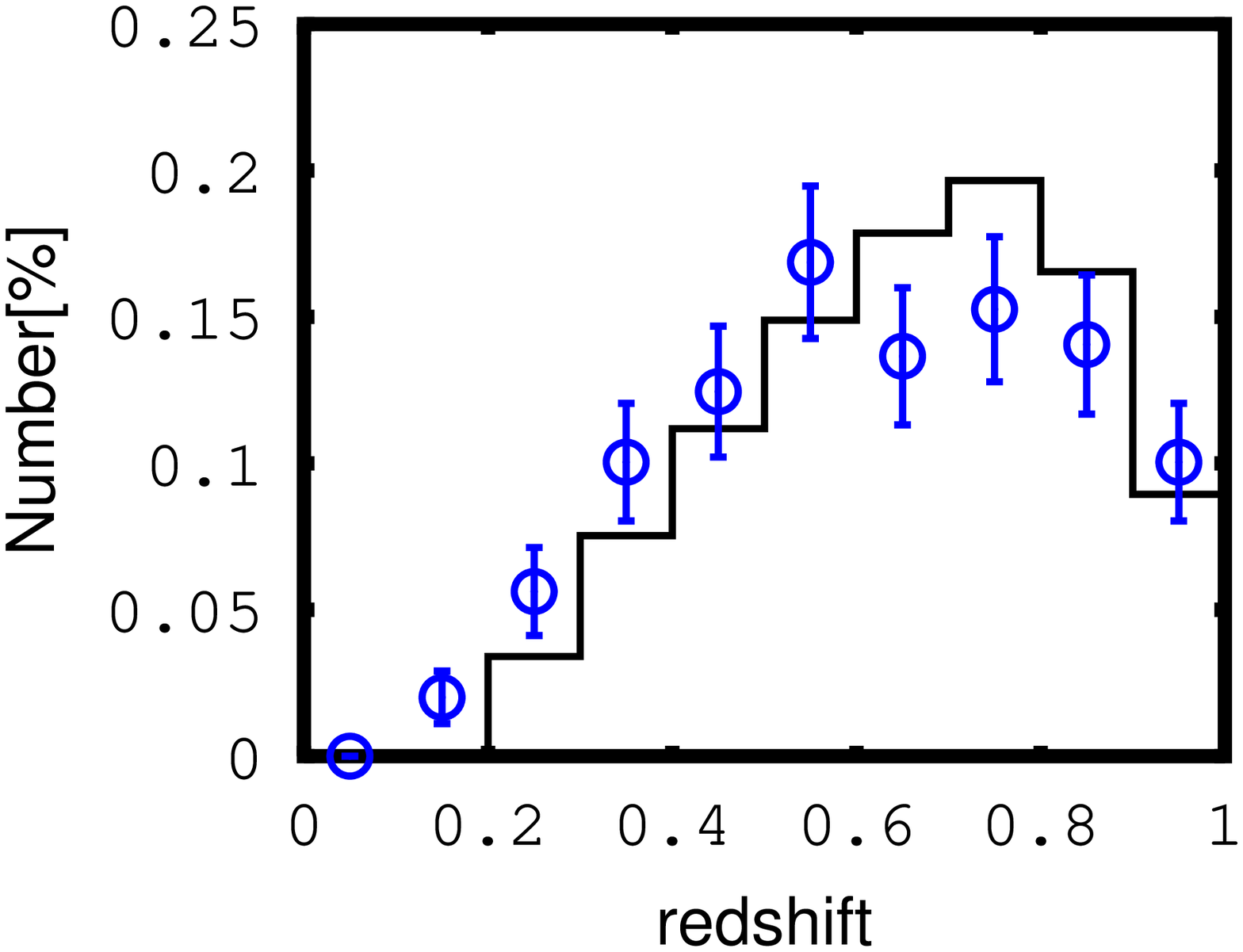}
 \includegraphics[width=0.22\textwidth]{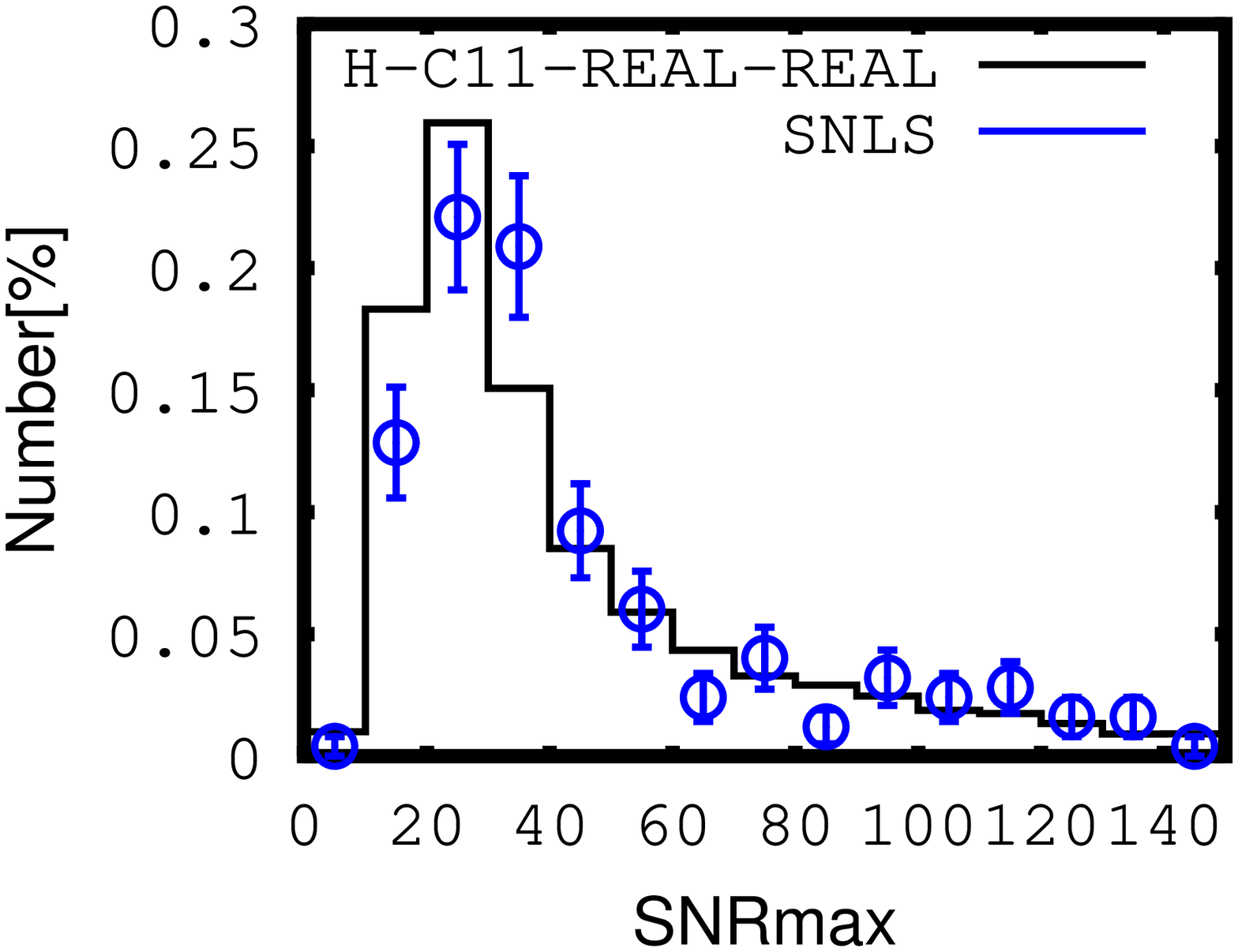}\\
 \includegraphics[width=0.22\textwidth]{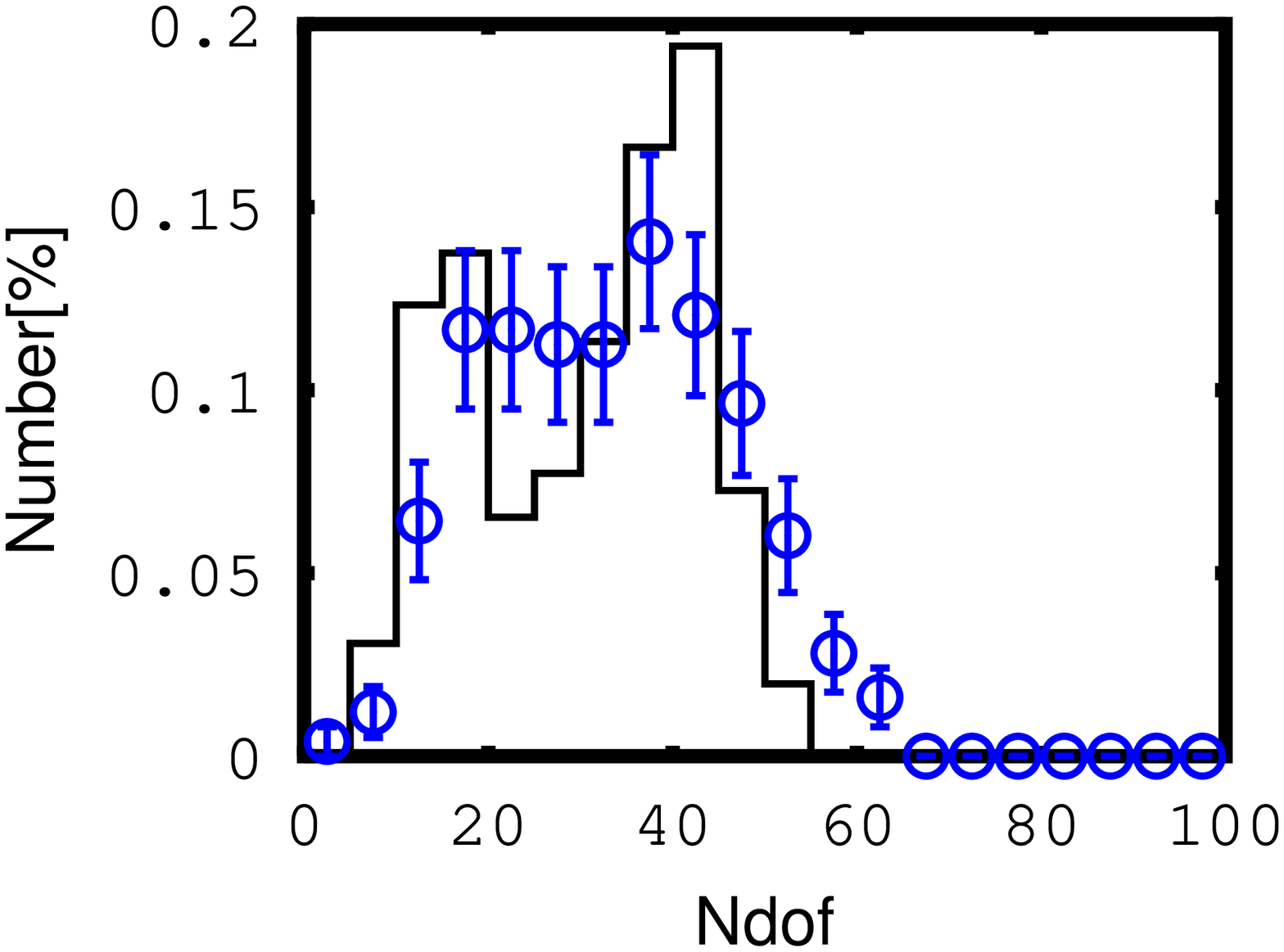}
 \includegraphics[width=0.22\textwidth]{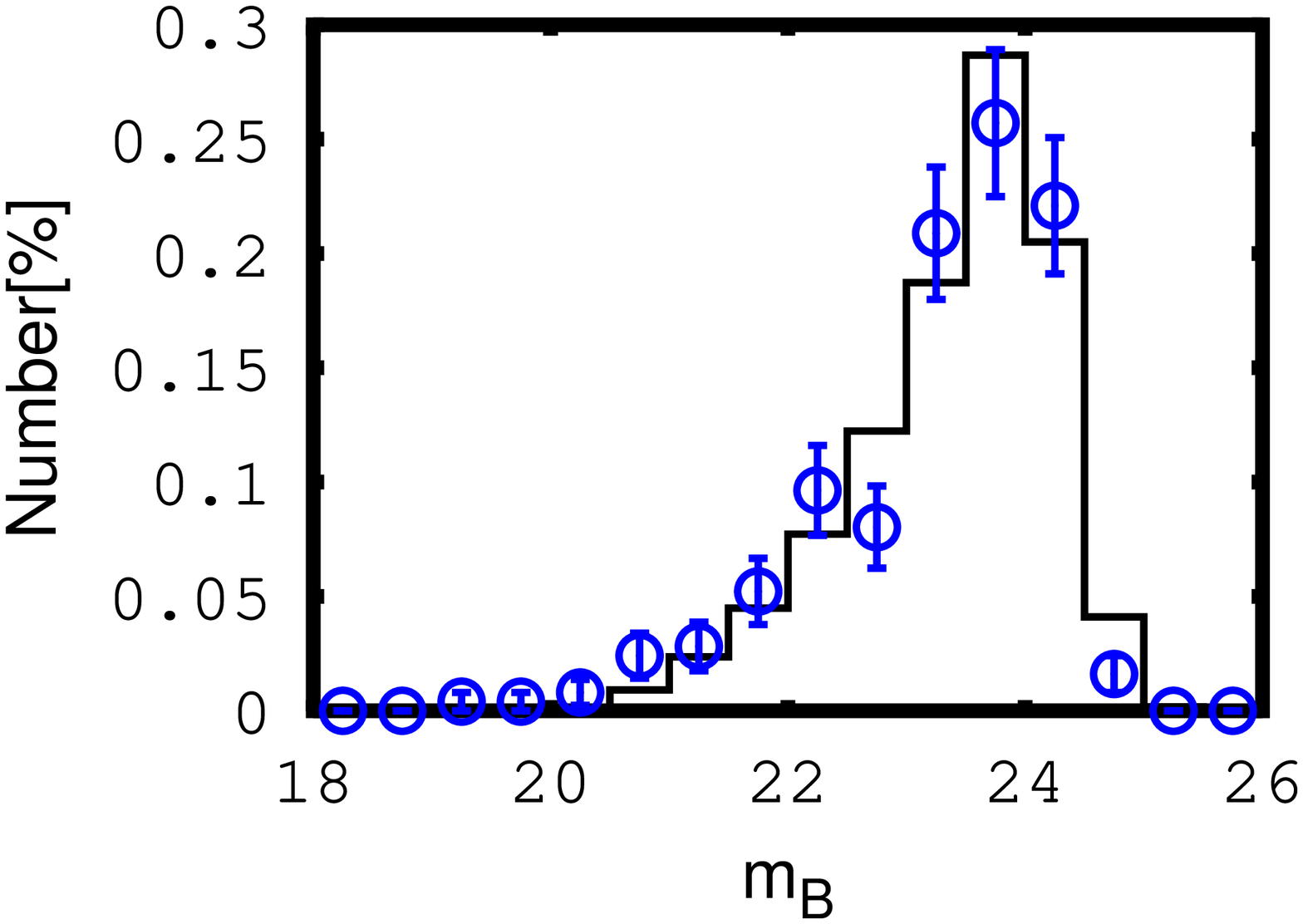}\\
 \includegraphics[width=0.22\textwidth]{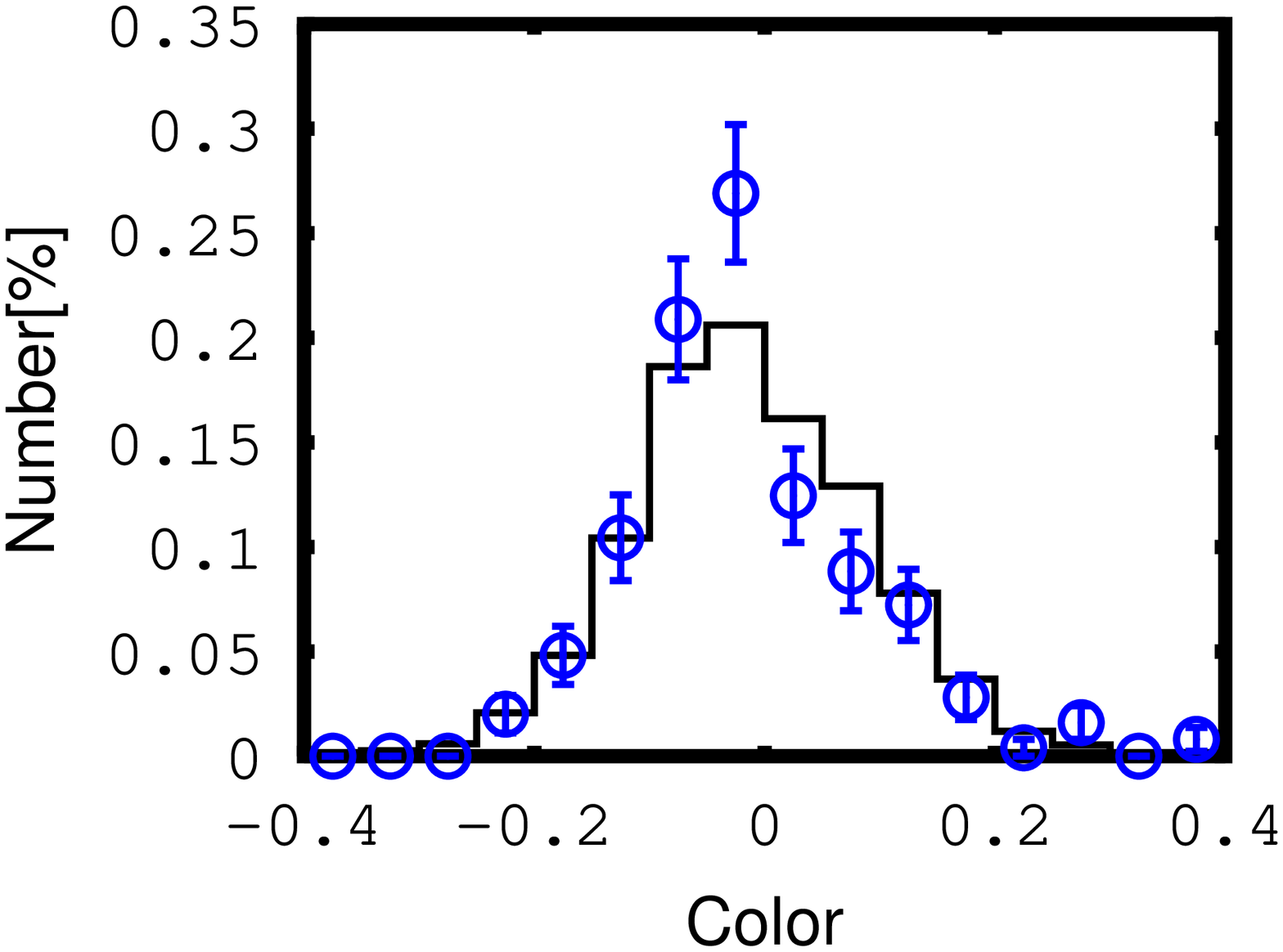}
 \includegraphics[width=0.22\textwidth]{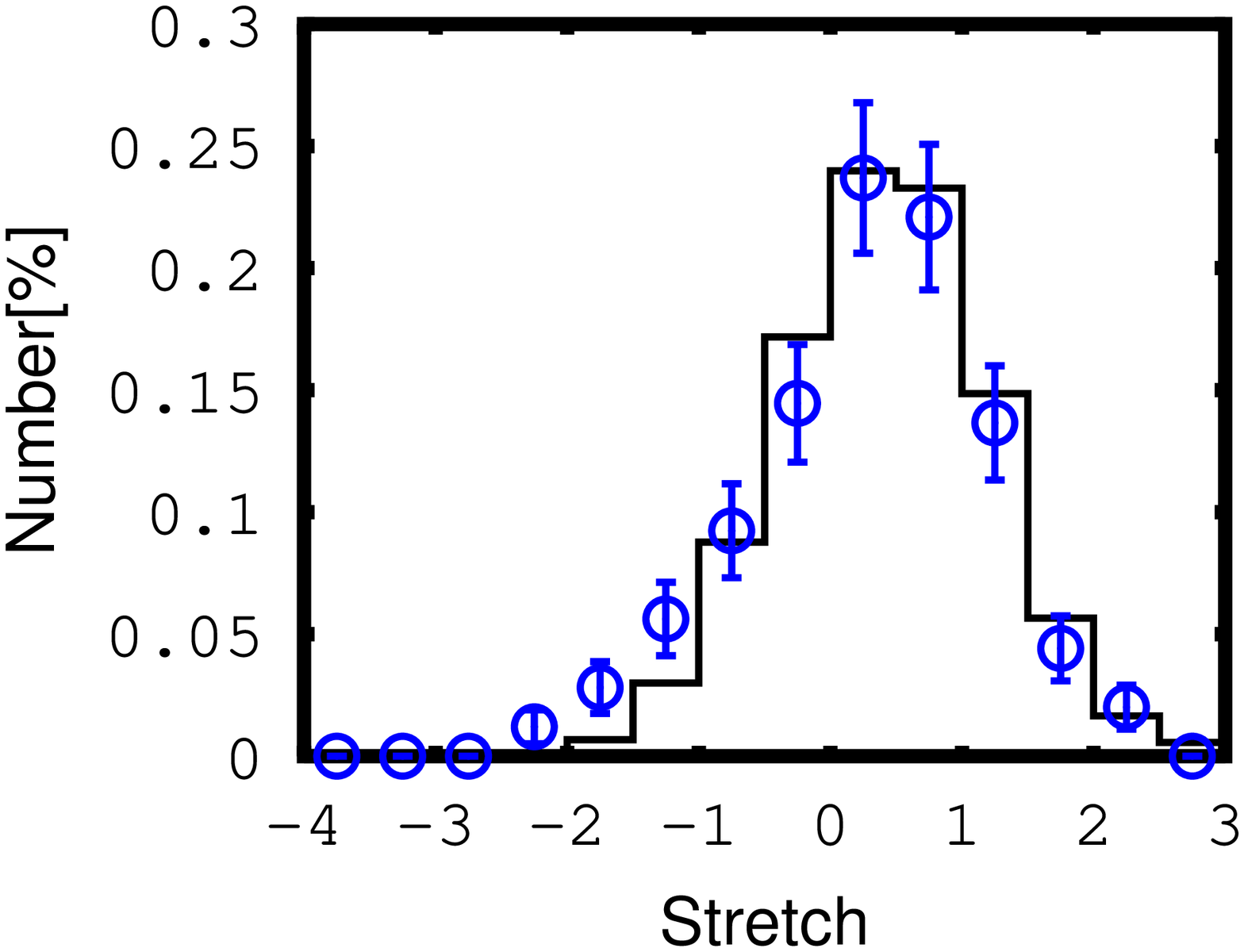}\\
 \includegraphics[width=0.22\textwidth]{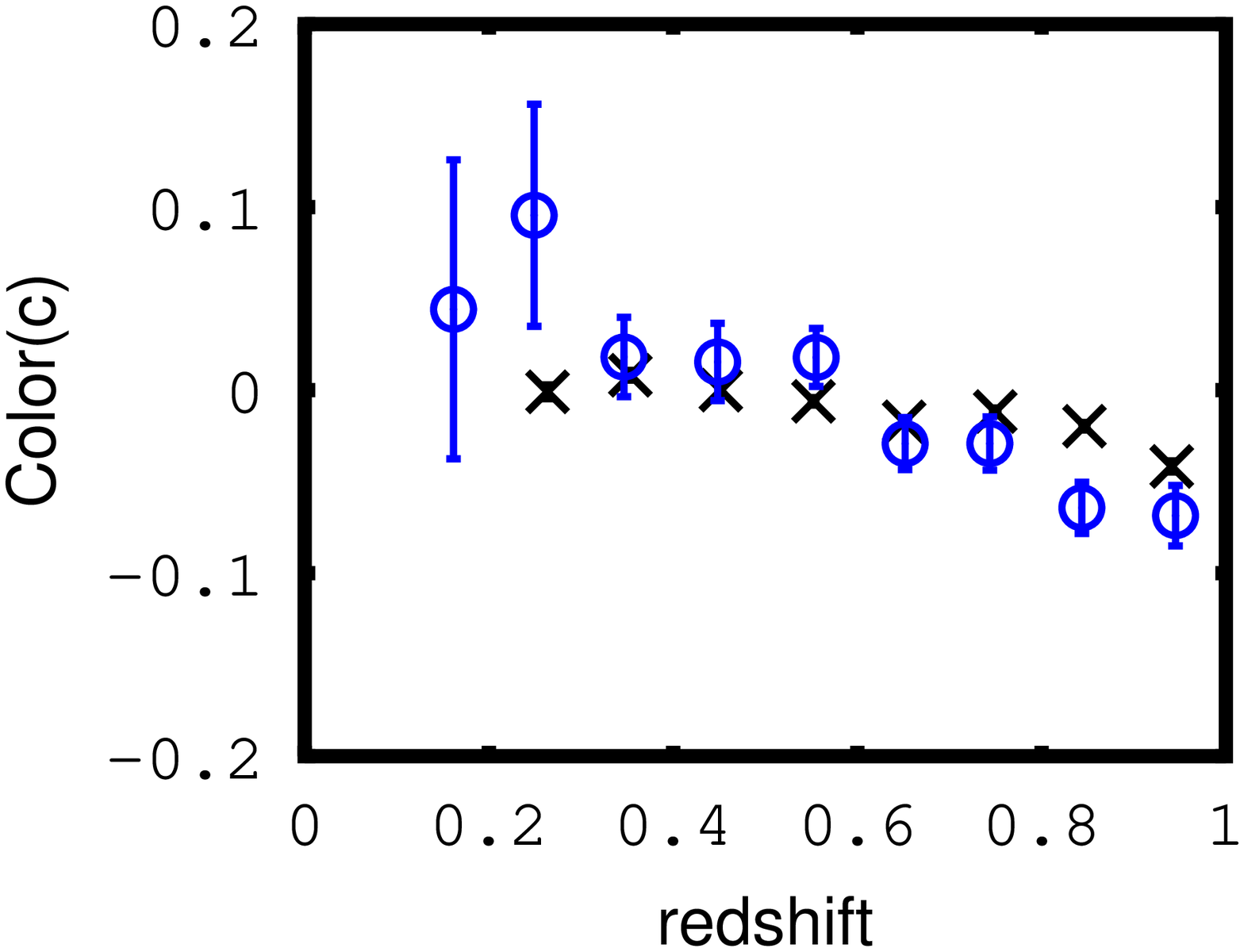}
 \includegraphics[width=0.22\textwidth]{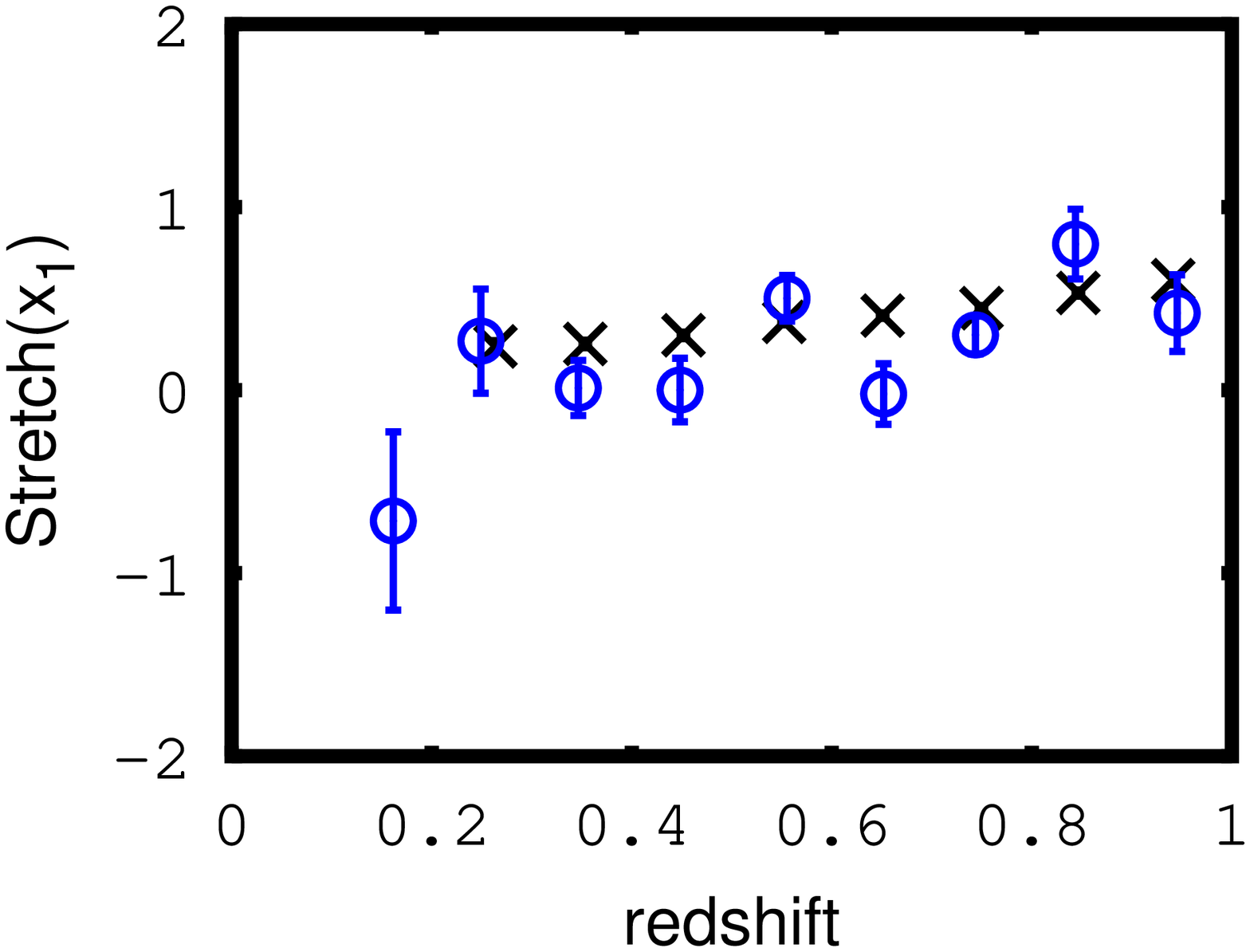}\\
\end{tabular}
\caption{Same as Fig~\ref{fig:REALDATA_SDSS}, with SNLS3-Megacam data replacing the SDSS-II data.}
\label{fig:REALDATA_SNLS}
\end{figure}

The quality of the simulations for the nearby, SDSS-II, and SNLS3-Megacam samples is illustrated with 
several data/MC comparisons in Figures~\ref{fig:REALDATA_LOWZ} -- ~\ref{fig:REALDATA_SNLS}.

Most aspects of the SDSS-II and SNLS3 MCs agree well with the observed data. 
The largest difference comes in fitted SALT-II color ($c$) and stretch ($x_1$) distributions, 
particularly for the nearby sample (e.g. Figure~\ref{fig:REALDATA_LOWZ}).  
As stated in Section~\ref{sec:gensimstuff}, to simplify the implementation of the training tests
we have chosen to use the same stretch and color parameter distributions for all SNe within a 
given training test, regardless of survey. 
Despite this simplification, our fitted MC mean color and stretch as a function of redshift
agree reasonably well with their SDSS and SNLS data counterparts.
\begin{center}
\begin{deluxetable}{cccccc}[h]
\tablewidth{0pt}
\tabletypesize{\normalsize}
\tablecaption{MC $x_1$ and $c$ Parent Gaussian Distribution Parameters\label{table:gausspars}}
\tablehead{
  \colhead{}&
  \colhead{mean}&
  \colhead{$\sigma_-$}&
  \colhead{$\sigma_+$}&
  \colhead{gen. range}
}
\startdata
\sidehead{Input Model = \NZJG}
$x_1$ & 0.0 & 1.0 & 0.7 & $-$2, $+$2\\
$c$ & 0.0 & 0.07 & 0.1 & $-$0.3, $+$0.3\\
\sidehead{Input Model = \HRK}
$x_1$ & 0.5 & 1.0 & 0.7 & $-$2.5, $+$2.5\\
$c$ & 0.0 & 0.07 & 0.1 & $-$0.3, $+$0.3\\
\enddata
\parbox{6in}{\tablecomments{
The same distributions are used for all training
data simulations with a given input model. }}
\end{deluxetable}
\end{center}

\subsection{Determining Distances}\label{sec:fitting}
The simulated test set is fit $N$ times, once by each realization of the trained model 
(see Figure~\ref{fig:trainschema}). 
For each SN in the test set, the best-fitting scale ($x_0$), stretch ($x_1$), 
and color ($c$) parameters are determined by minimizing a $\chi^2$ based on the 
difference between the SN photometry and synthetic photometry of the model flux. 
To ensure good quality fits to the light curves, the following selection criteria are applied.
\begin{itemize}
\item At least one observation before $-2$ days.
\item At least one observation after $+10$ days.
\item At least three filters with an  observation that has a S/N above five.
\item At least five observations in the fitted epoch range $-15.0$ to $+45.0$ days.
\end{itemize}

The initial distance modulus \rawmu ~for each test set SN is given by 
\begin{equation}
\protect \label{eq:DM}
\mu_{\rm fit}= m_B - M_B + \alpha x_1 - \beta c
\end{equation}
where the effective $B$-band magnitude $m_B$ is defined as 
$m_B = -2.5~log_{10}(x_0) + 10.635$ and the global parameters 
$\alpha$, $\beta$, and $M_B$ are determined by a fit of the 
entire test set using the {\tt SALT2mu} program 
described in ~\citet{2011ApJ...740...72M}.  

\subsection{Redshift-dependent Bias Corrections}\label{sec:MBCORR}
To obtain accurate cosmology parameters, flux-limited SN surveys must 
account for the impact of selection effects on distance modulus measurements. 
Typically this bias is determined by simulating SN Ia light curves from parameter distributions (color, stretch, $M_B$) 
consistent with the observed data, evaluating selection biases from these simulations, and correcting the
initial distance moduli accordingly. 
Slight variations exist in the ways the simulated light curves are deployed and the biases are evaluated.
 
\subsubsection{Bias Correction Method}\label{sec:TOTALdetails}
In most previous analyses, bias corrections are calculated by analyzing the simulated bias data
in a manner identical to the real data. In other words, the simulated light curves are generated 
and fit from the chosen SN Ia model, and subjected to any additional processing 
(i.e., global parameter fitting if using SALT) necessary to obtain distance measurements. 
The fitted distances are then compared with the underlying distances 
to obtain the recovered distance modulus bias as a function of redshift, and 
the real data are corrected accordingly. 
In terms of the fitted $\mu_{\rm fit}$ and simulated $\mu_{\rm sim}$ distances, the 
bias correction  $\Delta \mu(z)$ is
\begin{equation}
\protect \label{eq:TOTALcorr}
\Delta \mu(z) = \langle \mu_{\rm fit} - \mu_{\rm sim} \rangle_{z}
\end{equation}
and the corrected distances $\mu'(z)$ are
\begin{equation}
\mu(z) = \mu_{fit}(z) - \Delta \mu(z).
\end{equation}

This method has been used in 
~\citet{WVEssence1yr:2007} and ~\citet{KesslerSDSScosmo:2009}.
In these works the corrections are usually described as 
``Malmquist bias'' or ``selection  bias'' corrections. 
However, by including light curve fitting as part of the process, this technique
implicitly corrects for fitting biases as well as selection biases.

\subsubsection{Bias Correction Simulation and Fitting}\label{subsec:LCSIMS}

Following the schematic laid out in Figure~\ref{fig:trainschema}, 
we generate our bias correction sets directly from the trained models;
no lookup tables are created for these simulations.  
Each simulation's intrinsic scatter function is taken from its trained model's
color dispersion $k(\lambda)$ applied as per the \Ktw ~prescription 
(Section~\ref{subsub:Gtenscatter}). 

Because the bias correction simulations should match the test set simulations as 
closely as possible, the same proportions of nearby, SDSS-II, and SNLS3 SNe are simulated
with the same observing conditions as the test set. The number of SNe in the bias correction
set is approximately the same as that of the test set.

Naively, one might expect that because we know the underlying ideal model parameter distributions,
we can use those same distributions for bias correction simulations generated from our trained models. 
However, this is not the case. First, during the SALT-II model training,
the stretch parameter mean and rms are rescaled from their initial distribution to a final 
distribution of $\langle x_1 \rangle=0.0$ and $\sigma_{x_1}=1.0$. Therefore, the trained model $x_1$
distribution will not be the same as the input model $x_1$ distribution. 
Second, we wish to approximate a data-based light curve analysis as closely as possible. 
When performing bias corrections on real SN Ia distances, the underlying stretch and color
parameter distributions are unknown and must be backed out of the observed parameter 
distributions, which themselves suffer from redshift-related biases. 
Examples of this calculation can be found in ~\citet{1995NIMPA.362..487D} and 
~\citet{KesslerSDSScosmo:2009}.

To simplify our light curve analysis pipeline, 
we fit a high S/N realistic simulation with all $N$ trained model realizations and average
the results to obtain an approximation of the underlying trained model stretch and color 
parameter distribution.  
In this same vein, the individual $\alpha$ and $\beta$ values fitted by the trained models 
from the corresponding test set are used for the bias correction simulation.

Simulated bias correction SN light curves are fit with the trained model
in a manner identical to the test set.
Finally, {\tt SALT2mu} is used to determine the best-fit $\alpha$, $\beta$, and $M_0$
and to calculate distances \rawmu ~for each set.

\begin{figure}[h]
\begin{center}
\includegraphics[scale=0.37]{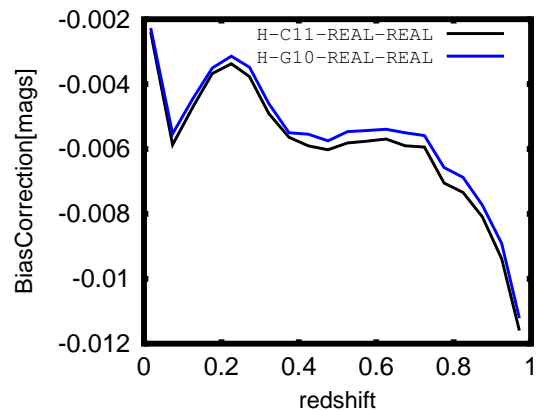} 
\caption[]{Bias correction (Equation~\ref{eq:TOTALcorr}) as a function of redshift for H-C11-REAL-REAL (black) and 
H-G10-REAL-REAL (blue) training tests. This correction is subtracted from the initial distances 
before determining best-fit cosmology parameters.
}
\label{fig:R13biases}
\end{center}
\end{figure}

As an illustration, two sets of bias corrections are shown in Figure~\ref{fig:R13biases}. 

\subsection{Cosmology}\label{sec:cosmo}
To obtain best-fit cosmology parameters, we fit the test set HD in a manner similar to the 
$Fw$CDM fits described in ~\citet{KesslerSDSScosmo:2009}. 
We assume spatial flatness $\Omega_k = 0$, but allow $w$ to differ from $-1$.
SN distance moduli are combined with Baryon Acoustic 
Oscillation (BAO) and cosmic microwave background (CMB) constraints as described below. 

For the BAO constraint, we use the quantity $A$ defined by ~\citet{2005ApJ...633..560E}, 
\begin{multline}
A(z_1;w,\Omega_M,\Omega_{\rm DE}) = 
    \frac{\sqrt{\Omega_M}}{E(z_1)^{1/3}}\\\times\left[ \frac{1}{z_1} \int_0^{z_1} \frac{dz'}{E(z')}\right]^{2/3},
\end{multline}
and for the CMB we use the shift parameter $R$
\begin{equation}
R(z_{\rm CMB}; w,\Omega_M,\Omega_{\rm DE}) = \sqrt{\Omega_M}\int_0^{z_{\rm CMB}}\frac{dz'}{E(z')} .
\end{equation} 
In both cases, 
\begin{equation}
E(z') = \left[\Omega_{M}(1+z)^3+\Omega_{\rm DE}(1+z)^{3+3w}\right]^{1/2}.
\end{equation}
Rather than taking the best-fit $A$ and $R$ values from the data
~\citep[e.g.,][]{2005ApJ...633..560E, 2009ApJS..180..330K}, 
we determine $A_{\rm calc} = 0.487$ and $R_{\rm calc} = 1.750$ 
from our SN simulation cosmology parameters 
($H=70, \Omega_M = 0.3, \Omega_{\rm DE} = 0.7, w = -1.0, \Omega_k = 0.0$)
and the experimentally determined redshifts $z_1 = 0.35$ and $z_{\rm CMB} = 1090$.
To account for our much larger test sample, we scale the 
experimentally determined uncertainties by the equivalent number of 
\Gtenmodel ~data sets  ($N \sim 70$), such that
\begin{equation}
\sigma_{\rm BAO} = 0.017/\sqrt{N}
\end{equation}
and
\begin{equation}
\sigma_{\rm CMB} = 0.019/\sqrt{N}.
\end{equation}
Tests have been performed to ensure that scaling the BAO and CMB uncertainties by the number 
of equivalent data sets $N$ produces the same result as averaging best-fit cosmology
parameters from $N$ data sets. 

Taken all together, the constraints on our best-fit cosmology are
\begin{equation}
\chi_{\rm BAO}^2 = \left[(A(z_1; w,\Omega_M) - A_{\rm calc})/\sigma_{\rm BAO}\right]^2
\end{equation}
and 
\begin{equation}
\chi_{\rm CMB}^2 = \left[(A(z_{\rm CMB}; w,\Omega_M) - R_{\rm calc})/\sigma_{\rm CMB}\right]^2 .
\end{equation}

In the absence of input SN data, these constraints yield the best-fit cosmology parameters 
$\Omega_M = 0.299 \pm 0.052$ and $w = -1.010 \pm 0.3$.

\subsection{Final Distances and Best-fit Cosmology}
Using the bias correction simulations and fits described previously (Section~\ref{sec:TOTALdetails}), 
bias corrections as a function of redshift are calculated using Equation~\ref{eq:TOTALcorr} 
and applied to the recovered test set distances \rawmu ~to obtain corrected distances \finalmu. 
The corrected HD is fit to obtain final best-fit cosmology parameters \finalw ~and \finalOMM.

\section{Training Test Evaluations}\label{sec:anal}

In this section we describe the quantities that will be used to 
evaluate the results of the training tests. 

\subsection{Quantities Derived From Training}\label{sec:anal-train}
To evaluate the performance of the training procedure, 
we begin by examining the training products themselves.  
The training process and products have been outlined in Figure~\ref{fig:pcafitschema} 
and described in Section~\ref{sec:TrainS2}. 
Color laws and color dispersions $k(\lambda$) will be compared
directly with those of the input model. 
The color dispersion residual \SCATTERresid ~is defined as
\begin{equation}
\Delta k(\lambda) = k(\lambda)_{\rm train} - k(\lambda)_{\rm input}.
\end{equation}

Similarly, the color law residual \CLresid ~is defined as 
\begin{equation}
\protect \label{eq:CLRESID}
\Delta {\rm CL}(\lambda) = {\rm CL}_{\rm train} - CL_{\rm input}.
\end{equation}
By definition, the color law is defined for SNe with $c=1.0$ (e.g., Equation~\ref{eq:S2function}). 
Because the observed SN~Ia color dispersion $\sigma_c$ is 0.1, the quantity $0.1 \times$\CLresid
~is the typical magnitude difference resulting from \CLresid, and is the quantity
we evaluate from here forward. 
 
We also calculate residuals \FLUXresid ~of the model $M_0$ components.
First, the trained and input model surface fluxes $F(\lambda)$ are integrated
over 1000-\AA ~wide box filters centered on 3500~\AA ~($U$), 4500~\AA ~($B$), and 5500~\AA ~($V$). 
The fractional flux difference is calculated for each filter independently, e.g., the $U$-band
residual $\Delta M_{0,U}$ is
\begin{equation}
\protect \label{eq:M0RESID}
\Delta M_{0,U} = \frac{U_{\rm train}-U_{\rm input}}{U_{\rm input}}.
\end{equation}
For purposes of comparison, we take the \HRK 
~model ``$M_0$'' surface to be the spectral time series with stretch 
and color both equal to zero.

\subsection{Quantities Derived from Light Curve Fitting}\label{sec:anal-fit}
For a given training test, it is interesting to compare the recovered $\alpha$,$\beta$ 
and distances $\mu$ with their input values. 
Various quantities may also be constructed from the distance moduli 
\rawmu ~and \finalmu.
These actions are described in the following sections. 

\subsubsection{Hubble Scatter}
Hubble scatter is defined as the dispersion on $\Delta\mu$, the difference between the fitted
distance modulus $\mu_{fit}$ (e.g., Equation~\eqref{eq:DM}) and the distance modulus calculated from 
the best-fit cosmological parameters.

This quantity is often used to characterize the quality of a light curve model. For a given 
SN~Ia data set, the ``best'' model is the one that yields an HD with the smallest Hubble scatter.
When calculated from HD of simulated data, the Hubble scatter may also be used as a test of the 
quality of the simulation. A high-quality simulation will yield Hubble scatter values similar 
to those observed with real data, typically from $0.10$ to $0.15$ for SALT-II light curve fits. 

For our cosmology results, we calculate Hubble scatter from the uncorrected SN distances 
\rawmu ~as described above. 
However, when evaluating regularization schemes (Section~\ref{sec:Regularization}) 
we simplify the calculation of this quantity slightly by computing the dispersion of 
$\mu_{\rm fit} - \mu_{\rm sim}$.

\subsubsection{Hubble Bias}
Hubble bias is the difference between the recovered HD and the true HD as a function of redshift. 
In this work, we use simulated data to measure Hubble bias stemming from SALT-II light curve analyses.
In other words, our Hubble bias indicates the accuracy of a measured SN~Ia distance modulus in a 
particular redshift bin.  
In the limit of a ``perfect training,''\footnote{A training performed with an infinitely large, 
high cadence, and high S/N training set} we expect that the Hubble bias should go to zero for all redshifts. 
In Section~\ref{sec:TCideal} we check that this is indeed the case. 
We also measure Hubble bias for models trained from simulations designed to mimic the \Gtenpaper
~training set in composition, S/N, and cadence.
These measurements may be used as estimates of the systematic uncertainty as a function of redshift
for HD made with the \Gtenmodel ~SALT-II model.  

For a single test supernova   SN$_i$, we calculate the average fitted distance modulus 
over all training realizations $N$ as
\begin{equation}
\protect \label{eq:meanDM}
\langle \mu_{i} \rangle \equiv \frac{1}{N} \sum_{j=1}^{N}\mu_{i,j},
\end{equation}
and define the Hubble bias as the difference between the average
distance modulus ~\eqref{eq:meanDM} and the actual distance modulus 
$\mu_{\rm sim}$
\begin{equation}
\protect \label{eq:HUBBbias}
\Delta \mu_i \equiv \langle \mu_{i} \rangle - \mu_{\rm sim,i}. 
\end{equation}

\subsubsection{Training Model Scatter}\label{sec:tmscatter}
This quantity, defined below, quantifies the reproducibility of an 
individual SN's measured distance modulus as a function of redshift. 
In this sense, training model scatter is a measure of the uncertainty in the trained 
model due to the training set sample size. 
In the limit of a perfect training, the training model scatter should approach zero in 
all redshift bins. 
Our training model scatter calculations will be used to evaluate the SALT-II model 
statistical uncertainty estimations described in \Gtenpaper. 

For a single test SN$_i$, we define the training model scatter $\sigma_{\mu_i}$ as
the dispersion of the fitted distance moduli $\mu_{(fit,i),j}$ 
(where $j$ runs over training realizations 1 to $N$) 
about the mean fitted distance modulus $<\mu_i>$:

\begin{equation}
\protect \label{eq:MUEM}
\sigma_{\mu_i} \equiv \sqrt{\frac{1}{N-1} \sum_{j=1}^N (\mu_{i,(fit,j)}- \langle \mu_{fit,i}\rangle)^2}.
\end{equation}

\section{Results}\label{sec:RESULTS}

\subsection{Test Case 1: Ideal Training}\label{sec:TCideal}

We are working under the hypothesis that given enough input data, the SALT-II training process will produce an
accurate copy of the input model,  and that light curve fits made with such a trained model will recover 
input cosmology to high accuracy. For this work, we define an ``IDEAL'' training as one in which a 
comprehensive, high-quality training set is used as input. 
In this section, we describe the test set used for these trainings and the results of our training tests.

\subsubsection{Ideal Training Set}\label{sec:idealTrainingset}
Our IDEAL trainings are based on the \HRK ~and \NZJG ~input models paired with no 
intrinsic scatter (i.e., scatter model NONE). 
We generated $N=$\Nideal ~independent realizations of the IDEAL training set, 
each with its own set of independent random seeds. 
These training sets consist of two samples: 
110 simulated nearby SNe drawn from a flat redshift distribution spanning $z\in[0.001,0.13]$, 
and 110 simulated SDSS SNe drawn from a flat redshift distribution spanning $z\in[0.01,0.4]$. 
For both samples, the photometry is of high quality, with two-day observer-frame cadence 
and S/N$\sim 2000$ at peak $B$ or $g$ band regardless of redshift. 
Spectra are of similarly high quality. 
Seven spectra are generated for each simulated supernova, 
with the first observation date randomly chosen between rest frame epochs 
$-$14 and $-$4~days, and the remaining spectra spaced every 10 rest-frame days thereafter. 
This selection mechanism results in a 
uniform distribution in phase of the observations of supernova spectra.
All spectra span rest frame wavelengths 2000 to 9200~\AA, and have S/N$\sim 1000$ 
defined on a bin size of 100~\AA. 

Given the high-quality training sets the training configuration is adjusted accordingly. 
Bands in the ultraviolet region are not de-weighted.
Because our training set spectra have no calibration errors, we turn off photometry-based spectral 
rescaling. Finally, although we implement nominal regularization, the high sampling 
of the input data makes regularization irrelevant. 

\subsubsection{Ideal Test Set}\label{sec:idealTestset}
Once training is complete, the new IDEAL models are evaluated by using them to fit large, 
similarly high-quality test sets. As with the training sets, our test sets are generated from
either the \HRK ~or the \NZJG ~input model paired with no intrinsic scatter (i.e., scatter model NONE).
These test sets (which we also call IDEAL) consist of 4000 SDSS and 4000 SNLS SNe. 
The relevant spectroscopic selection efficiencies have been applied to the data set, 
but the exposure times of the simulations have been adjusted to give photometric data with 
S/N$\sim 1000$ in the $B$-band at peak. 
Light curve fits of the SDSS test sample are performed with $ugri$ photometry; fits of the SNLS test sample are
performed with $griz$ photometry. 
The fitted light curve parameters are then used to determine the global best-fit $M_B$, $\alpha$ and $\beta$ values, and
initial SN distances \rawmu, as described earlier in Section~\ref{sec:cosmo}. 
Finally, bias corrections are applied to derive the corrected SN distances \finalmu, 
and best-fit cosmology parameters \finalOMM ~and \finalw ~are obtained. 

\subsubsection{Trained Model Residuals}
Because we have tailor-made our fiducial \NZJG ~SN~Ia model to 
match SALT-II training capabilities (see Section~\ref{sec:NZJG}), 
synchronized our simulation software with our training software, 
and trained with extensive, high S/N training sets, 
we expect that the resulting \NZJG-NONE-IDEAL-IDEAL trained models will match 
the input model well in all bands. 
Our input \HRK ~model has no input from or connection to the SALT-II model, 
so we do not  expect 
\HRK-NONE-IDEAL-IDEAL trained surfaces to reach the same level of agreement. 

\begin{figure}[th]
\epsscale{0.43}
\centering
\begin{tabular}{@{}cc@{}}
  \includegraphics[width=0.23\textwidth]{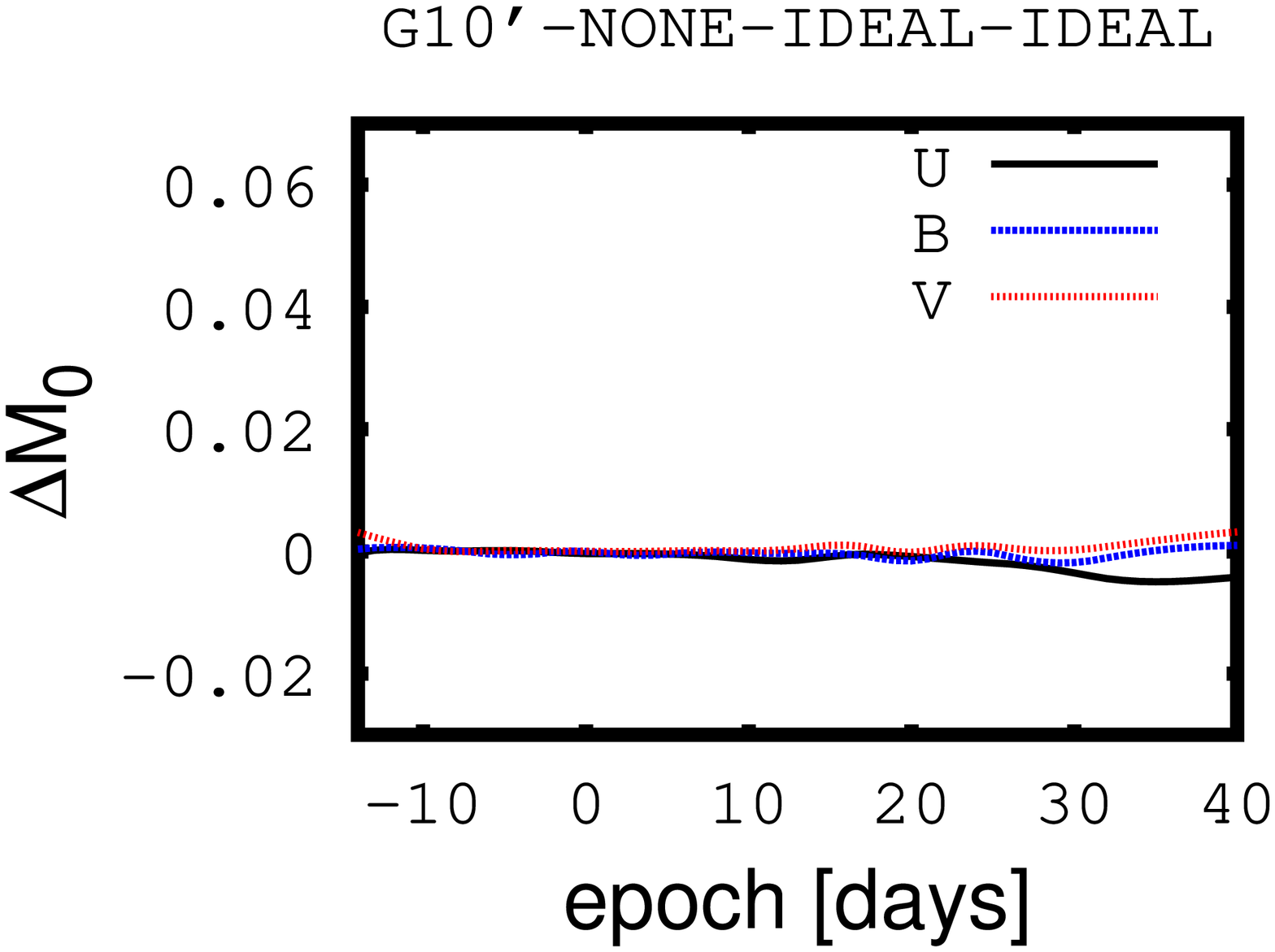} &
  \includegraphics[width=0.23\textwidth]{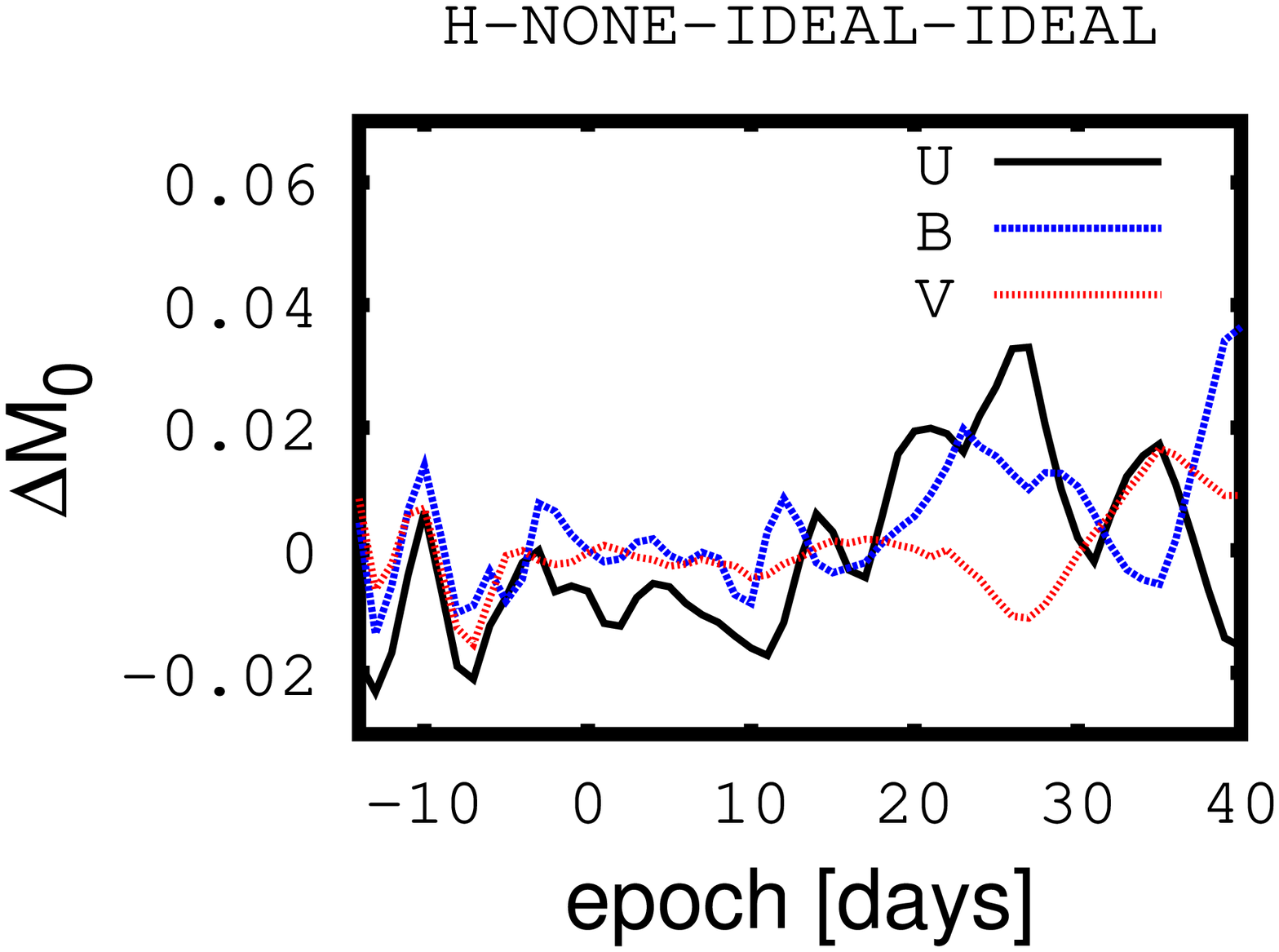} \\
  \includegraphics[width=0.23\textwidth]{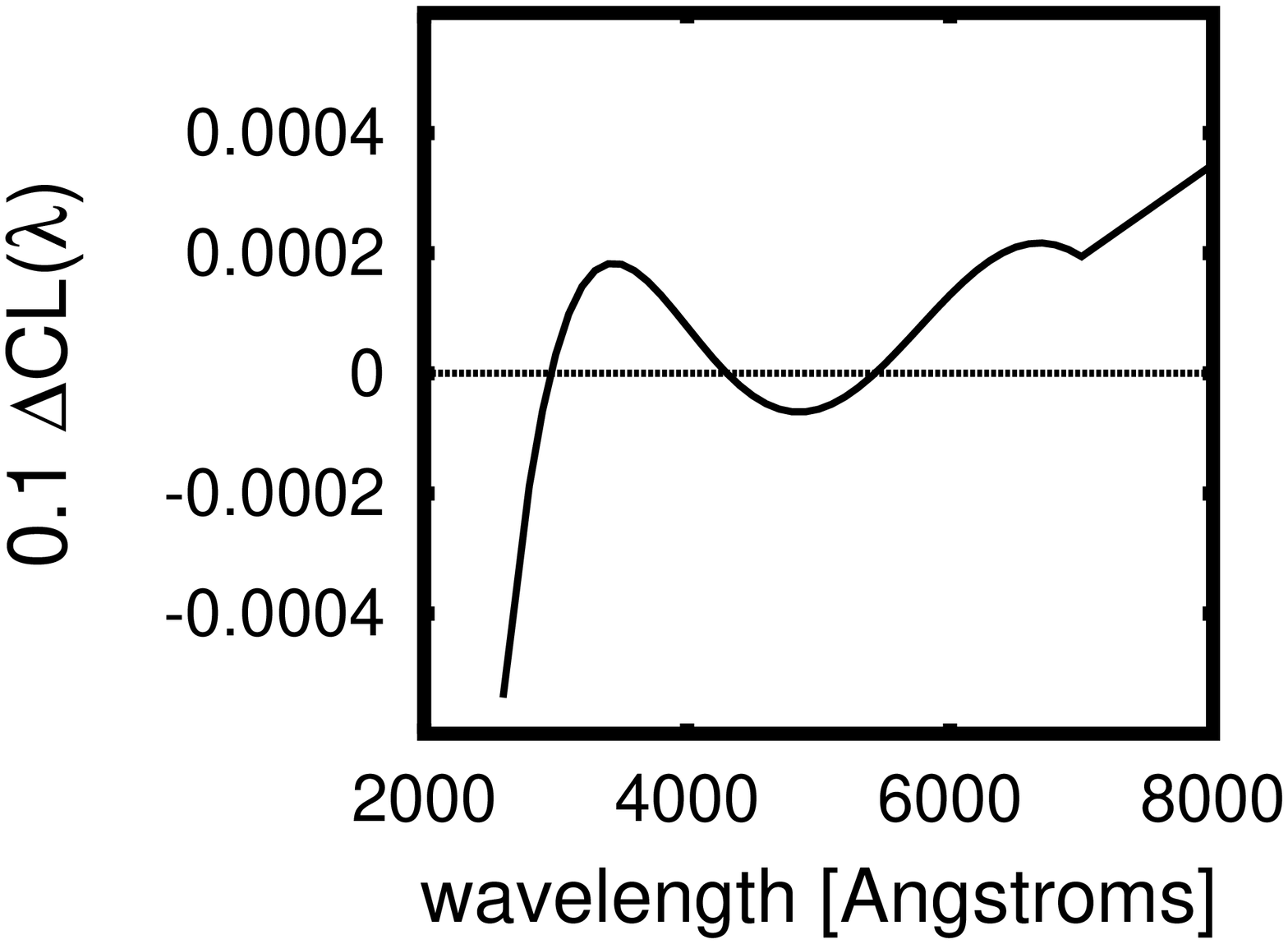} &
  \includegraphics[width=0.23\textwidth]{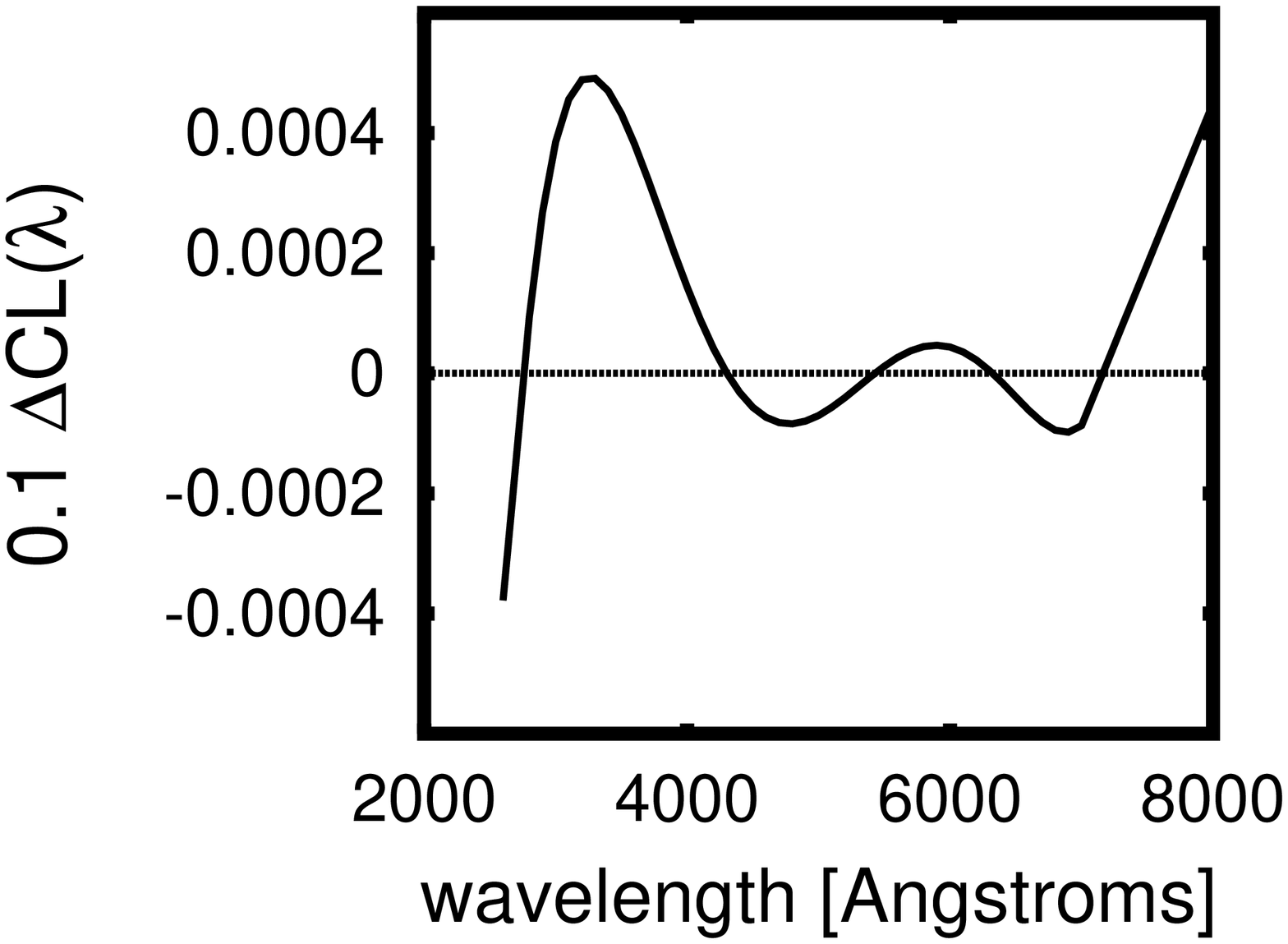} \\
\end{tabular}
\caption[../fig/IDEAL_resid.pdf]{
Mean flux (\FLUXresid, Equation~\ref{eq:M0RESID}) and color law(\CLresid, Equation~\ref{eq:CLRESID}) 
residuals from the \NZJG ~and \HRK ~ideal training tests. 
The top row shows flux residuals from the $U$(solid), $B$(dashed), and
$V$(dotted) bands. The bottom row shows color law residuals ($c=0.1$).
}
\label{fig:IDEALresid}
\end{figure}

Plots of surface residuals bear out these expectations. 
Figure~\ref{fig:IDEALresid} shows the mean integrated flux residuals and 
mean color law residuals for our ideal training tests.  

The \NZJG ~integrated flux residuals \FLUXresid ~are smaller than $0.5\%$ in flux in all bands 
except $U$ (which agrees to $1.0\%$ in flux), indicating excellent agreement between 
the input and trained model surfaces. 
Near peak luminosity, the \HRK ~model \FLUXresid ~values are similarly small ($\sim$$1.0\%$ in $UBV$ flux),
but increase to $>$$2.0\%$ in flux at early and late epochs, particularly in the bluest
wavelength ranges.  We attribute this difference to the inability of the SALT-II training code
to exactly reproduce the stretch features of the \HRK ~model. 
Color law residuals \CLresid ~are small for both models, 
no larger than 0.0005~mags for $c=0.1$. 

\subsubsection{Cosmology Residuals}

\begin{center}
\begin{deluxetable*}{lccccc}[th]
\tablewidth{0pt}
\tabletypesize{\footnotesize}
\tablecaption{Recovered Fit and Cosmology Parameters --- IDEAL TRAININGS\label{table:IDEALTRAINS}}
\tablehead{
        \colhead{\trainop}&
        \colhead{$\alpha$\tablenotemark{a}} &
        \colhead{$\beta$\tablenotemark{b}} &
        \colhead{\finalw} &
        \colhead{$\sigma_{\rm int}$} &
        \colhead{$\tilde{\chi}^2$} 
        }
\startdata
\NZJG-NONE-IDEAL-IDEAL & $0.094 \pm 0.002$ & $3.201 \pm 0.007$ & $-1.000 \pm 0.005$ & 0.000 & 0.080  \\ 
H-NONE-IDEAL-IDEAL & $0.090 \pm 0.002$ & $3.202 \pm 0.009$ & $-1.003 \pm 0.005$ & 0.000 & 0.163  \\ 
\enddata
\tablenotetext{a}{The simulated value of $\alpha$ is 0.11. 
As described in Section~\ref{sec:traindistrib} and Appendix~\ref{appxalpha}, the expected values of $\alpha$ 
are model-dependent, and typically $~\sim0.1$. }
\tablenotetext{b}{The simulated value of $\beta$ is 3.2.}
\parbox{4in}{\tablecomments{Mean cosmology parameters recovered by ideally trained models on ideal data sets. 
Model naming conventions are described in Table~\ref{table:trainopts}. All errors are errors in the mean.}}
\end{deluxetable*}
\end{center}

Each of the \Nideal trained model realizations was used to fit a single ideal set of \Nidealtest test SNe light curves.
{\tt SALT2mu} was applied to the resulting light curve parameters to obtain \Nideal sets of best-fit model values 
$M$, $\alpha$ and $\beta$, and \Nideal sets of distance moduli to use for cosmology fits. 
Assuming a flat $\Lambda$CDM cosmology, a chi-squared minimization was used to determine \finalw ~and 
\finalOMM ~for each realization. The mean recovered values of $\alpha$, $\beta$, and \finalw 
~are shown in Table ~\ref{table:IDEALTRAINS}. 

As described in Section~\ref{sec:traindistrib} and Appendix~\ref{appxalpha}, 
the rescaling of the $x_1$ parameter that occurs during SALT-II model training 
changes the value of $\alpha$ from its input value of $0.11$ to an expected
value closer to $0.10$. Both trainings recover mean $\alpha$ values somewhat smaller than the expected value:
0.094 for the \NZJG ~model and 0.090 for the \HRK ~model. 
A discrepancy in $\alpha$ is expected for the \HRK ~model, since the stretch function 
used to create the SED width--magnitude relation cannot be exactly recreated with the SALT-II linear 
combination of surfaces. 
The two training tests recovered $\beta$ and \finalw ~with high precision and within $1\sigma$ of
the input values. For our ideal test sets, the $\mu$ bias is flat as a function of redshift: both 
the raw and the bias-corrected \finalw ~are recovered within the $1\sigma$ limit of 0.005.

\begin{figure}[h]
\includegraphics[scale=0.35]{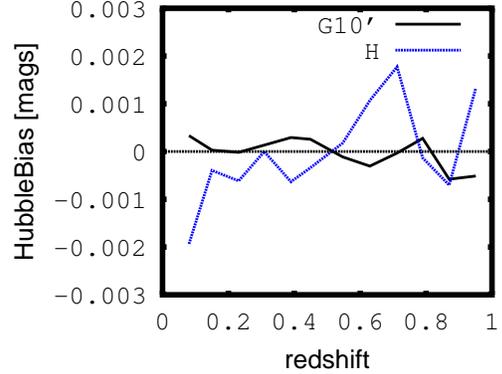}
\caption[../fig/IDEAL_TRAIN_BIAS.eps]{
Hubble bias for the \NZJG ~and \HRK ~ideal trainings evaluated with ideal test sets. 
Both sets of trainings show small bias with minimal redshift dependency. 
}
\label{fig:IDEALbias}
\end{figure}

Figure~\ref{fig:IDEALbias} shows the Hubble bias as a function of redshift for the \HRK ~and \NZJG ~trainings. 
The average distance modulus \finalmu ~differs from the true distance modulus $\mu_{\rm sim}$ by less than 
0.003~mags in all redshift bins. As with the surface residuals \FLUXresid, 
the \NZJG ~input model has smaller Hubble bias
than the \HRK ~input model. However, in either case, the bias is small 
compared to the overall precision of the measurements.

\subsubsection{Ideal Training Test Conclusions}
With the exception of slightly low recovered $\alpha$ values,
training the SALT-II model with a complete, well-sampled, high quality set of data 
results in the recovery of the input model with high accuracy. 
For both models, the mean SED broadband-integrated flux is recovered to better than $2\%$ in all 
bands for phases between $-$10 and $+$10~days of the $B$-band maximum. At late times, the maximum flux 
difference seen for the \HRK ~model is $3\%$, and for the \NZJG ~model, only $0.5\%$. 

When applied to an ideal set of test data, the ideally trained models yield Hubble biases 
smaller than 0.002~mags at all redshifts and recover \finalw ~with high accuracy. 
After applying bias corrections, the final $w$ bias is $0.003 \pm 0.005$ for the \HRK ~model 
and $0.000 \pm 0.005$ for the \NZJG ~model.

\subsection{Test Case 2: Realistic Training}\label{sec:TCreal}

Having established that the training procedure is able to reproduce the input model under ideal
conditions (i.e., full phase-space coverage by the training set, no intrinsic scatter)
we proceed to test SALT-II training with more realistic training and test data.
We define a ``REAL'' training as one in which the simulated training and test set SN~Ia 
redshift distributions, cadences, and S/N closely match the SALT-II \Gtenpaper 
~training and test set data.  
We examine the impact of regularization on trainings with realistic S/N and cadence 
but without intrinsic scatter, and select the best regularization settings to use 
for our realistic training tests. 
We then describe the results of our full realistic training tests, 
including trained model residuals, Hubble scatter, and cosmology results.

\subsubsection{Regularization Tests}\label{sec:Regularization}

As described in Section~\ref{sec:TrainS2}, ``regularization'' 
refers to the addition of extra terms to the $\chi^2$, with the
goal of reducing the amount of artificial structure on small
wavelength scales trained into 
regions of the best-fit model where input data is sparse. 
Ringing in the best-fit model adds extra scatter to light curve fits, 
thereby reducing the precision with which distances and cosmology parameters
can be measured. On the other hand, the addition of regularization terms can
systematically bias the best-fit model.
Therefore, uncertainties due to regularization must be measured and 
accounted for in the distance modulus error budget.

The forms of regularization used in this work are described in 
Section~\ref{sec:pcafitreg}. The overall strength of each term is 
determined by an arbitrary weight. 
\Gtenpaper ~evaluated the change in mean distance modulus as a function of redshift 
for two different regularization weights. 
They found the change to be small, less than or equal to 0.005~mags for redshifts below 1. 
Using our simulation techniques we reevaluate the impact of regularization on $\mu$ bias.

\begin{figure}[th]
\epsscale{1.4}
\plottwo{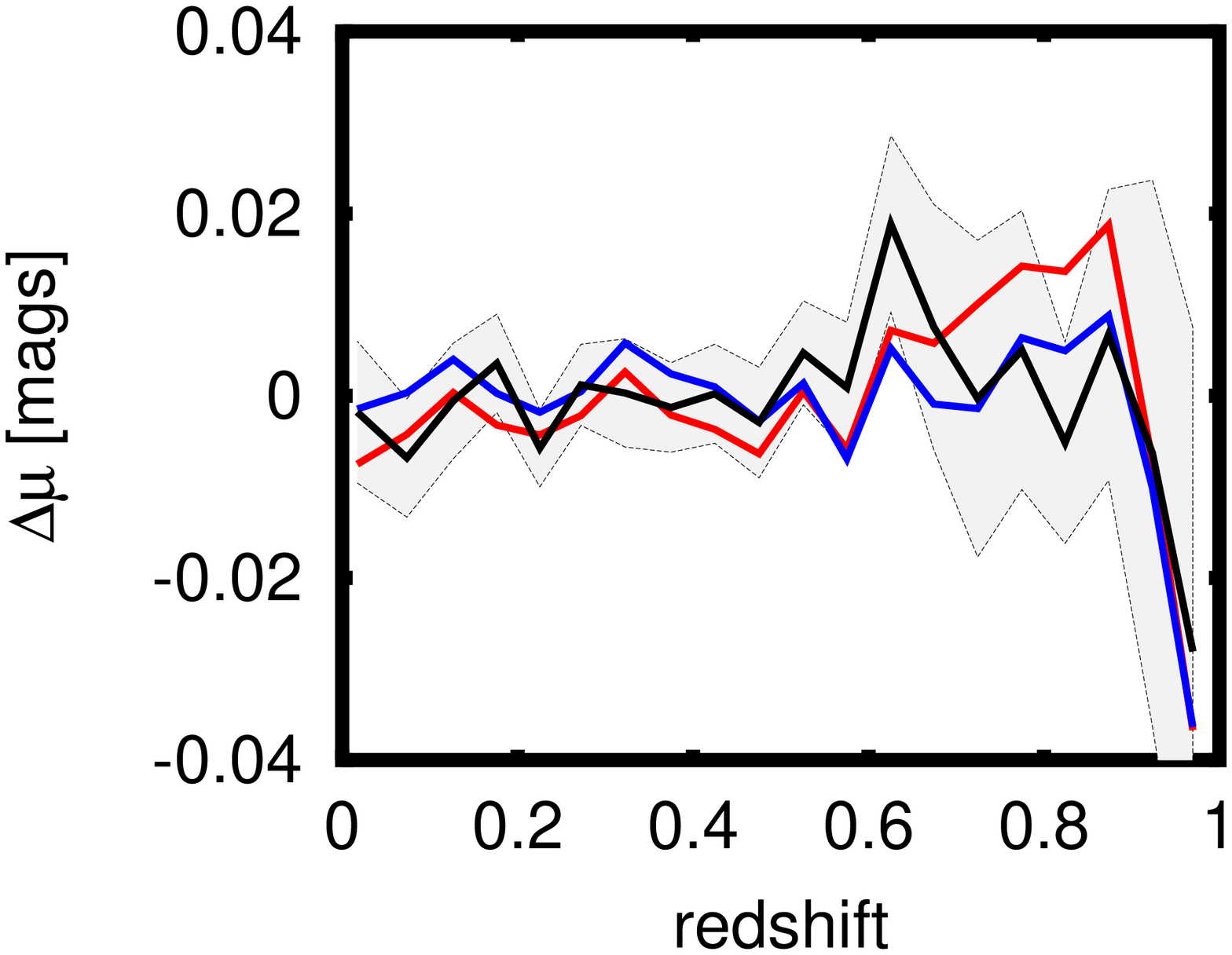}{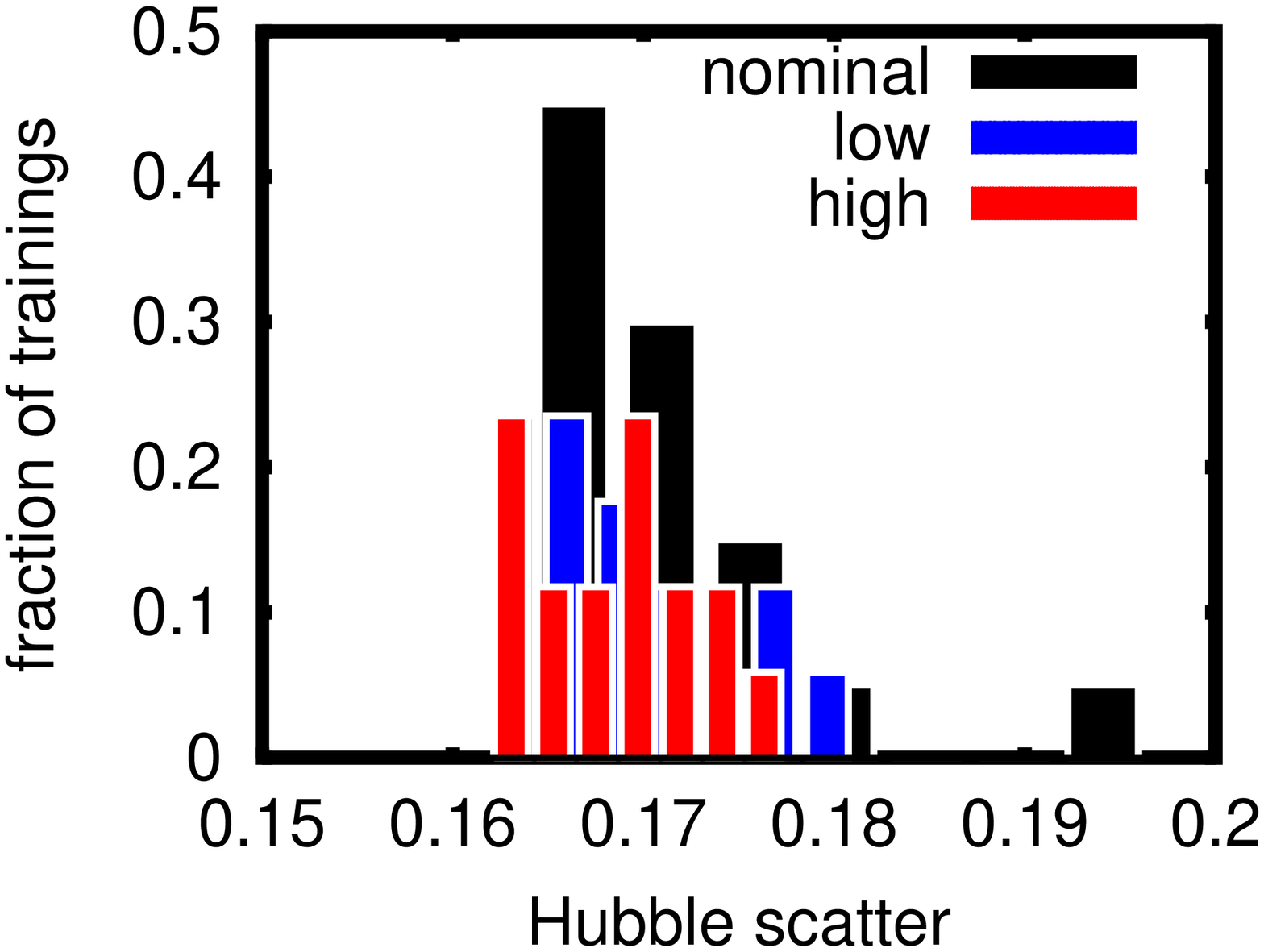}
\caption[../fig/REGTESTS_bias.eps]{ 
\HRK-C11-REAL-REAL Hubble bias and Hubble scatter (Section~\ref{sec:anal-fit}) 
as a function of regularization weight.  Top: Hubble bias $\Delta\mu$ of nominal (black), 
high ($10\times$nominal, red), and low~($0.1\times$nominal, blue)
regularization as a function of redshift. 
The gray shaded region indicates $1\sigma$ uncertainty for the nominal Hubble bias. 
Uncertainties for high and low biases are similar, and have been omitted to increase readability. 
Bottom: Hubble scatter distributions for \NZJG -NONE-REAL-REAL trainings with 
differing levels of regularization. Line colors are identical to the top plot. 
Distance modulus and Hubble scatter have been calculated with simulated values 
of $\alpha$ and $\beta$.
Model naming conventions are described in Table~\ref{table:trainopts}. 
}
\label{fig:REGSYSTEST}
\end{figure}

\paragraph{Methods}
We define nominal regularization weights as those used in the \Gtenmodel ~training, 
i.e., 10 for the gradient term and 1000 for the dyadic term.
\HRK-C11-REAL-REAL trainings have been run for three sets of regularization weights:
 ``nominal,''  ``high'' ($10\times$nominal), and ``low'' ($0.1\times$nominal).
The same regularization terms and weights are applied to both $M_0$ and $M_1$.
To determine the optimal regularization, we compare average Hubble biases and 
Hubble scatter (see Section~\ref{sec:anal-fit}) for the three trainings. 
As mentioned previously, these Hubble scatter results have been calculated with simulated 
values of alpha and beta (0.11 and 3.2, respectively), rather than with the {\tt SALT2mu} 
fitted values which change from realization to realization. 
We have checked that this simplification does not alter our conclusions. 
Results of these comparisons are shown in Figure~\ref{fig:REGSYSTEST} and Table~\ref{table:NZREGTESTS}. 

\begin{center}
\begin{deluxetable*}{lccccc}[th]
\tablewidth{0pt}
\tabletypesize{\footnotesize}
\tablecaption{Regularization Tests - Recovered Fit and Cosmology Parameters\label{table:NZREGTESTS}}
\tablehead{
        \colhead{\trainop}&
        \colhead{$\alpha$} &
        \colhead{$\beta$} &
        \colhead{\finalw} &
        \colhead{$\sigma_{\rm int}$} &
        \colhead{$N$} 
        }
\startdata
NOMINAL & $0.090 \pm 0.019$ & $2.549 \pm 0.118$ & $-1.003 \pm 0.007$ & 0.125 & 20  \\ 
$0.1\times$ & $0.094 \pm 0.016$ & $2.572 \pm 0.100$ & $-1.008 \pm 0.007$ & 0.124 & 17  \\ 
$10\times$ & $0.093 \pm 0.017$ & $2.581 \pm 0.102$ & $-1.020 \pm 0.010$ & 0.123 & 17  \\ 
\enddata
\parbox{4in}{\tablecomments{
Mean cosmology parameters recovered by realistically trained models on realistic data sets.
All errors are errors in the mean.
}}
\end{deluxetable*}
\end{center}

\paragraph{Results}

Minimal differences are observed between $0.1\times$, nominal-regularization, and $10\times$ 
Hubble biases $\Delta\mu$ (e.g., top panel of Figure~\ref{fig:REGSYSTEST}). 
Recovered $\Delta\mu$ are equivalent to within 0.005~mags for most redshifts, with the only
small discrepancy ($\sim$1-$\sigma$) seen in the redshift region $0.6-0.9$.
This finding is consistent with the regularization tests 
described in \Gtenpaper. 
As shown in the lower panel of Figure~\ref{fig:REGSYSTEST}, regularization has little effect on Hubble scatter. 
Finally, $0.1\times$ and nominal regularizations recover entirely consistent $w$ results, 
consistent with each other and consistent with the input value at the $< 1.2~\sigma$ level. 
The $10\times$ regularization yields a less consistent value of $w$, differing from the input at $2\sigma$.

Based on the $w$ results, we choose nominal regularization as the best training choice. 
Nominal regularization produces the same Hubble bias and Hubble scatter as low regularization, 
and results in a slightly more consistent recovery of $w$. 

\subsubsection{Simulations and Training Configuration}
Our realistic trainings incorporate intrinsic scatter, described by the scatter models COH, G10, and C11, 
into both the training sets and the test sets. 
For the six possible combinations of input model and scatter model (e.g., Table~\ref{table:trainopts}), 
$N=$~\Nreal independent training set realizations have been generated, 
each with its own set of random seeds. 
Recall that our REAL training sets consist of 220 simulated SNe with 
redshift distributions, cadences and S/N pattered after the low-redshift sample and the SNLS3 sample. 
Details of the light curve and spectral simulations are presented in 
Sections~\ref{subsubsec:LCSIMS} and \ref{subsubsec:SPECSIMS}. 

Training configurations are similar to those used in the \Gtenpaper ~paper. 
We use nominal regularization (see Section~\ref{sec:Regularization}) 
and a spectral recalibration step size of 800~\AA ~corresponding to the
typical width of a broadband filter. We do not de-weight bands in the ultraviolet region. 

As with the IDEAL training tests, we evaluate our REAL trained models by using them to fit large 
test sets of simulated SN~Ia light curves generated from the same input models and parameter distributions 
as the training sets. We continue to fit IDEAL test sets (as described in Section~\ref{sec:idealTestset}), 
but add to these fits and analyses of REAL test sets, which include not only realistic spectroscopic 
selection efficiencies, but also realistic photo-statistics. 

\subsubsection{Model Residuals}
Analogous to our presentation of the ideal training tests, we start by examining residuals of the trained 
surfaces and color laws with respect to the input models. 
We compare the observed spread in $\mu$ (\TMscatter, Equation~\ref{eq:MUEM}) with 
the analytic estimates made in \Gtenpaper. Finally, we examine how well input cosmology parameters 
are recovered and determine the recovered distance modulus biases as a function of redshift.

\begin{figure}[h]
\epsscale{0.3}
\centering
\begin{tabular}{@{}cc@{}}
  \includegraphics[width=0.23\textwidth]{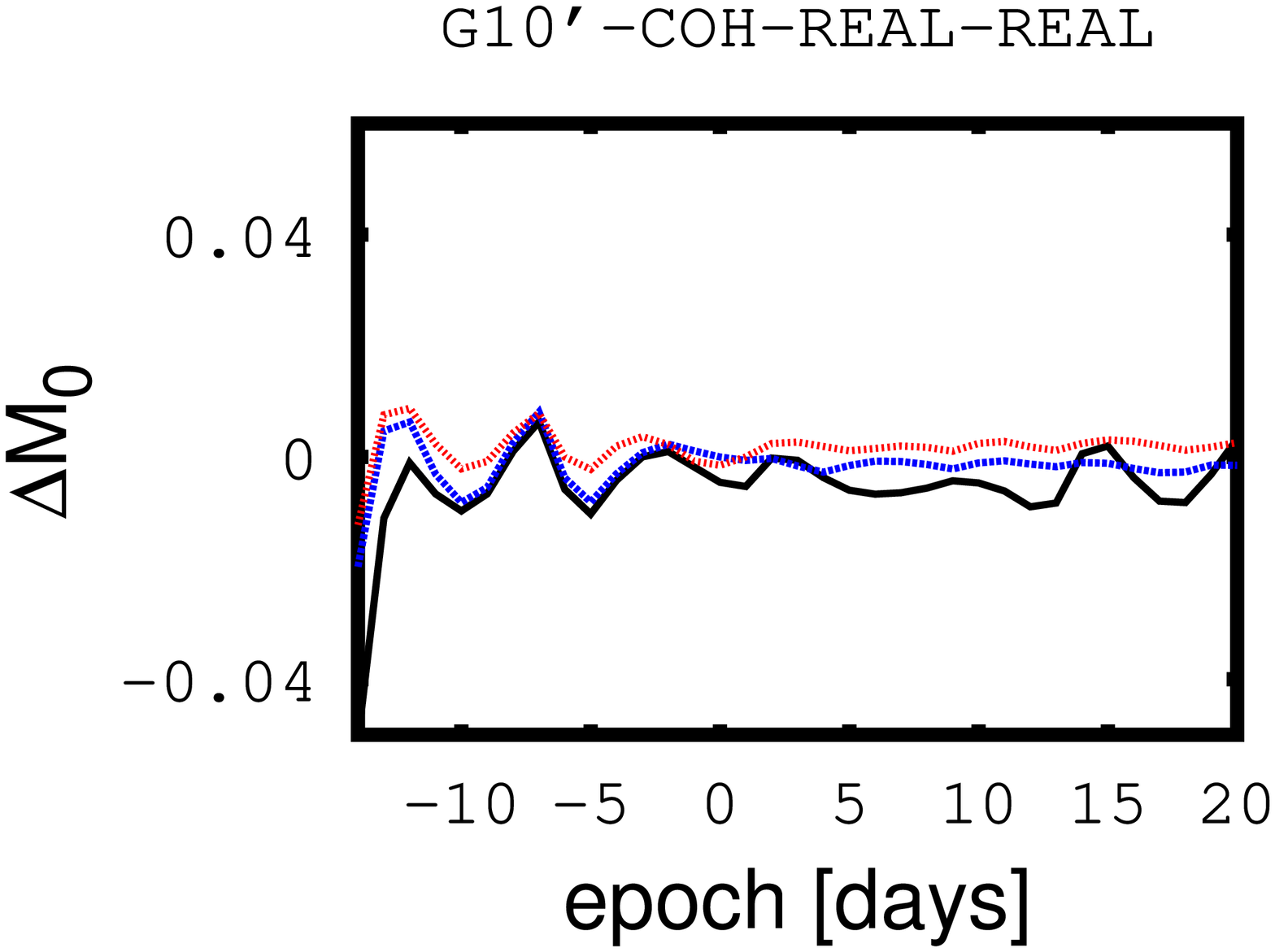} &
  \includegraphics[width=0.23\textwidth]{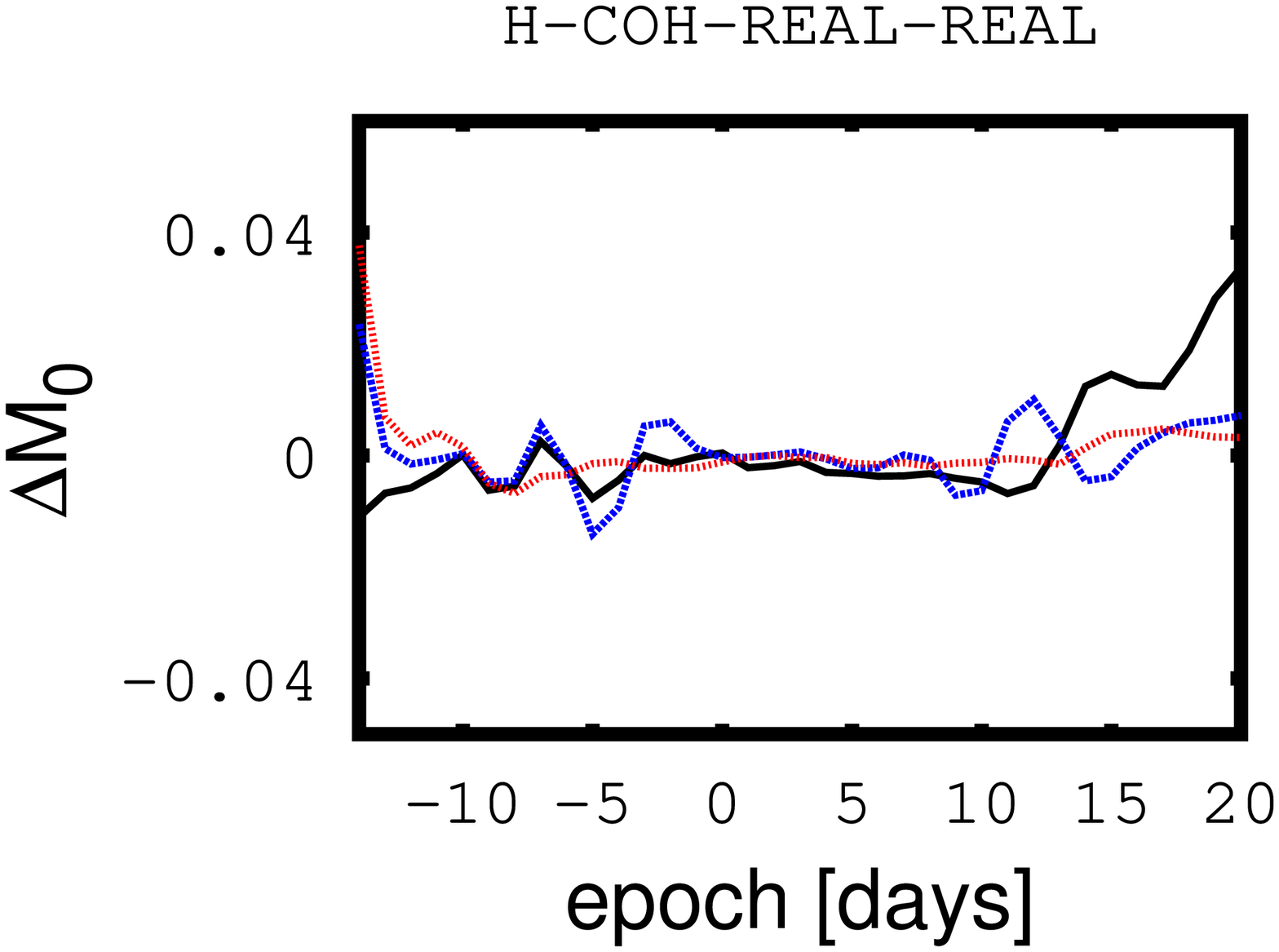} \\
  \includegraphics[width=0.23\textwidth]{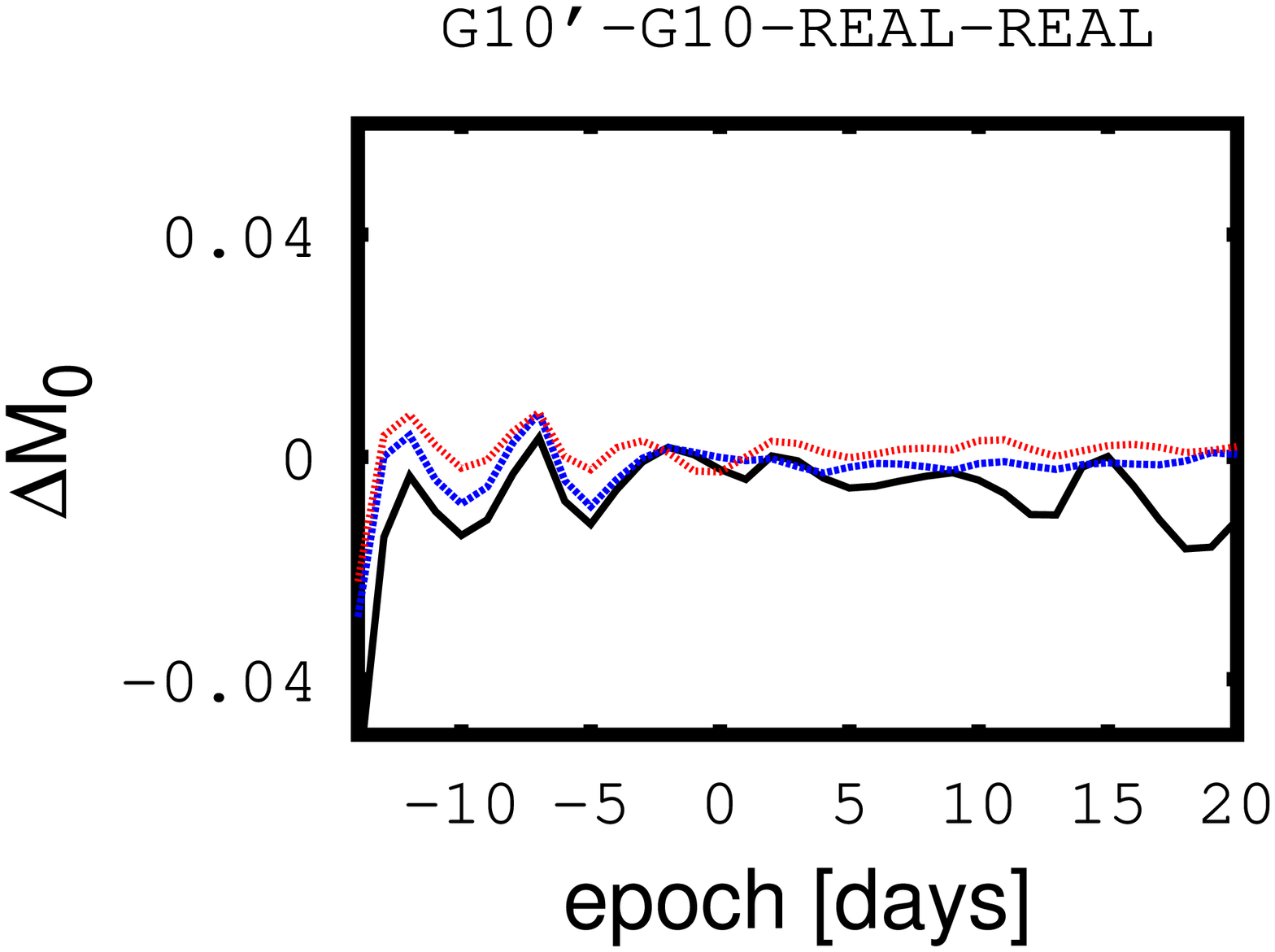} &
  \includegraphics[width=0.23\textwidth]{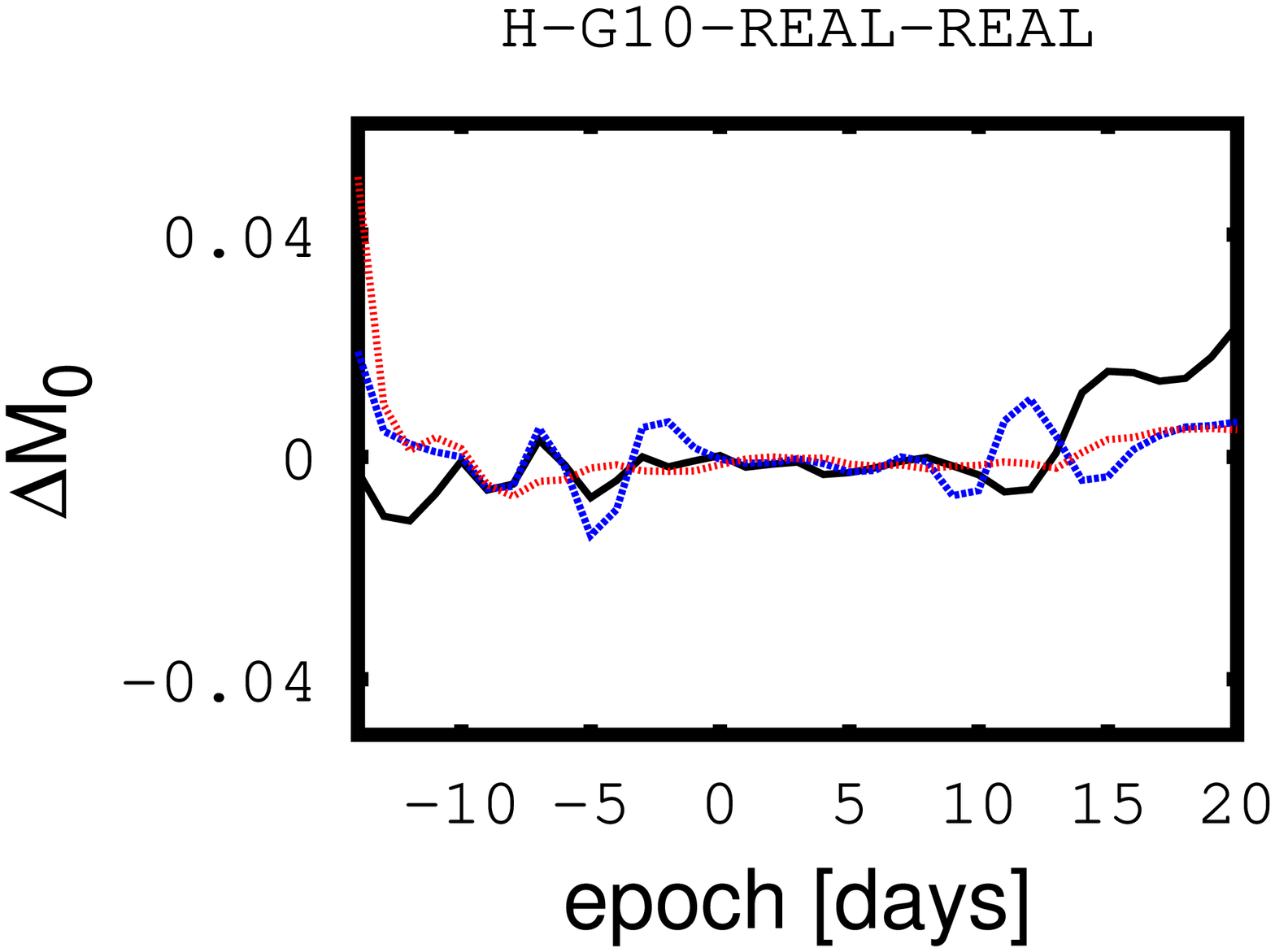} \\
  \includegraphics[width=0.23\textwidth]{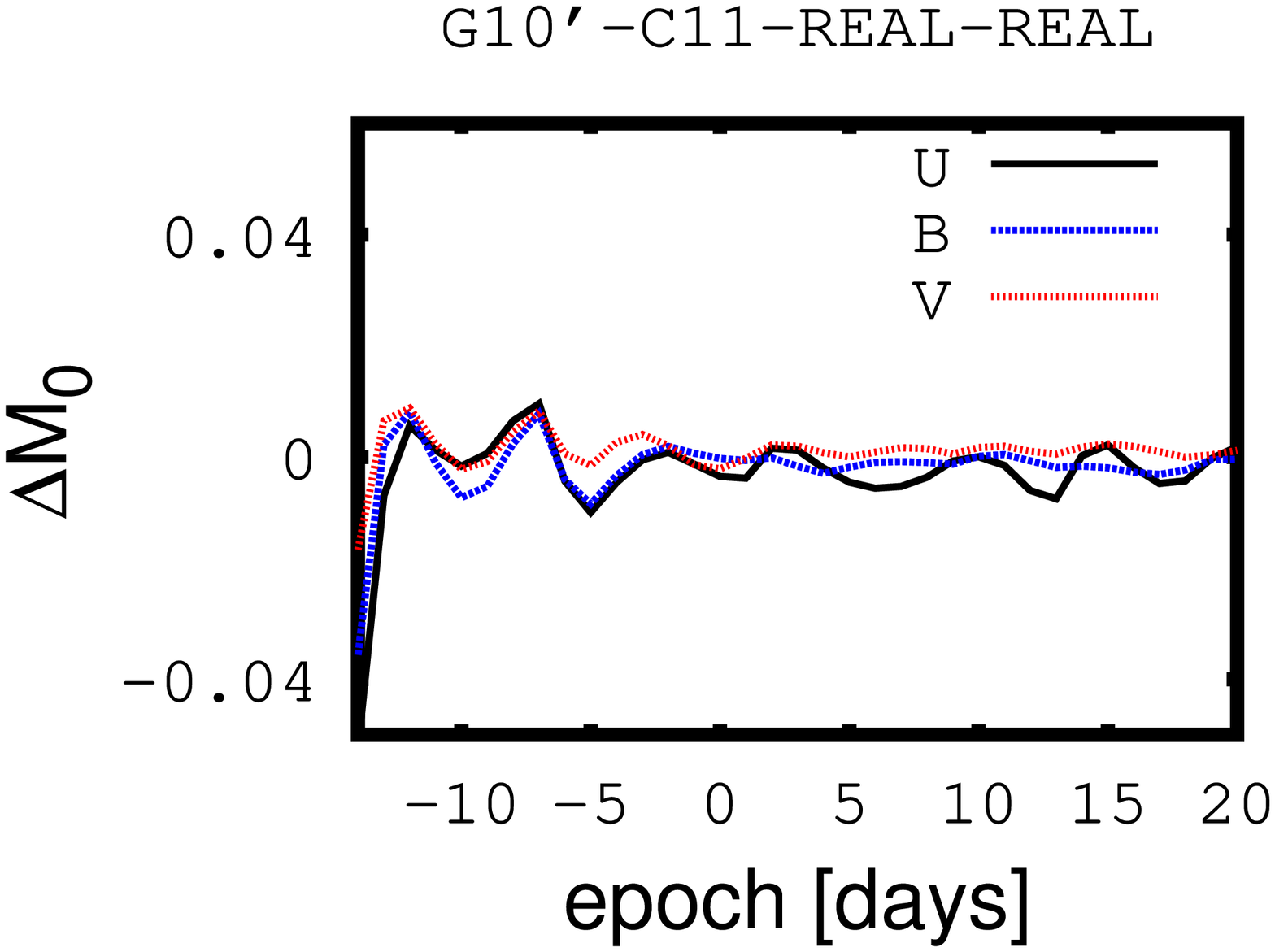} &
  \includegraphics[width=0.23\textwidth]{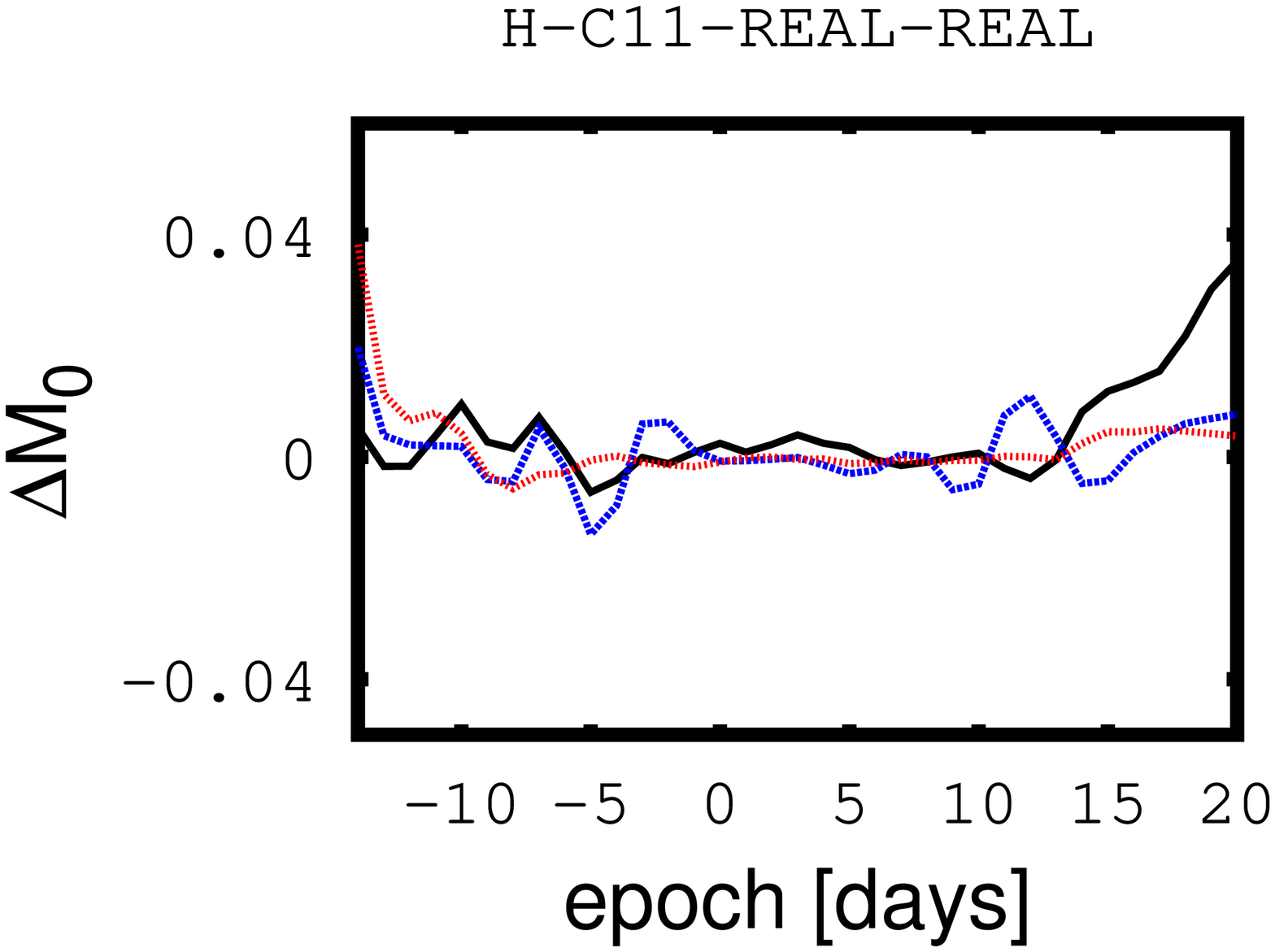} \\
\end{tabular}
\caption[]{
Mean $M_0$ surface residuals as a function of epoch. 
Plot labels indicate the input and scatter models used for each 
training test. Model naming conventions are described in 
Table~\ref{table:trainopts}.
}
\label{fig:NZ_REAL_FLUXRESID}
\end{figure}

In general, all six trainings recover the $M_0$ surface reasonably well. 
Figure~\ref{fig:NZ_REAL_FLUXRESID} shows light curve residuals 
$\Delta M_0$ as a function of phase for the six main realistic trainings.
The largest residuals, up to a $5\%$ difference in flux, are seen at early and later phases.
All six models have $5\%$ $\Delta M_0$ residuals at $p<-12$~days: the \NZJG ~input models in $UBV$, and 
the \HRK ~input models in $BV$. At later phases, $p>+15$~days, all six models agree well in $BV$. Two of 
the \NZJG ~models, \NZJG-COH-REAL-REAL and \NZJG-C11-REAL-REAL, also agree well in $U$, 
whereas the other four input models have $\Delta M_{0,U} \sim 2-4 \%$.

\begin{figure}[h]
\begin{center}
\epsscale{1.1}
\plottwo{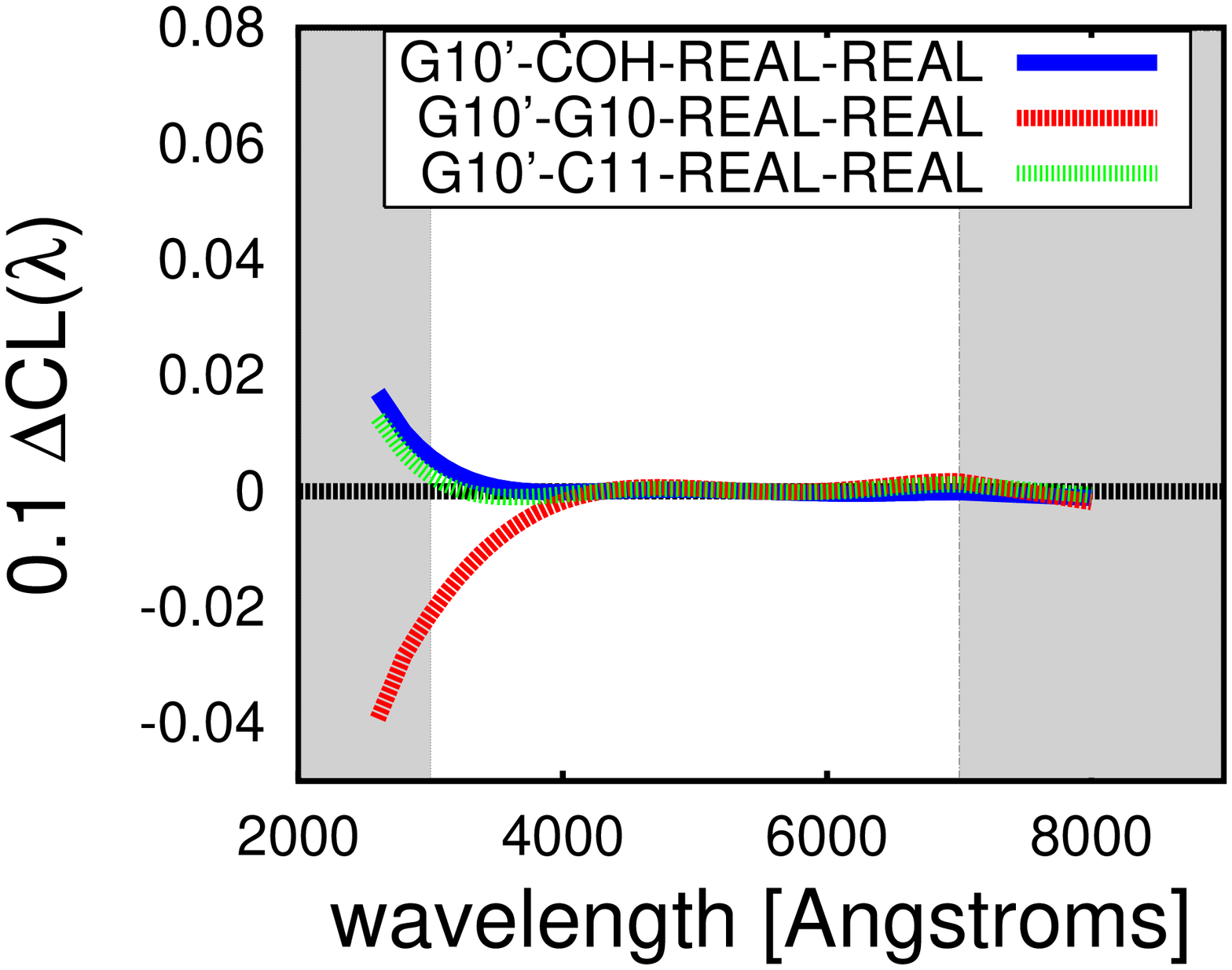}{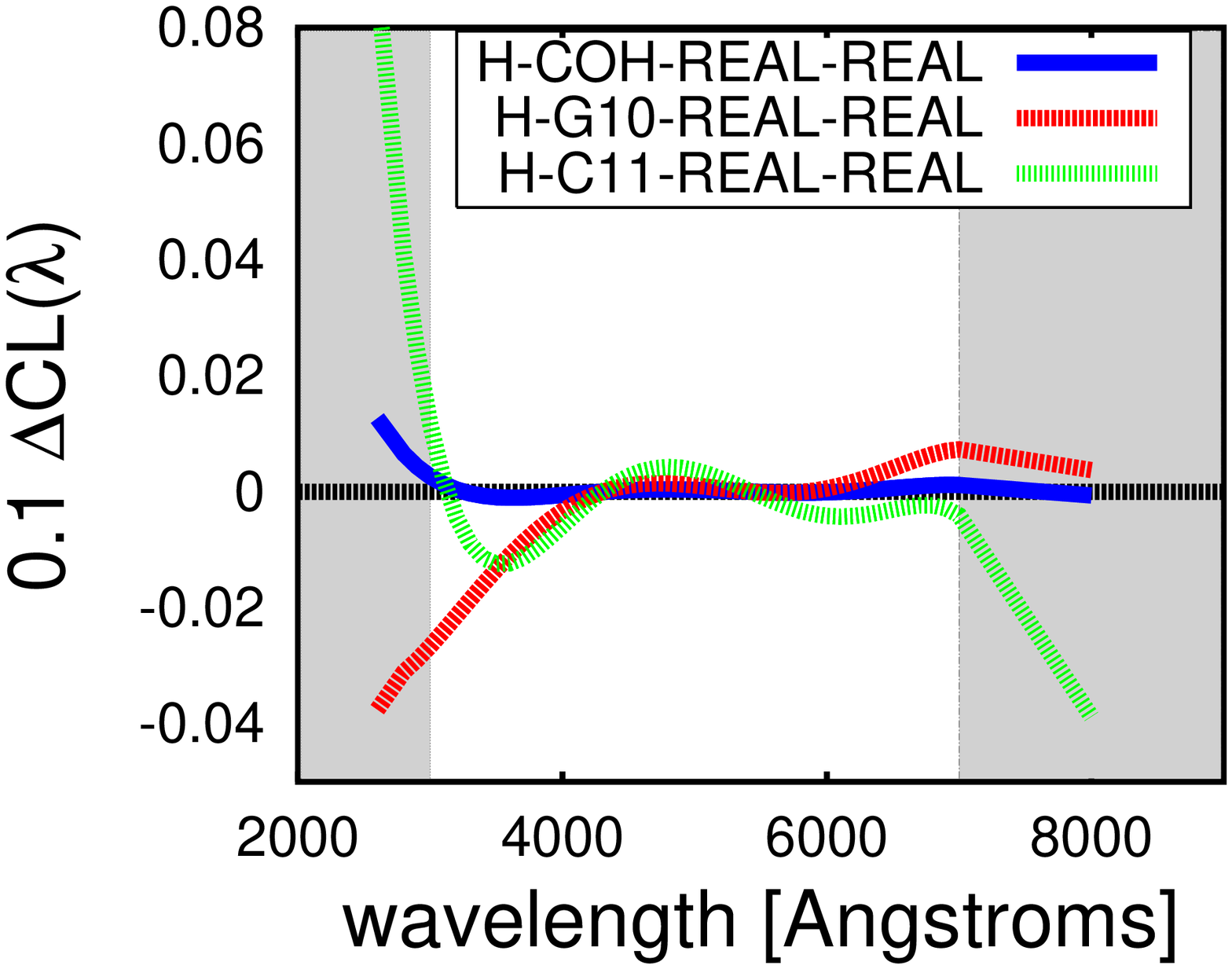}
\caption[]{
Mean color law residuals from the \NZJG (left) and \HRK (right) realistic training tests, for $c=0.1$.
Light curve data are included in test-data fits 
if observed in filters whose mean rest-frame 
wavelengths are within the unshaded wavelength region (3000-7000~\AA).
}
\label{fig:REAL_COLORLAWS}
\end{center}
\end{figure}

Color law residuals \CLresid ~are not uniformly well recovered.
As seen in Figure~\ref{fig:REAL_COLORLAWS}, the coherent scatter trainings return color laws in excellent agreement 
with the input model. However, the other scatter models show deviations from the input at wavelengths below 
3500~\AA ~and, for the C11 scatter model in particular, above 7000~\AA. 
Overall, the shapes of the color law residuals are similar for all trainings using the C11 scatter model
regardless of the input model chosen: at the smallest wavelengths, 
the recovered color laws have 15\%--25\% more extinction than the input color laws. 
The \NZJG-G10-REAL-REAL color law residual has a similar wavelength dependence as 
the \NZJG-C11-REAL-REAL color law residual, 
whereas the \HRK-G10-REAL-REAL color law residual is quite different in shape from the 
\HRK-C11-REAL-REAL color law residual; of the four trainings shown in Figure~\ref{fig:REAL_COLORLAWS}, 
\HRK-G10-REAL-REAL alone results in a trained model with
smaller blue extinction than the input model. 

\begin{figure}[h]
\epsscale{1.1}
\begin{tabular}{@{}ll@{}ll@{}}
  \includegraphics[width=0.205\textwidth]{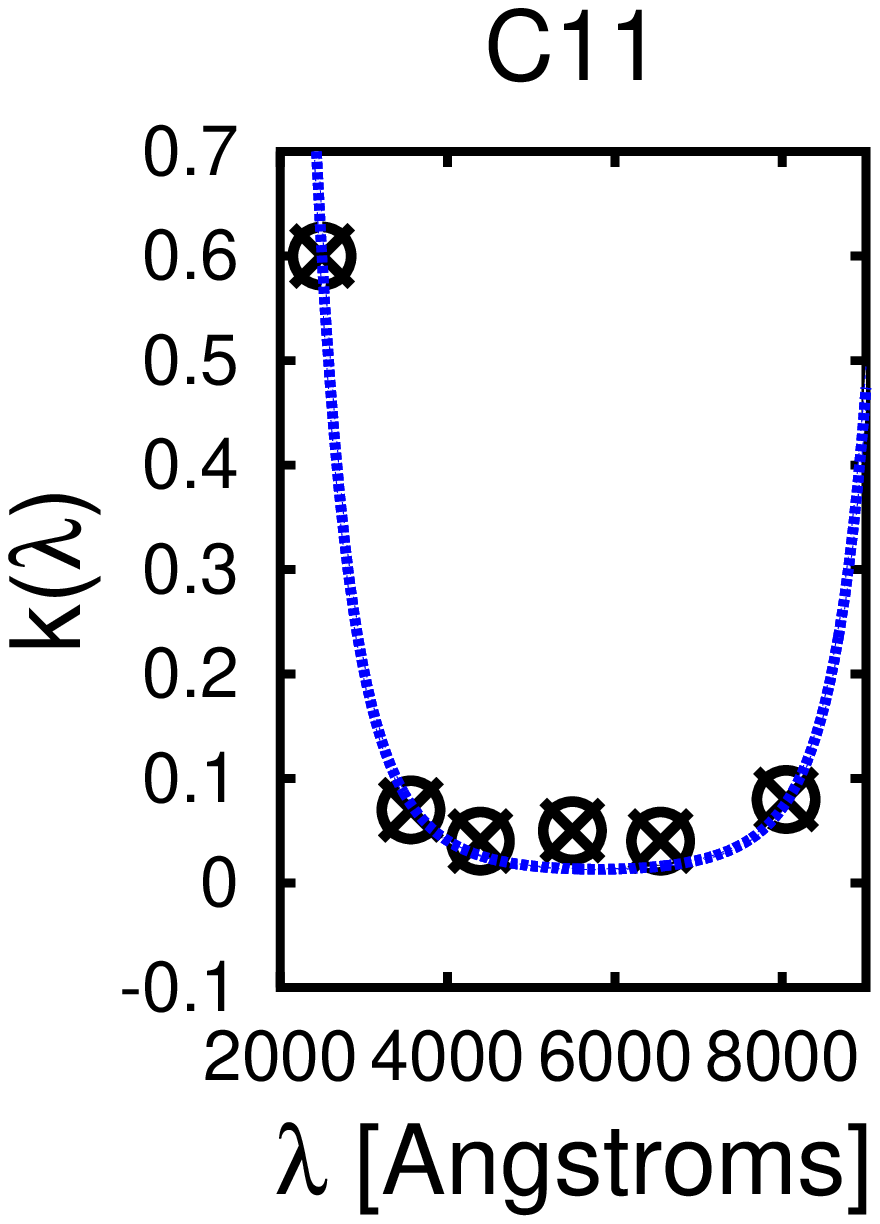}&\hspace*{-3.9em}
  \includegraphics[width=0.205\textwidth]{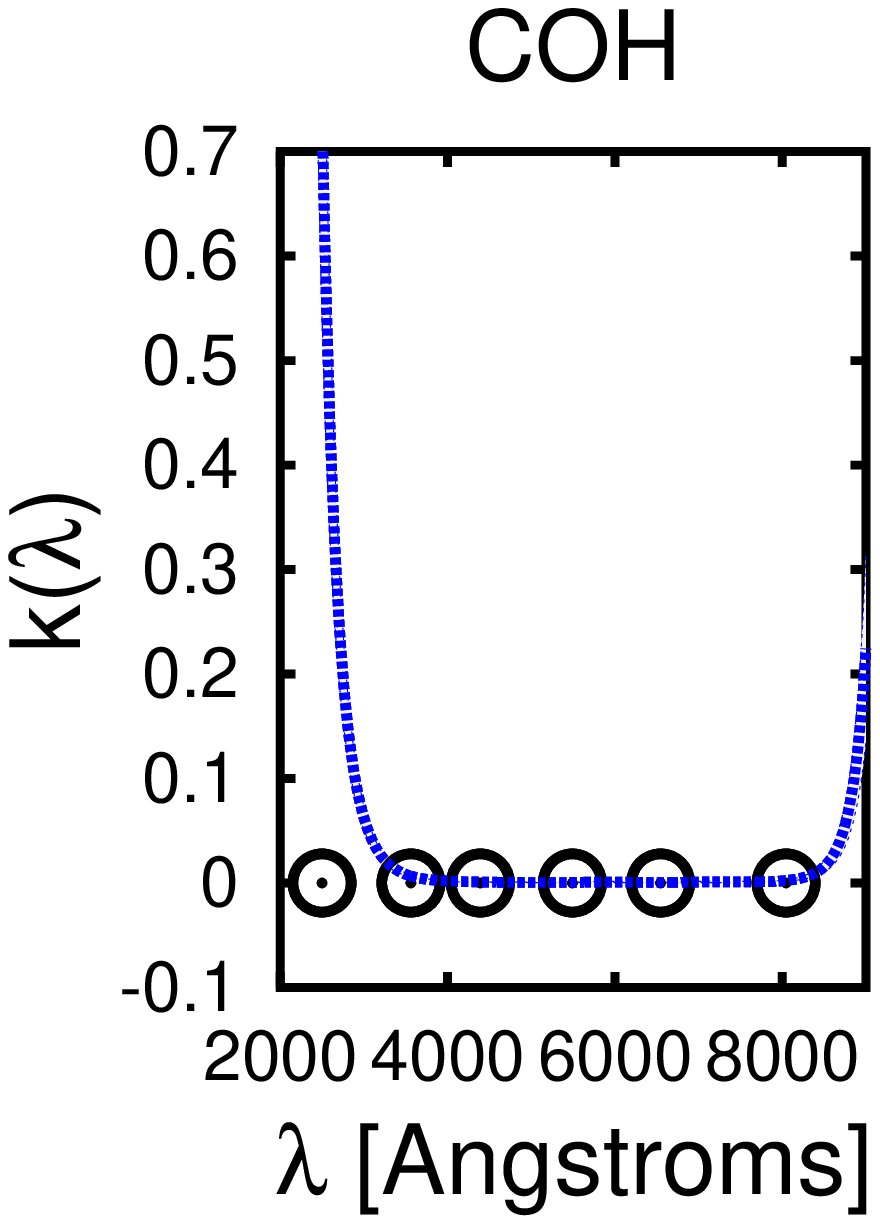}&\hspace*{-2.9em}
  \includegraphics[width=0.205\textwidth]{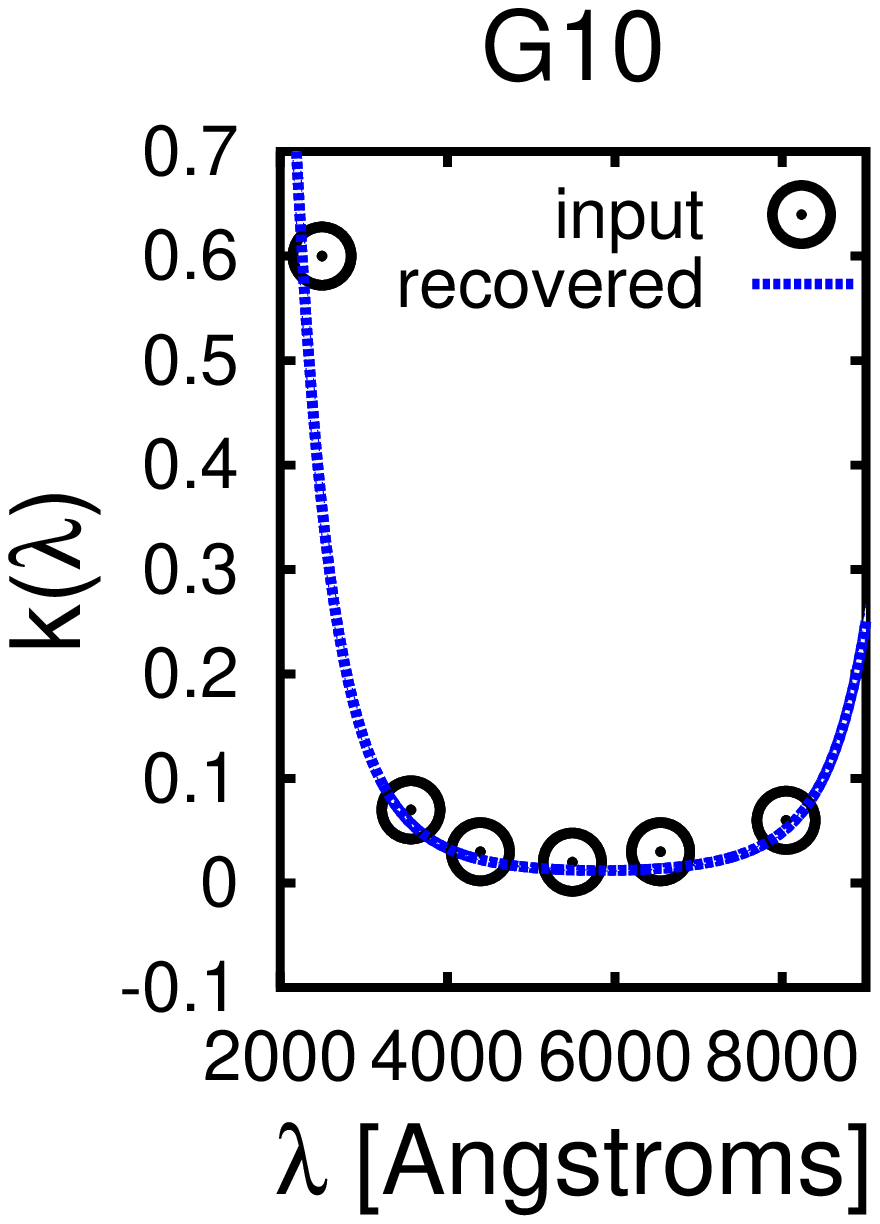} \\
\end{tabular}
\caption[]{
Input(symbols) and mean recovered broadband dispersions $k(\lambda)$ (lines) 
for the G10-scatter-based~(left), COH~(center), 
and C11-scatter-based~(right) realistic training tests. 
Only the diagonal term of the input C11 dispersion is shown. 
Model naming conventions are described in Table~\ref{table:trainopts}.
}
\label{fig:COLORDISP_COMP}
\end{figure}

Finally, we can compare the recovered broadband color dispersions $k(\lambda)$ with the input models. 
The input COH scatter model does not change SN colors (i.e., $k(\lambda) = 0$). 
For the input C11 scatter model, the diagonals of the reduced correlation matrix 
are used as a proxy for its dispersion as a function of wavelength.\footnote{See Table 4 of \Ktw.} 
Analogous quantities for the input G10 scatter model are the \Gtenpaper ~$k(\lambda)$ values at 
2500~\AA ~and at $UBVRI$ central wavelengths. The recovered dispersions are the trained 
$k(\lambda)$ functions. 
Figure~\ref{fig:COLORDISP_COMP} shows the input and recovered dispersions as a function of wavelength for our
three realistic intrinsic scatter models C11, COH, and G10. The choice of input model (\NZJG ~or \HRK)
makes very little difference to the recovered $k(\lambda)$; therefore only \HRK-based training results are displayed in
Figure~\ref{fig:COLORDISP_COMP}. 

The input broadband dispersion is recovered reasonably well at all wavelengths for the C11 scatter models. 
The recovered G10 and COH dispersions match the input dispersions above 3500~\AA ~but diverge
in the near-UV.
The $k(\lambda)$ divergences at the extreme blue and red regions 
are from inadequate spectral coverage and regularization effects. 
Although the divergences in G10 and COH are significant, the wavelength regions of the 
discrepancies are so blue (and red, in the case of COH) that their relevance to 
SN~Ia light curve fits should be minimal. 

\subsubsection{Statistical Uncertainty Estimates}
\begin{figure}[h]
\begin{center}
\includegraphics[scale=0.45]{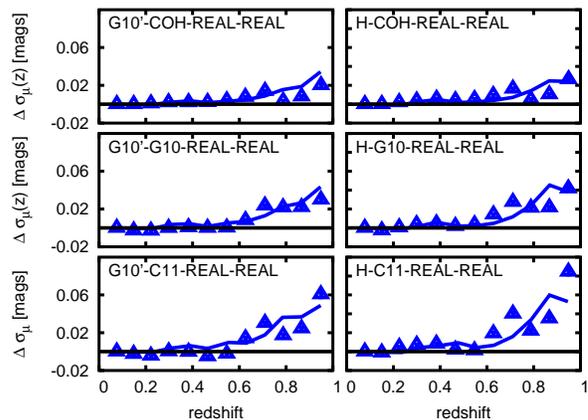} 
\caption[../fig/allREALSCAT_SCAT.eps]{
Relative training model scatter $\Delta$\TMscatter ~as a function of redshift.
Triangular symbols show scatter recovered from our training tests, and
solid lines show the mean SALT-II scatter estimates produced during the model training 
process. 
Plot labels indicate the input model and training type. 
Model naming conventions are described in Table~\ref{table:trainopts}.
}
\label{fig:REALSCAT_SCAT}
\end{center}
\end{figure}

Using multiple training set realizations, we have been able to directly measure the scatter in SN~Ia
distance measurements as a function of redshift due to training statistics. 
In Figure~\ref{fig:REALSCAT_SCAT}, we compare our \TMscatter results with estimates calculated by \Gtenpaper.
Our distance scatter measurements include both the wavelength dispersion uncertainty and 
training statistical uncertainty; thus, we have combined the analogous components of the \Gtenpaper ~estimates.
Because cosmology depends only on relative distances, we consider only the relative scatter  
$\Delta \sigma_{\mu} = \sigma_{\mu(z)} - \sigma_{\mu(z=0)}$.

Intrinsic scatter model and redshift are the key determinants of $\Delta$\TMscatter.
In general, the COH scatter model has the smallest $\Delta$\TMscatter ~and C11 has the largest 
$\Delta$\TMscatter. 
For all input scatter models $\Delta$\TMscatter ~is flat for redshifts below 0.6, and ranges from 0.00 to 0.02 mags. 
At higher redshifts, $\Delta$\TMscatter ~increases up to a maximum of 0.03 to 0.11~mags at $z=1$,
with the \NZJG-COH-REAL-REAL model having the lowest scatter and the \HRK-C11-REAL-REAL model having the most scatter.
The SALT-II ~estimates agree well with our results for all intrinsic scatter models. 

\subsubsection{Hubble Bias and Cosmology}

Fits of realistically trained models on realistic test data yield the mean recovered values of $\alpha$, 
$\beta$, and \finalw ~shown in the first six rows of Table~\ref{table:NZREAL}. 
\begin{deluxetable*}{lccccc}[th]
\tablewidth{0pt}
\tabletypesize{\footnotesize}
\tablecaption{Recovered Fit and Cosmology Parameters --- Bias Corrections Use TRAINED Model\label{table:NZREAL}}
\tablehead{
        \colhead{\trainop}&
        \colhead{$\alpha$\tablenotemark{a}} &
        \colhead{$\beta$\tablenotemark{b}} &
        \colhead{\finalw} &
        \colhead{$\sigma_{\rm int}$} &
        \colhead{$N$} 
        }
\startdata
\NZJG-COH-REAL-REAL & $0.085 \pm 0.009$ & $3.154 \pm 0.034$ & $-0.986 \pm 0.007$ & 0.133 & 13  \\ 
\HRK-COH-REAL-REAL & $0.091 \pm 0.007$ & $3.126 \pm 0.031$ & $-1.004 \pm 0.007$ & 0.131 & 12  \\ 
\NZJG-G10-REAL-REAL & $0.084 \pm 0.017$ & $2.798 \pm 0.107$ & $-1.005 \pm 0.012$ & 0.125 & 16  \\ 
H-G10-REAL-REAL & $0.089 \pm 0.016$ & $2.721 \pm 0.075$ & $-1.015 \pm 0.009$ & 0.119 & 16  \\ 
\NZJG-C11-REAL-REAL & $0.107 \pm 0.027$ & $2.587 \pm 0.059$ & $-1.024 \pm 0.010$ & 0.124 & 18  \\ 
H-C11-REAL-REAL & $0.090 \pm 0.019$ & $2.549 \pm 0.118$ & $-1.010 \pm 0.009$ & 0.125 & 20  \\ 
\enddata
\tablenotetext{a}{The simulated value of $\alpha$ is 0.11. 
As described in Section~\ref{sec:traindistrib} and Appendix~\ref{appxalpha}, the expected values of $\alpha$ 
are model-dependent, and typically $~\sim0.1$. }
\tablenotetext{b}{The simulated value of $\beta$ is 3.2.}
\parbox{4in}{\tablecomments{
Mean cosmology parameters recovered by realistically trained models on realistic data sets.
Model naming conventions are described in Table~\ref{table:trainopts}.
All errors are errors in the mean.
}}
\end{deluxetable*}

As described in Appendix~\ref{appxalpha} and shown in Table~\ref{table:expalpha}, the expected values 
of $\alpha$ are roughly $10\%$ smaller than the input value of 0.11, and vary with input model. 
Taking this effect into account, our recovered $\alpha$ parameters agree reasonably well with expectations. 

Only one of our six training tests --- \NZJG-COH-REAL-REAL --- recovers $\beta$ consistent ($<2\sigma$) with 
the input value of 3.2. Because our realistic test data suffers from selection effects, 
and because $\beta$ is fit prior to applying bias corrections, some decrease in $\beta$ is to be expected. 
However, our $\beta$ values vary with input scatter model. With an average $\beta \sim 3.14$, the COH intrinsic 
scatter trainings come closest to the input value. G10 intrinsic scatter tests are the next closest, with an average
$\beta$ of $\sim2.76$, whereas the trainings with C11 color scatter have the most discrepant values with average 
$\beta \sim 2.6$. As we will discuss further in Section~\ref{sec:biasbeta}, 
our finding is consistent with other results, including \Ktw ~and \citet{2011A&A...529L...4C}.

Also shown in Table~\ref{table:NZREAL} are our recovered  \finalw ~values. 
After corrections, all recovered \finalw ~values are slightly smaller than the input value of $w=-1.0$.
Four of the six training tests recover \finalw ~consistent with the input $w$ at the $2\sigma$ level;
the two that do not are \NZJG-COH-REAL-REAL, which recovers $w=-0.986\pm0.007$ for a 2.0-$\sigma$ difference,
and \NZJG-C11-REAL-REAL, which recovers $w=-1.024\pm0.010$ for a 2.4-$\sigma$ difference. 

\begin{figure}[h]
\includegraphics[scale=0.43]{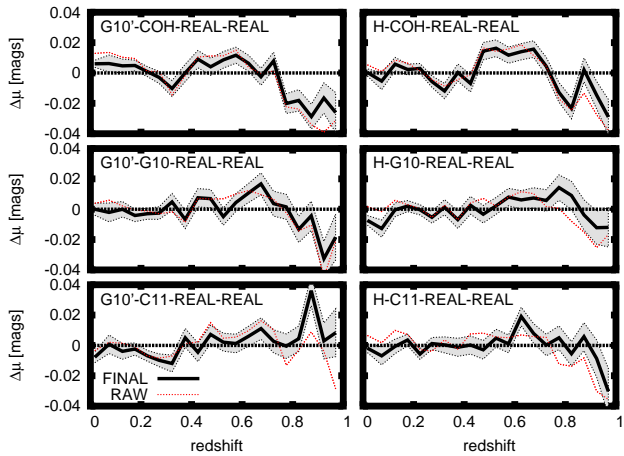}
\caption[../fig/NZJGall_REALdat_BIAS_R13.eps]{
Hubble bias plots for realistically trained models applied 
to realistic test data. 
Plot labels indicate the input model used for each training test. 
The solid line shows the final bias values, with uncertainties indicated by
the shaded gray region;
the dashed line indicates the raw (uncorrected) bias values. 
Model naming conventions are described in Table~\ref{table:trainopts}.
}
\label{fig:NZREALdatmatrix_R13}
\end{figure}

Figure~\ref{fig:NZREALdatmatrix_R13} shows distance bias $\Delta \mu$ as a function of redshift for each 
of our training configurations. 
The distance bias $\Delta \mu$ has been binned in redshift steps of $\sim0.05$. 
Both the final bias and the raw bias (before applying bias corrections) are shown. 
As expected, bias corrections reduce $\Delta \mu$, with the impact of the reductions being most significant in the 
low- and high-redshift regions ($z<0.2$ and $z>0.7$). The exception is the training \NZJG-C11-REAL-REAL, for which
bias corrections make the distance bias worse at redshifts larger than 0.8.

\section{Discussion}\label{sec:Disc}

In the following sections, we will discuss some of the implications of our residuals
and $w$ biases, examine procedures for redshift-bias correction, and remark on 
implications for optical SN~Ia cosmology and future model training.  

\subsection{Trained Model Biases}
Adding realistic intrinsic scatter to the simulated SALT-II training set yields biased models.
In general, these biases are small and are concentrated in the UV ($\lambda<4000$~\AA).  
As seen in Figure~\ref{fig:NZ_REAL_FLUXRESID}, all flux residuals \FLUXresid ~exhibit
oscillations in all bands at early and late epochs ($p<-10$~days, $p>+10$~days).
\NZJG ~$U$ flux residuals oscillate slightly at all epochs. 
In addition, \NZJG ~flux residuals decrease to $-0.04\%$ in $UBV$ at epochs less than $-$10 days, 
whereas the \HRK ~$U$ residuals increase to $+0.04\%$ at epochs greater than $+$10 days. 
Trained color laws and color dispersions also show biases in the UV. 
Three of the four realistic scatter model trainings have biased color
laws at wavelengths bluer than 4000~\AA ~(Figure~\ref{fig:REAL_COLORLAWS}), and 
the G10 scatter model training tests underestimate broadband dispersion $k(\lambda)$ 
at wavelengths bluer than 3000~\AA ~(Figure~\ref{fig:COLORDISP_COMP}). 

\begin{figure}[h]
\epsscale{1.15}
\plotone{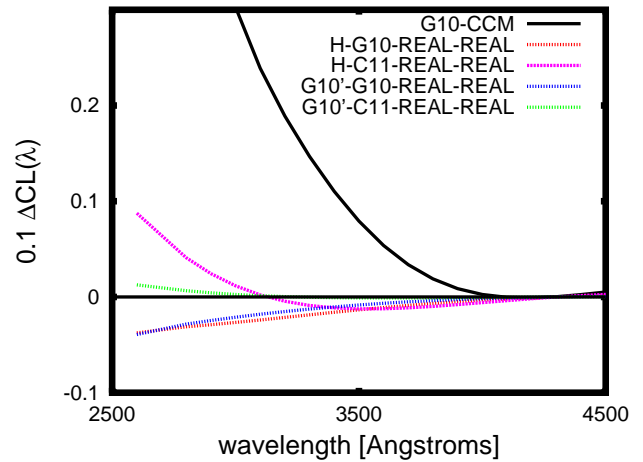} 
\caption{Mean color law differences for the four realistic-scatter trainings. 
Each set of data points shows output $-$ input CL as a function of wavelength. The solid black line
represents the difference between the G10 color law and the CCM ($R_V = 3.1$) color law as a reference. 
Model naming conventions are described in Table~\ref{table:trainopts}. 
}
\label{fig:deltaCLS}
\end{figure}

The colored lines in Figure~\ref{fig:deltaCLS}, a magnified version of the low-wavelength regions 
of Figure~\ref{fig:REAL_COLORLAWS}, present the difference between the recovered and the input color laws 
for the four realistic-intrinsic-scatter models. In the same figure, the solid black line shows the
difference  between the G10 model color law and the CCM extinction law. 
The observed color law biases from our training tests are less than 0.03 mag in the wavelength 
region most relevant for light curve fitting ($\lambda>3000$~\AA), compared with a G10--CCM 
difference of $\sim0.3$ mag in the same wavelength region.

Based on this result, we conclude that the training process itself is not responsible
for the  observed difference between G10 and CCM color laws. However, we have not eliminated the 
 possibility that the SN Ia training set is a source of this difference:
as currently configured, our simulated training set does not include systematic calibration offsets, 
nor have we attempted to simulate differences in SN Ia populations at low and high redshifts.

Both C11 scatter model training tests exhibit positive \CLresid ~biases below $\lambda=3000$~\AA.
This trend is in the same direction as the observations of \citet{2012MNRAS.426.2359M}, who found that 
the \Gtenmodel ~color law overcorrected UV spectral fluxes in comparison to the CCM color law. However, the size and
wavelength region of our bias is smaller and bluer than the effect \citet{2012MNRAS.426.2359M} describe.

\subsection{Biased Color and Biased $\beta$}\label{sec:biasbeta}

Biases in the model affect its ability to correctly fit for the color parameter 
$c$ in redshift regimes where rest-frame UV and $U$ data are important. 
In turn, incorrectly fitted $c$ parameters as a function of redshift yield 
recovered $\beta$ values that differ significantly from the input values. 
All of our recovered $\beta$ values are smaller than the input value of 
$\beta=3.2$. Training tests based on COH intrinsic scatter models recover
$\beta\sim3.14$, a value within 2-$\sigma$ of the input $\beta=3.2$
On the other hand, training tests performed with G10 intrinsic scatter models 
result in $\beta$ values of $\sim2.76$, and those based on C11 scatter models 
yield $\beta\sim2.57$, both significantly lower than the input value. 

A similar effect has been described by \Ktw, who reported scatter-model-dependent 
recovered $\beta$ values when fitting simulated light curves with 
fixed intrinsic scatter $\sigma_{M_B}$. 
\Ktw ~added realistic intrinsic scatter to SN~Ia light curves simulated from the
\Gtenmodel ~$M_0$, $M_1$, and CL. The resulting light curves were fit with  
the \Gtenmodel ~model. As we do here, \Ktw ~evaluated three types of scatter:
coherent, \Gtenmodel, and C11\footnote{\Ktw ~used two C11 variants. Their ``C11\_0''
is the one we have used in this work.}. They recovered $\beta$ 
values equal to or smaller than the input value: for coherent scatter $\beta=3.18\pm0.02$,
for \Gtenmodel ~scatter $\beta=3.23\pm0.02$, and for C11 scatter $\beta=2.86\pm0.01$. 
In other words, \Ktw ~found $\beta$ to be recovered correctly for those simulations 
generated either with COH or with the same scatter model as the model used to fit the data, 
and to be recovered as significantly smaller for those simulations generated with a different scatter
model than the model used to fit the data. 

\begin{figure}[hb]
\epsscale{1.31}
\plotone{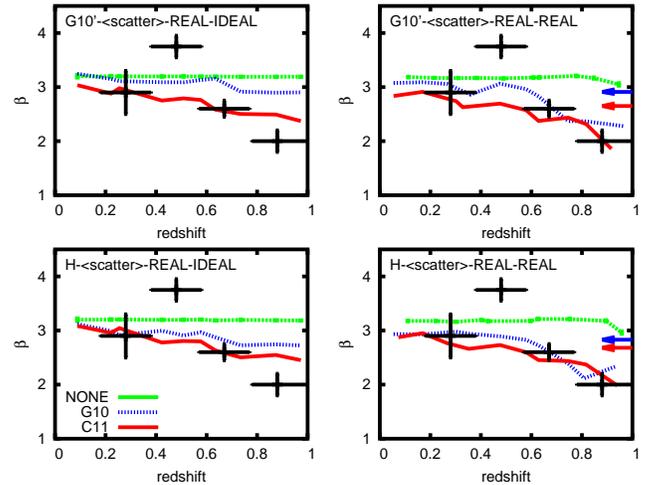}
\caption[../fig/ZBETA_ALL.eps]{
Dependence of $\beta$ on redshift for the SNLS3 data
(\Gtenpaper ~Figure 17) and one subset of our simulation-based 
training tests. The training tests for a particular panel are 
indicated by the label. The patterned lines correspond to different 
choices of scatter model (``$\langle$scatter$\rangle$''): 
``NONE'' (dotted), ``G10'' (dashed), and ``C11'' (solid). 
The $\beta$ values for ideal test sets are shown on the left;
$\beta$ values for realistic test sets are shown on the right.
Uncertainties on $\beta$ are comparable to the width of the line. 
Arrows on the right panels show the $\beta$ values recovered if 
one assumes $\beta$ is constant with redshift;
dark gray indicates G10, light gray indicates C11. 
Model naming conventions are described in Table~\ref{table:trainopts}.
}
\label{fig:ZPAR_ALL}
\end{figure}

A key difference between our work and \Ktw 
~is that we retrain the SALT-II model. Thus our test set light curves are \emph{always} 
simulated with a model different from the one used to fit them. 
In this sense, our results --- in which we find both G10 and C11 scatter model $\beta$ to be 
significantly lower than their input values --- are entirely consistent with the results of \Ktw. 
As discussed by K09 and \Gtenpaper, color law biases can result in redshift-dependent values of $\beta$
even if the underlying value of $\beta$ is constant. 
By allowing our $\beta$ values to float with redshift, this effect appears in our test samples.
We divide our TEST SNe Ia into redshift bins of width=0.1 and fit each bin individually for $\beta$. 
The resulting values of $\beta(z)$, along with the $\beta(z)$ values observed by \Gtenpaper ~from real
SNLS3 data, are shown in Figure~\ref{fig:ZPAR_ALL}.
Training tests with no added intrinsic scatter show minimal~(no) change in $\beta$ when applied to 
ideal~(realistic) test samples. When intrinsic scatter is added, 
both ideal and realistic training tests have identical $\partial\beta$/$\partial z$
until a redshift of 0.65 when Malmquist bias causes the realistic test sample $\partial\beta$/$\partial z$ 
to decrease more rapidly. With the exception of one redshift bin, 
the mean $\beta(z)$ obtained from our simulated samples is in good agreement 
with G10 SNLS3 results ~\citep[see Figure 17,][]{Guy:20103yr}. 

Because the recovered value of $\beta$ is sensitive to small biases in the color law 
and broadband dispersion,
the effects of training on the recovered value of $\beta$ should be taken into account to
determine the underlying value. 
\Gtenpaper ~recovered $\beta=3.2$; our results suggest the bias-corrected value of $\beta$ is
3.65--3.82. 

Furthermore, if light curve analysis produces the correct cosmology parameters, biases in the recovered beta values
are somewhat irrelevant. In terms of our ability to recover cosmological parameters, the significance of
color law biases is small. As shown in Table~\ref{table:NZREAL}, after simulated bias corrections are applied,
the best-fit \finalw ~values are underestimated by 0.01--0.02
depending on the choice of input scatter model. 

\begin{deluxetable*}{lccccc}[th]
\tablewidth{0pt}
\tabletypesize{\footnotesize}
\tablecaption{Recovered Fit and Cosmology Parameters -- Bias Corrections Use INPUT model\label{table:NZREAL_PRIVATE}}
\tablehead{
        \colhead{\trainop}&
        \colhead{$\alpha$} &
        \colhead{$\beta$} &
        \colhead{\finalw} &
        \colhead{$\sigma_{\rm int}$} &
        \colhead{$N$} 
        }
\startdata
\NZJG-COH-REAL-REAL & $0.085 \pm 0.009$ & $3.154 \pm 0.034$ & $-1.000 \pm 0.007$ & 0.133 & 17  \\ 
H-COH-REAL-REAL & $0.091 \pm 0.007$ & $3.126 \pm 0.031$ & $-1.018 \pm 0.007$ & 0.131 & 19  \\ 
\NZJG-G10-REAL-REAL & $0.084 \pm 0.017$ & $2.798 \pm 0.107$ & $-0.999 \pm 0.007$ & 0.125 & 16  \\ 
H-G10-REAL-REAL & $0.089 \pm 0.016$ & $2.721 \pm 0.075$ & $-1.003 \pm 0.007$ & 0.119 & 16  \\ 
\NZJG-C11-REAL-REAL & $0.107 \pm 0.027$ & $2.587 \pm 0.059$ & $-0.996 \pm 0.007$ & 0.124 & 18  \\ 
H-C11-REAL-REAL & $0.090 \pm 0.019$ & $2.549 \pm 0.118$ & $-1.003 \pm 0.007$ & 0.125 & 20  \\ 
\enddata
\parbox{4in}{\tablecomments{
Mean cosmology parameters recovered by realistically trained models on realistic data sets.
Model naming conventions are described in Table~\ref{table:trainopts}.
All errors are errors in the mean.
}}
\end{deluxetable*}

As shown in Table~\ref{table:NZREAL_PRIVATE},
generating the bias correction simulations from the input model rather than the trained 
model causes the observed $w$ biases to disappear. The sole exception is the \HRK-COH-REAL-REAL model, 
whose $w$ bias becomes worse rather than better. However, the COH scatter model is clearly unrealistic.
Recovering the correct values of $w$ for the more realistic G10 and C11
scatter models  validates our bias correction procedure, 
and points to differences between the trained and input model as being the source of the 
observed bias in $w$. In other words, if we knew the true underlying SN~Ia model and were
able to generate bias correction simulations from that model directly, our bias correction
technique would correctly compensate for redshift-dependent biases induced by the training.
Instead, we generate bias correction simulations from our imperfect trained model. 
These corrections are not quite right, resulting in a slightly biased $w$ value. 

\subsection{SNLS3 Bias Corrections}

The SNLS3 cosmology analysis ~\citep{2011ApJ...737..102S} uses a bias correction technique
which is slightly different from the TOTAL method used in this work. 
As described by ~\citet{2010AJ....140..518P}, the SNLS3 analysis inserts simulated SN light curves
 directly into SNLS survey search images and runs them through the discovery pipeline in the 
same manner as their real data. The simulations are configured such that each simulated SN~Ia's 
actual $\Delta m_B$ (including both stretch and intrinsic dispersion variations) are known a priori. 
For the set of detected real SNe Ia, the applied selection bias correction arises from the recovered 
$\langle \Delta m_B \rangle$ as a function of redshift. The simulated light curves are never fit.  

We performed our training tests using an approximation of this bias correction method
(described in Appendix ~\ref{appxMALM}). For each scatter model tested, recovered $w$ values are
indistinguishable from those determined with the TOTAL correction, and the average $w$ bias
$-1.006\pm0.005$ is consistent with our nominal result. 

\subsection{Impact of Scatter Models on HD Biases and Cosmology}\label{SNLS3bias}

After omitting results from the unphysical COH model,
we see no strong differences in the $w$ bias as a function of intrinsic scatter model. 
The $w$ bias obtained from our remaining four realistic scatter-model trainings are 
indistinguishable within the uncertainty. 
For the C11 training tests alone, the average $w$ bias is $-0.016\pm0.007$, which 
is fully consistent with the ``Nearby + SDSS-II + SNLS3'' C11\_0 
Malmquist-corrected $w$-bias of $-0.017\pm0.003$
reported by \Ktw. The average $w$ bias of our G10 training tests is $-0.011\pm0.007$, 
smaller than the $+0.001\pm0.003$ observed by \Ktw.
As stated earlier, \Ktw ~did not retrain the SALT-II model for their tests. The training process introduces
biases in the trained model which result in slight biases in $w$, even after application of the bias corrections. 
The results presented here supercede the results in \Ktw. 

~Combining the results from the remaining four realistic trainings,
we find an average $w$ bias of \resultw.

\subsection{Using these Results to Improve Constraints on Dark Energy}
We have used a series of MC samples to train the SALT-II model and measure HD biases as a function
of redshift. Our MC samples were specifically designed to match the SNLS3 SN~Ia cosmology sample 
~\citep[e.g.,][]{Conley:SNLS3yrSYSSERR2011}. 
\begin{figure}[h]
\includegraphics[scale=0.43]{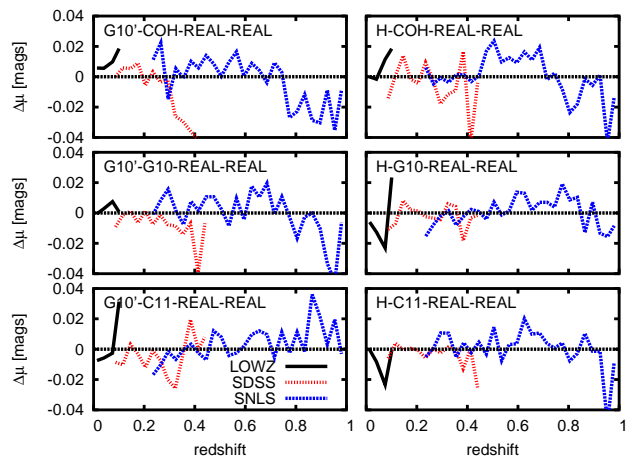}
\caption[../fig/MIDWAY_realdata_onebias_bysurvey_R13.eps]{
Hubble bias as a function of survey for realistically trained models applied 
to realistic test data. 
Plot labels indicate the input model used for each training test. 
Surveys are indicated by line type: low-z(solid), SDSS(dotted), and SNLS(dashed). 
Model naming conventions are described in Table~\ref{table:trainopts}.
}
\label{fig:NZREALdatmatrix_BYSURVEY_R13}
\end{figure}

HD bias as a function of survey is shown in Figure ~\ref{fig:NZREALdatmatrix_BYSURVEY_R13}. 
Because our HD bias measurements incorporate uncertainties 
due to the SN~Ia model, intrinsic scatter, regularization, global parameter values (e.g., $\beta$), and bias 
corrections in a self-consistent way, these survey-dependent HD bias measurements are used by
~\citet{Betoule:SNLS3cosmoupdate} (\Markpaper) 
to update SN~Ia model-related systematic error estimates 
from the most recent SNLS3 cosmology analysis ~\citep{2011ApJ...737..102S}.  

While the \Markpaper ~training and cosmology samples are not identical to the ones used here --- 
 the training sample includes SDSS photometry and the cosmology sample includes SN Ia 
light curves from the full 3-year SDSS survey --- they are similar enough to enable the transfer of 
HD bias results (e.g., Figure ~\ref{fig:NZREALdatmatrix_BYSURVEY_R13}). Our SDSS-II simulation parameters 
(e.g., cadence, imaging noise, and selection function) are drawn from the full three-year sample and the 
addition of SDSS-II light curves to the training set does not significantly alter the trained model.
Note that different training samples and/or surveys will 
produce different biases in the cosmology, thus this calculation must be adapted to 
each specific analysis.

\section{Conclusions}\label{sec:conc}

We have used simulated SN Ia samples to determine that SALT-II model training, fits, 
and bias correction of the SNLS3 cosmology sample 
published in ~\citet{2011ApJ...737..102S}, and
introduce a redshift-dependent HD bias, 
resulting in a \resultw ~bias on $w$.
Here, we have excluded the COH scatter model results, as this scatter model is clearly unphysical.
The uncertainty on $w$ is given by the rms of the remaining four input and scatter model results.
In order to perform these tests, we have upgraded
the SN analysis software package {\tt SNANA} to enable spectrum simulations incorporating realistic
photo-statistics, galaxy contamination, and intrinsic scatter. 

When the SALT-II model is trained on an SN Ia spectral time series with no intrinsic scatter 
and a high-quality MC training set, we are able to recover the input model components 
$M_0$, $M_1$, and CL, and obtain an HD with no biases. 
When the training set is made more realistic via the addition of intrinsic scatter 
to the input model and a reduction in the training set size and S/Ns, 
we predict a $\partial\beta/\partial z$ bias that agrees well with the $\partial\beta/\partial z$
profile reported in \Gtenpaper, and conclude that there is no evidence for a redshift-dependent
value of $\beta$. We have also verified that the estimates of \Gtenpaper ~accurately predict the 
statistical uncertainty in SN Ia distances $\mu$ due to the training sample size. 

We confirm the findings of \Ktw ~that the recovered value of $\beta$ depends on the intrinsic 
scatter model assumed. By adding the training process, 
we extend the work of \Ktw ~to discover that training biases \emph{both} G10 and C11 results: 
both intrinsic scatter models result in smaller recovered $\beta$ values and slightly biased, consistent
$w$ values. Our G10 training tests yield a $w$ bias ($w_{\rm input} - w_{\rm recovered}$)
of $-0.011\pm0.007$; our C11 training tests produce a $w$ bias of $-0.017\pm0.003$. 

Differences between the trained model and the underlying model alter
the measurement of individual SN Ia color parameters $c$ as a function of redshift, 
and cause the recovered color correction parameter $\beta$ to be systematically underestimated. The 
extent to which the recovered $\beta$ deviates from the input $\beta$ depends on the form of the intrinsic
scatter. 
Using G10 scatter results in $\beta$ biases of $0.45\pm0.06$; 
using C11 scatter results in $\beta$ biases of $0.62\pm0.05$.
The underlying value of $\beta$ requires a bias correction, and this correction depends
on the nature of intrinsic variations. 
\Gtenpaper ~recovered $\beta=3.2$; our results suggest that the bias-corrected value of 
$\beta$ is 3.65--3.82. 

Although the scatter models used in this work have been thoroughly tested (\Ktw) and found to 
reproduce key photometric observables such as photometric redshift residuals, Hubble scatter, and color dispersion,
they are solely dependent on wavelength, and as such are somewhat unrealistic. Intrinsic scatter 
most likely results from a combination of effects, including metallicity, viewing angle, and progenitor properties. 
Thus, in addition to the wavelength dependence assumed in our models, the intrinsic scatter may 
also depend on phase, stretch, color, and redshift. 
An improved understanding of the origins and nature of intrinsic scatter will be important for reducing 
systematic uncertainties in future SN Ia distance measurements. 

The procedure we have used in this work provides a recipe for evaluating 
systematic errors due to light curve analysis. 
Although our HD bias measurements are model and survey specific, 
our procedure is general and may be adapted to test other SN Ia models and data samples. 
Given a training set and a cosmology set of SNe Ia, a corresponding training sample simulation can be used to train 
a light curve model. Using the trained model, distances may be fitted from a corresponding simulated cosmology sample. 
Finally, HD and $w$ biases can be determined. These biases may then be incorporated into the systematic error budget of 
the data sample.

The HD biases measured in this work have been used to evaluate model-related systematic uncertainties in
a joint SDSS+SNLS cosmology analysis reported in B14.

\section{Acknowledgements}\label{sec:Ackn}

  J.F. and R.K. are grateful for the support of
  National Science Foundation grant 1009457,
  a grant from ``France and Chicago Collaborating in the Sciences''
  (FACCTS), and support from the
  Kavli Institute for Cosmological Physics at the University of Chicago.

  This work was completed in part with resources provided by the University of Chicago
  Research Computing Center.

  M.S. is supported by the Department of Energy grant DE-SC-0009890.

\begin{appendices}

\section{Expected alpha determination}\label{appxalpha}

Mixing between the input $x_1$ and $c$ parameters during the training process is not unexpected,
and will lead to predictable changes in the recovered $\alpha$ model parameter. 

The training process separates color from width by assuming that the peak $B-V$ color will be zero 
for all SNe with $c=0$, regardless of $x_1$ value. 
If the input training set shows some variation in color with $x_1$, such that the observed color
follows the form $c_{\rm obs} = c_0 + b x_1$ , the width-varying part of the color will be incorporated into the 
SALT-II model $\alpha$ term as follows:
\begin{equation}
m_B = M_B + K + \mu(z) - (\alpha - \beta b) x_1 + \beta c_0 , 
\end{equation}
where $K$, $\mu(z)$, $x_1$, and $c$ are the $k$-correction, distance modulus, width, 
and color of a specific SN~Ia, and $M_B$, $\alpha$, and $\beta$ are global SN Ia parameters.

By linearly fitting peak $B-V$ color as a function of $x1$ for each of our input models, 
and normalizing this slope with respect to the base G10 model such that $b\equiv  b - b_{G10}$ ,
we can measure the slope and predict the expected alpha value $\alpha_{\rm exp}$:
\begin{equation}
\alpha_{\rm exp} = \alpha - \beta b .
\end{equation}

We have used this technique to calculate expected values for $\alpha$ as a function of input model;
these values are shown in Table ~\ref{table:expalpha}.

\begin{center}
\begin{deluxetable}{cc}[h]
\tablewidth{0pt}
\tablecaption{Expected $\alpha$ Values\label{table:expalpha}}
\tablehead{
        \colhead{\trainop}&
        \colhead{$\alpha_{\rm exp}$}
}
\startdata
GP-NONE & 0.100 \\
H-NONE & 0.102 \\
GP-G10 & 0.079 \\
H-G10 & 0.106 \\
GP-C11 & 0.099 \\
H-C11 & 0.100 \vspace{0.2cm}
\enddata
\parbox{4in}{}
\end{deluxetable}
\end{center}

\section{Redshift-dependent bias corrections}\label{appxMALM}

As an alternative to the TOTAL technique adopted earlier in this work, 
we consider the ``MALM'' bias correction technique used by the SNLS3
cosmology analysis. As described by ~\citet{2010AJ....140..518P}, 
simulated SNe light curves are generated independently from the SNLS3 fitting models (SALT-II and SIFTO), 
inserted directly into search images, and run through the discovery pipeline in the same manner as the actual data. 
The simulations are configured such that each simulated SN Ia's actual $\Delta m_B$ 
(including both stretch and intrinsic dispersion variations) is known a priori. 
For the set of detected SNe Ia, the applied selection bias correction comes from the recovered 
$\langle \Delta m_B \rangle$ as a function of redshift. The simulated data are never fit.  

Although we are unable to implement this technique in a fit-independent manner, we approximate it here
by simulating a realistic full SN Ia sample (i.e., no efficiency cuts are applied),
fitting the sample with minimum selection cuts, and comparing the distance moduli thus derived
to the distance moduli from a ``detected'' sub-sample (i.e., those passing our efficiency cuts). 
Within this framework, we define the MALM bias as:

\begin{equation}
\protect \label{MALM}
{\rm MALM} = \langle \mu_i^{\rm fit} \rangle _{\rm DETECTED} - \langle \mu_i^{\rm fit} \rangle _{FULL}.
\end{equation}

\begin{deluxetable*}{lcccccc}[th]
\tablewidth{0pt}
\tabletypesize{\footnotesize}
\tablecaption{Recovered Fit and Cosmology Parameters -- Bias Corrections Use TRAINED Model\label{table:NZREAL_BOTH}}
\tablehead{
        \colhead{\trainop}&
        \colhead{$\alpha$} &
        \colhead{$\beta$} &
        \colhead{\finalw(TOTAL)} &
        \colhead{\finalw(MALM)} &
        \colhead{$\sigma_{\rm int}$} &
        \colhead{$N$} 
        }
\startdata
\NZJG-COH-REAL-REAL & $0.085 \pm 0.009$ & $3.154 \pm 0.034$ & $-0.986 \pm 0.007$ & $-0.980 \pm 0.008$ & 0.133 & 13  \\ 
H-COH-REAL-REAL & $0.091 \pm 0.007$ & $3.126 \pm 0.031$ & $-1.004 \pm 0.007$ & $-1.000 \pm 0.007$ & 0.131 & 12  \\ 
\NZJG-G10-REAL-REAL & $0.084 \pm 0.017$ & $2.798 \pm 0.107$ & $-1.005 \pm 0.012$ & $-1.005 \pm 0.012$ & 0.125 & 16  \\ 
H-G10-REAL-REAL & $0.089 \pm 0.016$ & $2.721 \pm 0.075$ & $-1.015 \pm 0.009$ & $-1.004 \pm 0.008$ & 0.119 & 16  \\ 
\NZJG-C11-REAL-REAL & $0.107 \pm 0.027$ & $2.587 \pm 0.059$ & $-1.024 \pm 0.010$ & $-1.025 \pm 0.010$ & 0.124 & 18  \\ 
H-C11-REAL-REAL & $0.090 \pm 0.019$ & $2.549 \pm 0.118$ & $-1.010 \pm 0.009$ & $-0.997 \pm 0.008$ & 0.125 & 20  \\ 
\enddata
\parbox{4in}{\tablecomments{
Mean cosmology parameters recovered by realistically trained models on realistic data sets. 
Model naming conventions are described in Table~\ref{table:trainopts}.
All errors are errors in the mean.
}}
\end{deluxetable*}

Table~\ref{table:NZREAL_BOTH} shows the trained-model MALM-corrected $w$ results as a function of input model
(the TOTAL-corrected $w$ have been shown alongside as a reference). 
As stated in Section~\ref{SNLS3bias}, MALM-corrected results are indistinguishable from TOTAL-corrected results.

\end{appendices}

\bibliographystyle{apj}
\bibliography{myrefs}

\end{document}